\documentclass[12pt]{article}

\usepackage{lipsum}

\usepackage[utf8]{inputenc}
\usepackage{lmodern}

\usepackage[utf8]{inputenc}
\pdfoutput=1

\usepackage{amsmath,amssymb,amsfonts,color}
\usepackage{graphicx,psfrag,epsf}
\usepackage{enumerate}

\usepackage{float}

\usepackage{algorithm,algorithmic}

\usepackage{url} % not crucial - just used below for the URL 

\def\sS{{\mathcal S}}

\def\sQ{{\mathcal Q}}
\def\sB{{\mathcal B}}
\def\sD{{\mathcal D}}

\def\sM{{\mathcal M}}

\def\sG{{\mathcal G}}

\def\vectorfontone{\bf}
\def\vectorfonttwo{\boldsymbol}
\def\va{{\vectorfontone a}}                      %
\def\vb{{\vectorfontone b}}                      %
\def\vc{{\vectorfontone c}}                      %
\def\vd{{\vectorfontone d}}                      %
\def\ve{{\vectorfontone e}}                      %
\def\vg{{\vectorfontone g}}                      %
\def\vh{{\vectorfontone h}}                      %
\def\vm{{\vectorfontone m}}                      % number of basis functions
\def\vr{{\vectorfontone r}}                      %
\def\vs{{\vectorfontone s}}                      %
\def\vv{{\vectorfontone v}}                      %
\def\vx{{\vectorfontone x}}                      % Covariates/Predictors
\def\vy{{\vectorfontone y}}                      % Targets/Labels
\def\vz{{\vectorfontone z}}                      %

\def\vone{{\vectorfontone 1}}
\def\vzero{{\vectorfontone 0}}

\def\vbeta{{\vectorfonttwo \beta}}               % Unpenalized coefficients
\def\veta{{\vectorfonttwo \eta}}                 % Vector of natural parameters
\def\vtheta{{\vectorfonttwo \theta}}             % Vector of combined coefficients
\def\vlambda{{\vectorfonttwo \lambda}}           % Vector of smoothing parameters
\def\vmu{{\vectorfonttwo \mu}}                   % Vector of means
\def\vsigma{{\vectorfonttwo \sigma}}             %
\def\vomega{{\vectorfonttwo \omega}}             %

\def\matrixfontone{\bf}
\def\matrixfonttwo{\boldsymbol}
\def\mA{{\matrixfontone A}}                      %
\def\mB{{\matrixfontone B}}                      %
\def\mC{{\matrixfontone C}}                      % Combined Design Matrix
\def\mD{{\matrixfontone D}}                      % Penalty Matrix for \vu_J
\def\mE{{\matrixfontone E}}                      %
\def\mG{{\matrixfontone G}}                      % Penalty Matrix for \vu
\def\mH{{\matrixfontone H}}                      %
\def\mI{{\matrixfontone I}}                      % Identity Matrix
\def\mK{{\matrixfontone K}}                      %
\def\mL{{\matrixfontone L}}                      % Lower bound
\def\mM{{\matrixfontone M}}                      %
\def\mQ{{\matrixfontone Q}}                      %
\def\mR{{\matrixfontone R}}                      %
\def\mS{{\matrixfontone S}}                      %
\def\mT{{\matrixfontone T}}                      %
\def\mU{{\matrixfontone U}}                      % Upper bound
\def\mV{{\matrixfontone V}}                      %
\def\mW{{\matrixfontone W}}                      % Variance Matrix i.e. diag(b'')
\def\mX{{\matrixfontone X}}                      % Unpenalized Design Matrix/Nullspace Matrix
\def\mZ{{\matrixfontone Z}}                      % Penalized Design Matrix/Kernel Space Matrix

\def\mGamma{{\matrixfonttwo \Gamma}}             %
\def\mDelta{{\matrixfonttwo \Delta}}             %
\def\mTheta{{\matrixfonttwo \Theta}}             %
\def\mLambda{{\matrixfonttwo \Lambda}}           % Penalty Matrix for \vnu
\def\mSigma{{\matrixfonttwo \Sigma}}             %
\def\mOmega{{\matrixfonttwo \Omega}}             %
\def\mPsi{{\matrixfonttwo \Psi}}                 %

\def\bE{{\mathbb E}}                             % Expectation
\def\bS{{\mathbb S}}                             % Expectation
\def\bP{{\mathbb P}}                             % Probability
\def\bR{{\mathbb R}}                             % Reals
\def\bI{{\mathbb I}}                             % Reals
\def\bV{{\mathbb V}}                             % Reals

\def\bW{{\mathbb W}}

\def\ds{\displaystyle}
 
\def\given{\,|\,}

\newtheorem{result}{Result}

\newtheorem{proof}{Proof}

\definecolor{orange}{rgb}{1,0.5,0}
\definecolor{purple}{rgb}{0.75,0,1}
\definecolor{darkgreen}{rgb}{0,0.5,0}

\newcommand{\joc}[1]{{\color{blue}#1}}

\usepackage{rotating}
\usepackage{atkinson}
\usepackage{multirow}

\usepackage{graphicx}
\usepackage[colorlinks=true,urlcolor=blue,citecolor=blue]{hyperref}
\usepackage{listings}
\usepackage{algorithm}
\usepackage[svgnames]{xcolor}
\usepackage{authblk}
\usepackage[perpage]{footmisc}
\usepackage{setspace}
\usepackage{multirow}
\usepackage{longtable}
\usepackage{threeparttable}
\usepackage[toc,page]{appendix}

\usepackage[perpage]{footmisc}

\usepackage{caption}
\usepackage[toc,page]{appendix}
\usepackage{doi}
\usepackage{titlesec}
\usepackage{xcolor}
\usepackage{tabularray}
\usepackage{colortbl}
\usepackage{array}
\usepackage{xr} 
\newcommand{\blind}{1}
\newtheorem{Lemma}{Lemma}
\newtheorem{Theorem}{Theorem}

\providecommand{\keywords}[1]
{
  \small	
  \textbf{\textit{Keywords:}} #1
}

\usepackage{geometry}
\title{\textbf{Moment Propagation}}

\makeatletter
\def\BState{\State\hskip-\ALG@thistlm}
\makeatother

\author[1]{John T. Ormerod}
\author[2]{Weichang Yu}
\author[1]{Mohammad Javad Davoudabadi\thanks{mohammad.davoudabadi@sydney.edu.au}}
\author[1]{Yuhao Li}

\affil[1]{School of Mathematics and Statistics, University of Sydney, Australia;}
\affil[2]{ School of Mathematics and Statistics, University of Melbourne,  Parkville, Australia.}

\date{}

\usepackage[
  backend=biber,
  style=authoryear,   % or whatever style the journal requires
  natbib=true,
  giveninits=true,
  maxcitenames=2,
  mincitenames=1,
  uniquename=false,
  useprefix=true
]{biblatex}

\begin{document}

\maketitle
\doublespacing

\begin{abstract}
Mean-field variational Bayes is a fast and scalable approach to approximate Bayesian inference, but its independence assumptions often lead to underestimated posterior uncertainty. We introduce moment propagation (MP), a framework for improving marginal posterior approximations by propagating conditional posterior moment information between parameter blocks and matching these moments within convenient approximating families. In conjugate settings, MP identifies variance terms omitted by mean-field approximations and uses them to construct corrected marginal updates. We develop both mean-field and Gaussian variants of MP, the latter using conditional Gaussian structure to obtain low-dimensional local updates for non-conjugate models. We establish consistency and asymptotic variance correctness for two-component models, and show that MP recovers exact marginal posteriors for linear regression and multivariate normal models. We also derive algorithms for multivariate normal models with missing data, probit regression, generalized linear models, and Bayesian Lasso regression. Numerical experiments show that MP, particularly Gaussian MP, can substantially improve marginal posterior accuracy over standard variational approximations while remaining much faster than MCMC.
\end{abstract}


\section{Introduction}
\label{sec:intro}

Bayesian inference provides a coherent framework for uncertainty quantification, but exact posterior inference is available only in relatively simple models. Markov chain Monte Carlo (MCMC) methods remain a default benchmark for posterior approximation because, under suitable conditions, they can generate asymptotically exact samples from the posterior distribution. However, MCMC can be computationally expensive for large datasets, high-dimensional or complex models, and repeated analyses. This has motivated approximate Bayesian inference (ABI) methods that trade some accuracy for substantial computational gains.

Variational Bayes (VB), also known as variational inference (VI), is one of the most widely used families of ABI methods. VB converts posterior approximation into an optimization problem by selecting a tractable class of approximating densities and finding the member of this class that is closest to the posterior distribution in Kullback--Leibler divergence \parencite{KullbackLeibler1951}; see, for example, \textcite{Bishop2006}, \textcite{Ormerod2010}, \textcite{murphy2013machine}, and \textcite{BleiEtAl2017}. Among VB methods, mean-field variational Bayes (MFVB) is especially popular because it leads to simple coordinate ascent algorithms and can scale well to large problems. MFVB assumes that the posterior approximation factorizes across a partition of the model parameters, often yielding analytically tractable updates in conjugate  models.

Despite these advantages, MFVB is well known to produce poorly calibrated uncertainty estimates. Empirical and theoretical studies have shown that MFVB often underestimates posterior variances and ignores important posterior dependence between parameters; see, for example, \textcite{WangTitterington2005} and 
\textcite{TurnerEtAl2011}. This can lead to overly narrow credible intervals, overconfident posterior rankings, and misleading downstream decisions. As a result, MFVB is often more reliable for fast model exploration and prediction than for inferential tasks requiring accurate marginal posterior uncertainty.

Several approaches have been proposed to improve deterministic Bayesian approximations. Fixed-form variational methods, including Gaussian variational Bayes (GVB), allow richer approximating families than standard mean-field assumptions \parencite{OpperArchambeau2009,JMLR:v14:challis13a}. Expectation propagation (EP) uses local approximations and moment matching to construct accurate global approximations for many models \parencite{Minka2001}. Laplace and integrated nested Laplace approximations use local curvature and low-dimensional numerical integration to approximate marginal posterior distributions \parencite{RueEtal2009}. Linear response variational Bayes (LRVB) and related covariance correction methods improve uncertainty quantification by studying the sensitivity of MFVB solutions to perturbations \parencite{GiordanoEtAl2015}. These methods have different strengths, but some are model-specific, require expensive optimization, are sensitive to implementation details, or provide corrected moments without directly constructing improved marginal approximating densities.

In this paper we introduce moment propagation (MP), a framework for improving marginal posterior approximations by propagating conditional posterior moment information between parameter blocks. The starting point is the observation that MFVB updates often behave like plug-in approximations: they replace uncertain quantities in a parameter block's full conditional distribution by their variational expectations. This is computationally convenient, but it omits terms arising from uncertainty in the conditioning variables. In conjugate block models, these missing terms can be identified directly using the laws of total expectation and total covariance. MP uses this insight to construct improved marginal approximations by matching moments obtained from local conditional posterior calculations.

The basic idea is simple. Suppose the parameter vector is partitioned into components. Rather than approximating the conditional distribution of one component given fixed expectations of the others, MP approximates the marginal posterior moments of a component by averaging conditional posterior moments over an approximation to the distribution of the relevant neighbouring components. The resulting moments are then matched to a convenient approximating family. In this way, MP retains much of the blockwise structure of MFVB while explicitly propagating information about posterior uncertainty between parameter blocks.

We develop two main versions of the method. Mean-field moment propagation (MFMP) is closest in spirit to MFVB: it approximates neighbouring components by a product of marginal approximations, but uses conditional posterior moments.
%rather than plug-in variational updates. 
This gives a direct correction to MFVB-style variance underestimation in conjugate settings. Gaussian moment propagation (GMP) starts from a joint Gaussian approximation to the posterior and uses its conditional structure to form low-dimensional local approximations to marginal posterior distributions. These local approximations are then used to update the Gaussian approximation by moment matching, providing a practical route for applying MP in non-conjugate models.

The contributions of this paper are as follows. First, we identify how MFVB can underestimate posterior variances in conjugate block models and use this analysis to motivate moment-based corrections. Second, we introduce MP as a general framework for constructing improved marginal posterior approximations from conditional posterior moments. Third, we develop MFMP and GMP algorithms, showing how they can be implemented using exact conditional moments, low-dimensional numerical integration, and delta-method approximations. Fourth, we establish theoretical properties of MP in two-component conjugate settings, including consistency and asymptotic correctness of posterior variances. Finally, we derive MP algorithms for several archetypal 
models, including linear regression, multivariate normal models, multivariate normal models with missing data, generalized linear models, and Bayesian Lasso regression.

The examples illustrate both conjugate and non-conjugate uses of MP. Linear regression and multivariate normal models provide exact-recovery benchmarks. Multivariate normal models with missing data demonstrate propagation of latent-variable uncertainty. Probit regression illustrates an auxiliary-variable setting in which not all approximating densities need to be specified explicitly. Generalized linear models and Bayesian Lasso regression demonstrate GMP in non-conjugate settings, including a model with a non-smooth prior.

Our numerical results compare MP-based methods with standard ABI methods. Across the examples considered, MP-based approximations often provide substantially more accurate marginal posterior approximations than MFVB, particularly for posterior standard deviations and marginal posterior densities, while remaining much faster than routine MCMC.

The remainder of the paper is organized as follows. Section \ref{sec:VB}
  reviews VB and identifies the sources of MFVB variance underestimation in conjugate and non-conjugate settings. Section \ref{sec:mp}
 introduces the moment propagation framework, including MFMP, GMP, and hybrid variants. Section \ref{sec:examples} derives MP algorithms for the example models. Section \ref{sec:Numerical_Experience} presents numerical experiments comparing MP with existing ABI methods. Section \ref{sec:conclusion} concludes. Technical derivations and proofs are provided in the Supplementary Material.

\section{Variational Bayes}
\label{sec:VB}

Let $\sD$ denote the observed data and let $\vtheta\in\bR^d$ denote the unknown parameters. Given a likelihood $p(\sD|\vtheta)$ and prior density $p(\vtheta)$, the posterior distribution is
$$
p(\vtheta|\sD)
=
\frac{p(\sD|\vtheta)p(\vtheta)}{p(\sD)},
\qquad
p(\sD)=\int p(\sD|\vtheta)p(\vtheta)\,d\vtheta.
$$

\noindent 
For most models of practical interest, the marginal likelihood $p(\sD)$ is unavailable in closed form and posterior approximation is required.

Variational Bayes (VB), also known as variational inference (VI), approximates $p(\vtheta|\sD)$ by choosing a tractable family of densities $\sQ = \{ q(\vtheta) \}$ and solving
$\ds q^\star = \arg\min_{q\in\sQ}
\mathrm{KL}[ q(\vtheta),p(\vtheta|\sD)]$,
where 
$
\mathrm{KL}[q(\vtheta),p(\vtheta|\sD)]
=
\bE_q[\log q(\vtheta)-\log p(\vtheta|\sD)]$.
For any $q\in\sQ$, we have
\begin{equation}\label{eq:elbo}
\log p(\sD)
=
\bE_q[\log p(\sD,\vtheta)-\log q(\vtheta)]
+
\mathrm{KL}[q(\vtheta),p(\vtheta|\sD)].
\end{equation}

\noindent 
The first term on the right-hand side of \eqref{eq:elbo} is the evidence lower bound (ELBO),
$\mbox{ELBO}_q \equiv 
\bE_q[\log p(\sD,\vtheta)-\log q(\vtheta)]$.
Since $\log p(\sD)$ does not depend on $q$, minimizing
$\mathrm{KL}[q(\vtheta),p(\vtheta|\sD)]$ is equivalent to maximizing
$\mbox{ELBO}_q$.

Two common choices of \(\mathcal Q\) are especially relevant here. The first is the mean-field family,
$$
q(\vtheta)
=
\prod_{k=1}^K q_k(\vtheta_k),
$$

\noindent 
where $\{ \vtheta_k\}_{k=1}^K$ is a partition of $\vtheta$. This leads to mean-field variational Bayes (MFVB). The second is a fixed-form family $q(\vtheta;\veta)$, where the approximating density is specified up to a finite-dimensional variational parameter $\veta$. A common fixed-form choice is a multivariate Gaussian density, often referred to as Gaussian variational Bayes. Intermediate choices, combining structured factorizations and parametric approximations, are also possible \parencite{WangBlei2013,RohdeWand2016}.

The mean-field assumption is computationally attractive because it often leads to simple coordinate ascent updates. However, the factorization removes posterior dependence between parameter blocks and is a major source of inaccurate posterior uncertainty. The purpose of this section is to make this loss explicit and to identify the moment corrections that motivate the moment propagation framework introduced in Section~\ref{sec:mp}.

Let $M(k)\subseteq\{1,\ldots,K\}\setminus\{k\}$ denote the indices of the parameter blocks in the Markov blanket of $\vtheta_k$. Then
$p(\vtheta_k|\sD,\vtheta_{-k})
=
p(\vtheta_k|\sD,\vtheta_{M(k)})$,
where $\vtheta_{-k}$ denotes all blocks except $\vtheta_k$. We refer to this conditional distribution as the full conditional distribution of $\vtheta_k$ and write it as
$p(\vtheta_k|\mathrm{rest})
\equiv p(\vtheta_k|\sD,\vtheta_{M(k)})$.
This conditional distribution plays a central role in both Gibbs sampling and MFVB coordinate ascent updates.

MFVB is particularly convenient when the full conditionals belong to exponential families. Suppose that
$p(\vtheta_k|\mathrm{rest})
=
h_k(\vtheta_k)
\exp\left[
\mT_k(\vtheta_k)^T \veta_k(\vtheta_{M(k)})
-
A_k\{\veta_k(\vtheta_{M(k)})\}
\right]$,
where $\mT_k(\vtheta_k)$ is a vector of sufficient statistics, 
$\veta_k(\vtheta_{M(k)})$ is the natural parameter, $A_k(\cdot)$ is the log-partition function, and $h_k(\vtheta_k)$ is the base measure. The natural parameter may depend on $\sD$, but we suppress this dependence in the notation.
For fixed $\{ q_j(\vtheta_j) : j\neq k\}$, the coordinate ascent variational update is
$q_k^\star(\vtheta_k)
\propto
\exp\left[
\bE_{q(\vtheta_{-k})}\{\log p(\sD,\vtheta)\}
\right]$.
In the exponential-family case  
\begin{equation}\label{eq:vb_update}
q_k^\star(\vtheta_k)
=
h_k(\vtheta_k)
\exp\left[
\mT_k(\vtheta_k)^T \widehat{\veta}_k
-
A_k(\widehat{\veta}_k)
\right],
\qquad \mbox{where} \qquad 
\widehat{\veta}_k
=
\bE_q[\veta_k(\vtheta_{M(k)})].
\end{equation}

\noindent 
Thus, the MFVB update replaces %the random natural parameter
$\veta_k(\vtheta_{M(k)})$ by its expectation under the current variational approximation. Sequentially applying \eqref{eq:vb_update} for $k=1,\ldots,K$ yields coordinate ascent variational inference and monotonically increases the ELBO under the usual regularity conditions.

\subsection{Variance underestimation in conjugate MFVB updates}

The plug-in nature of \eqref{eq:vb_update} helps explain why MFVB often underestimates posterior variances. Let
$\vg_k(\veta_k)=\nabla A_k(\veta_k)$, and 
$\mH_k(\veta_k)=\nabla^2 A_k(\veta_k)$.
For an exponential-family distribution with natural parameter 
$\veta_k$, these quantities are the mean and covariance of the sufficient statistics:
$\bE[\mT_k(\vtheta_k)] =\vg_k(\veta_k)$ and 
$\bV[\mT_k(\vtheta_k)] =\mH_k(\veta_k)$.
Therefore, under the MFVB update in \eqref{eq:vb_update},
$\bE_q[\mT_k(\vtheta_k)]
=
\vg_k\left[
\bE_q\{\veta_k(\vtheta_{M(k)})\}
\right]$, and 
\begin{equation}\label{eq:moments_vb}
\bV_q[\mT_k(\vtheta_k)]
=
\mH_k\left[
\bE_q\{\veta_k(\vtheta_{M(k)})\}
\right].
\end{equation}

By contrast, the exact posterior moments satisfy the laws of total expectation and total covariance. Conditional on $\vtheta_{M(k)}$, the full conditional distribution has natural parameter 
$\veta_k(\vtheta_{M(k)})$. Hence,
$\bE[\mT_k(\vtheta_k)|\sD]
=
\bE_{p(\vtheta_{M(k)}|\sD)}
\left[
\vg_k\{\veta_k(\vtheta_{M(k)})\}
\right]$,
and
\begin{equation}\label{eq:moments_exact}
\bV[\mT_k(\vtheta_k)|\sD]
=
\bE_{p(\vtheta_{M(k)}|\sD)}
\left[
\mH_k\{\veta_k(\vtheta_{M(k)})\}
\right]
+
\bV_{p(\vtheta_{M(k)}|\sD)}
\left[
\vg_k\{\veta_k(\vtheta_{M(k)})\}
\right].
\end{equation}

The first term in \eqref{eq:moments_exact} is the average conditional covariance of the sufficient statistics. The second term is the variation in the conditional mean induced by posterior uncertainty in the neighbouring blocks.
Comparing \eqref{eq:moments_vb} and \eqref{eq:moments_exact} identifies the source of variance underestimation. If 
$q(\vtheta_{M(k)})$ is a good approximation to $p(\vtheta_{M(k)}|\sD)$, then the MFVB quantities
$$
\vg_k\left[
\bE_q\{\veta_k(\vtheta_{M(k)})\}
\right]
\qquad\text{and}\qquad
\mH_k\left[
\bE_q\{\veta_k(\vtheta_{M(k)})\}
\right]
$$

\noindent 
can be interpreted as plug-in or first-order delta-method approximations to
$$
\bE_{p(\vtheta_{M(k)}|\sD)}
\left[
\vg_k\{\veta_k(\vtheta_{M(k)})\}
\right]
\qquad\text{and}\qquad
\bE_{p(\vtheta_{M(k)}|\sD)}
\left[
\mH_k\{\veta_k(\vtheta_{M(k)})\}
\right],
$$

\noindent 
respectively. However, the MFVB covariance in \eqref{eq:moments_vb} has no analogue of the second term in \eqref{eq:moments_exact}, i.e.,
$\bV_{p(\vtheta_{M(k)}|\sD)}
\left[
\vg_k\{\veta_k(\vtheta_{M(k)})\}
\right]$.
This missing term is nonnegative definite and represents the uncertainty propagated from the neighbouring parameter blocks into the conditional mean of $\mT_k(\vtheta_k)$. Whenever 
$\mT_k(\vtheta_k)$ contains $\vtheta_k$ itself, the same mechanism leads directly to underestimation of the marginal posterior variance of $\vtheta_k$.
This decomposition is the main conjugate motivation for moment propagation. Rather than treating the natural parameter 
$\veta_k(\vtheta_{M(k)})$ as fixed at its variational expectation, one can attempt to average conditional moments over an approximation to the posterior distribution of the neighbouring blocks. This restores, at least approximately, the second term in the total covariance decomposition.

\subsection{A conditional identity for non-conjugate updates}

The preceding argument relies on the full conditional distribution being available in a tractable exponential-family form. For non-conjugate models, the same plug-in interpretation is less direct, but a related identity suggests a useful route to improvement.
Let $q(\vtheta_{-k}|\vtheta_k)$ be any conditional density. Applying the decomposition in \eqref{eq:elbo}, now conditional on 
$\vtheta_k$, gives
\begin{equation}\label{eq:elbo2}
\log p(\sD,\vtheta_k)
=
\bE_{q(\vtheta_{-k}|\vtheta_k)}
\left[
\log p(\sD,\vtheta)-\log q(\vtheta_{-k}|\vtheta_k)
\right]
+
\mathrm{KL}[q(\vtheta_{-k}|\vtheta_k),
p(\vtheta_{-k}|\sD,\vtheta_k)].
\end{equation}

If $q(\vtheta_{-k}|\vtheta_k)$ is replaced by a marginal approximation $q(\vtheta_{-k})$ that does not depend on 
$\vtheta_k$, and if the KL term is ignored as a function of 
$\vtheta_k$, then \eqref{eq:elbo2} leads to the usual MFVB-style update. This approximation discards the posterior dependence of 
$\vtheta_{-k}$ on the block currently being updated.
Equation \eqref{eq:elbo2} therefore suggests a non-conjugate analogue of the conjugate correction above. If $p(\vtheta_{-k}| \sD,\vtheta_k)$ can be approximated well by a conditional density 
$q(\vtheta_{-k}|\vtheta_k)$, then the first term on the right-hand side of \eqref{eq:elbo2} provides an improved local approximation to the marginal log-density of $\vtheta_k$. In particular, retaining dependence on $\vtheta_k$ allows information about the remaining parameters to be propagated into the marginal approximation of 
$\vtheta_k$.

The calculations above identify two complementary ways that MFVB can lose posterior information. In conjugate models, MFVB omits the contribution of uncertainty in neighbouring blocks to the marginal posterior covariance. In non-conjugate models, MFVB-style updates ignore the conditional dependence of the remaining parameters on the current updated block. Moment propagation uses these observations to construct improved marginal posterior approximations by propagating conditional posterior moment information between parameter blocks.

\section{Moment propagation}
\label{sec:mp}

We now introduce moment propagation (MP), a framework for constructing improved marginal posterior approximations by propagating conditional posterior moment information between parameter blocks. Let
$\vtheta=\{ \vtheta_1,\ldots,\vtheta_K \}$,
$\vtheta_k\in\bR^{d_k}$,
denote a partition of the parameter vector. Our goal is to construct approximations
$q_k(\vtheta_k)\approx p(\vtheta_k|\sD)$,
$k=1,\ldots,K$,
with particular emphasis on accurate marginal posterior means and variances.

The key idea is to replace the plug-in updates used by MFVB with conditional moment averaging. For each block $\vtheta_k$, MP approximates marginal posterior moments by averaging conditional posterior moments of $\vtheta_k$ over an approximation to the posterior distribution of the neighbouring blocks. These moments are then matched to a convenient approximating family. This leads to algorithms that retain much of the blockwise structure of MFVB while restoring part of the posterior uncertainty lost through mean-field factorization.

\subsection{The basic marginal identity}

Let $M(k)\subseteq \{1,\ldots,K\}\setminus\{k\}$ denote the indices of the blocks in the Markov blanket of $\vtheta_k$. By the law of total probability,
$p(\vtheta_k|\sD)
= \bE_{p(\vtheta_{-k}|\sD)}
[p(\vtheta_k|\sD,\vtheta_{-k})]$.
%\tag{6}
Using the conditional independence encoded by the Markov blanket,
$p(\vtheta_k|\sD,\vtheta_{-k})
= p(\vtheta_k|\sD,\vtheta_{M(k)})$,
and therefore
\begin{equation}\label{eq7}
p(\vtheta_k|\sD) =
\bE_{p(\vtheta_{M(k)}|\sD)}
[p(\vtheta_k|\sD,\vtheta_{M(k)})].
\end{equation}
%\tag{7}

Suppose that $q(\vtheta_{M(k)}) \approx p(\vtheta_{M(k)}|\sD)$. Then \eqref{eq7} suggests the implicit marginal approximation
%\begin{equation}\label{eq8}
$\widetilde{q}_k(\vtheta_k)
=
\bE_{q(\vtheta_{M(k)})}
[p(\vtheta_k|\sD,\vtheta_{M(k)})]$.
%\tag{8}
%\end{equation}
Equivalently, define the local joint approximation
\begin{equation}\label{eq9}
r_k(\vtheta_k,\vtheta_{M(k)})
=
p(\vtheta_k|\sD,\vtheta_{M(k)})q(\vtheta_{M(k)}).
%\tag{9}
\end{equation}

When both factors on the right-hand side are normalized densities, $r_k(\vtheta_k,\vtheta_{M(k)})$ is also   normalized, and its marginal density in $\vtheta_k$ is 
$\widetilde{q}_k(\vtheta_k)$.
In general, direct calculation of $\widetilde{q}_k(\vtheta_k)$ may be difficult. However, its moments are often easier to approximate, e.g.,
\begin{equation}\label{eq10}
\bE_{\widetilde{q}_k}[\mT_k(\vtheta_k)]
=
\bE_{r_k}[\mT_k(\vtheta_k)]
=
\bE_{q(\vtheta_{M(k)})}
\left[
\bE\{\mT_k(\vtheta_k)|\sD,\vtheta_{M(k)}\}
\right].
%\tag{10}
\end{equation}

\noindent where $\mT_k(\vtheta_k)$ is a vector of functions.
Moment propagation uses \eqref{eq10}  to construct a tractable approximation $q_k(\vtheta_k;\veta_k)$ by choosing $\veta_k$ such that
$\bE_{q_k(\vtheta_k;\veta_k)}[\mT_k(\vtheta_k)]
=
\bE_{r_k}
[\mT_k(\vtheta_k)]$.
Thus, $\widetilde{q}_k$ denotes the implicit MP marginal induced by conditional averaging, whereas $q_k(\vtheta_k;\veta_k)$ denotes a tractable moment-matched approximation to that marginal.

\subsection{Moment propagation in conjugate block models}

We first consider the case where the conditional density $p(\vtheta_k|\sD,\vtheta_{M(k)})$
is available in normalized form and its relevant moments are known functions of $\vtheta_{M(k)}$. In this setting, MP can be implemented by moment matching without evaluating the full mixture density in \eqref{eq9}.

In particular, the posterior mean and covariance of $\vtheta_k$ can be approximated by
\begin{equation}\label{eq11}
\begin{array}{c}
\bE(\vtheta_k|\sD)
\approx
\bE_{r_k}(\vtheta_k)
=
\bE_{q(\vtheta_{M(k)})}
\left[
\bE(\vtheta_k|\sD,\vtheta_{M(k)})
\right], 
\qquad \mbox{and}
\\ [1ex]
\bV(\vtheta_k|\sD)
\approx
\bV_{r_k}(\vtheta_k)
=
\bE_{q(\vtheta_{M(k)})}
\left[
\bV(\vtheta_k|\sD,\vtheta_{M(k)})
\right]
+
\bV_{q(\vtheta_{M(k)})}
\left[
\mathbb E(\vtheta_k|\sD,\vtheta_{M(k)})
\right].
%\tag{11}
\end{array}
\end{equation}

The second term on the right hand side of the expression for $\bV_{r_k}(\vtheta_k)$ in \eqref{eq11} is the additional uncertainty induced by variation in the neighbouring blocks. This is the term that is typically absent in the corresponding MFVB update. Hence \eqref{eq11} provides a direct moment-based correction to the plug-in approximation discussed in Section~\ref{sec:VB}.

More generally, MP can be applied to any vector %of functions 
$\mT_k(\vtheta_k)$ for which the conditional moments
$\bE[\mT_k(\vtheta_k)|\sD,\vtheta_{M(k)}]$
can be evaluated or approximated. Given these moments, we choose a convenient approximating family $q_k(\vtheta_k;\veta_k)$ and solve the moment matching equations
$$
\mathbb E_{q_k(\vtheta_k;\veta_k)}
[\mT_k(\vtheta_k)]
=
\bE_{r_k}
[\mT_k(\vtheta_k)].
$$
 
When $q_k$ is an exponential-family density and $\mT_k$ is its vector of sufficient statistics, this moment matching step can be interpreted as a KL projection onto that family \parencite{minka2005divergence}. However, MP is not restricted to exponential-family approximations or to matching sufficient statistics. In later examples, we use multivariate $t$, inverse-Wishart, Gaussian, and inverse-gamma approximations, and in some cases match moments other than the sufficient statistics when this leads to tractable updates.

\subsection{Mean-field moment propagation}

Mean-field moment propagation (MFMP) is obtained by approximating the distribution of the neighbouring blocks by a product of their marginal approximations:
$$
q(\vtheta_{M(k)})
=
\prod_{j\in M(k)} q_j(\vtheta_j).
$$
This assumption is weaker than the MFVB factorization in the sense that  MP updates   block $k$ only requires a factorized approximation over the Markov blanket of $\vtheta_k$, rather than over all parameters. When $K=2$, no additional factorization is required inside the conditioning block, although the algorithm still constructs separate marginal approximations for the two components.
A generic MFMP update for block $k$ has the following form.

\begin{enumerate}
\item Given current approximations $\{q_j:j\in M(k)\}$, form
$q(\vtheta_{M(k)})=\prod_{j\in M(k)}q_j(\vtheta_j)$.

\item Compute the propagated moments
$\vm_k^{\mathrm{MP}}
=
\bE_{q(\vtheta_{M(k)})}
\left[
\bE\{\mT_k(\vtheta_k)|\sD,\vtheta_{M(k)}\}
\right]$.

\item Solve for $\veta_k$ such that
$\bE_{q_k(\vtheta_k;\veta_k)}
[\mT_k(\vtheta_k)]
=
\vm_k^{\mathrm{MP}}$.

\item Replace $q_k$ by the updated density $q_k(\vtheta_k;\veta_k)$, and cycle through the remaining blocks until convergence.
\end{enumerate}

The following results summarize the asymptotic behaviour of MFMP. Note that the results do not constrain the choice of $\sQ_k$ to the exponential family of distributions. Note that we provide only an intuitive condensed statement of each result here. Their full technical statements, required assumptions, and technical proofs are given in Supplementary Section~\ref{Suppl:profTheo1}. The first theorem provide sufficient conditions to guarantee the large-sample convergence of the MFMP solution to the fixed point of a limiting system of integral equations.
\begin{Theorem}
		\label{ConvergenceToLimitSolution}
		Assume (A1) to (A5) in Supplementary Section ~\ref{Suppl:profTheo1} holds. Then, the MFMP sequence of solutions $\{ \widehat{\vlambda}_{1:K}^{(n)} \}_{n \ge 1}$ converge in probability to the solution of a set of limiting integral equations $\vlambda_{1:K}^\star$ at rate $O_p(\epsilon_n)$, where $\epsilon_n$ is a deterministic sequence according that controls the rate of convergence of the exact posterior.
	\end{Theorem}
The following results establishes the consistency of the MFMP posterior.
\begin{Theorem}
		\label{MFMPconsistency}
		Assume (A1) to (A6) in Supplementary Section ~\ref{Suppl:profTheo1} holds. Then, for each $k=1,\ldots,K$ and any $\epsilon > 0$, 
		$\mathbb{P}_{q_k} \left \{ \lVert \vtheta_k - \vtheta_k^\star \rVert > \epsilon \right \} \xrightarrow{\bP^\star} 0,$
        where $\mathbb{P}_{q_k}$ is the probability measure corresponding to the $k$-th marginal MFMP posterior.
	\end{Theorem}
The following results establishes the asymptotic convergence of the MFMP posterior variance.
	\begin{Theorem}
		\label{asympNormality}
		Assume (A1) to (A9) holds and $K = 2$. Then, for each $k = 1, 2$, we have \\
		$\lVert \mSigma_{kn} - \mathbf{V}_{kn} \rVert = O_p(n^{-3/2})$,
		where $\mV_{kn} = \bV
        %_{p(\vtheta_k \mid \sD_n)} 
        ( \vtheta_k \mid \sD_n )$.
	\end{Theorem}

% \begin{Theorem}\label{Th:Theorem1}
% Consider a two-component model with $\vtheta=(\vtheta_1^T,\vtheta_2^T)^T$. Suppose that the MFMP fixed point is defined by matching moments of
% $$
% \widetilde{\mT}_k(\vtheta_k)
% =
% \{\vtheta_k^T,\mathrm{vec}(\vtheta_k\vtheta_k^T)^T,\mT_k(\vtheta_k)^T\}^T,
% \qquad k=1,2,
% $$

% \noindent 
% where $\mT_k$ contains any additional statistics required to identify the approximating family $q_k$. Under the regularity conditions stated in Supplementary Section~\ref{Suppl:profTheo1}, the MFMP approximations are consistent:
% $$
% q_{kn}(\vtheta_1)\Rightarrow \delta(\vtheta_k^\star), \qquad k = 1,2,
% % \qquad
% % q_{2n}(\vtheta_2)\Rightarrow \delta(\theta_2^\star),
% $$
% and their marginal covariance matrices are asymptotically equivalent to the corresponding posterior covariance matrices:
% $\sqrt n
% \left\|\mathbb V_{q_{kn}}(\vtheta_k)
% -\mathbb V_{p_n}(\vtheta_1|\sD_n)
% \right\|=
% o_p(1), \qquad k = 1,2$.
% % and
% % $$
% % \sqrt n
% % \left\|
% % \mathbb V_{q_{2n}}(\theta_2)
% % -
% % \mathbb V_{p_n}(\theta_2|\sD_n)
% % \right\|
% % =
% % o_p(1).
% % $$

% \end{Theorem}

Theorem~\ref{asympNormality} formalizes the main advantage of MFMP over MFVB in the two-component conjugate case: by propagating conditional moment information between blocks, MFMP can recover asymptotically correct marginal posterior variances. The result does not imply that MFMP exactly represents the joint posterior dependence between blocks, but it shows that accurate marginal variance estimation is possible even when the algorithm maintains separate marginal approximations.

MFMP is most useful when the conditional density $p(\vtheta_k|\sD,\vtheta_{M(k)})$ is known and the conditional moments in \eqref{eq10} can be calculated analytically or by low-dimensional numerical integration. This requirement limits its direct application in non-conjugate models. We next introduce a Gaussian alternative that is designed for such settings.

\subsection{Gaussian moment propagation}
\label{sec:GMP}

Gaussian moment propagation (GMP) starts from a joint Gaussian approximation
$q(\vtheta)
= \phi(\vtheta-\vmu;\mSigma)$,
where $\phi(\cdot;\mSigma)$ denotes the multivariate Gaussian density with mean zero and covariance matrix $\mSigma$. This Gaussian approximation is used as a propagation device: its conditional distributions are used to construct local marginal approximations, and the first two moments of these local approximations are then used to update the joint Gaussian approximation.

Consider updating block $\vtheta_k$. After a possible permutation of indices, write
$$
\vmu=
\begin{bmatrix}
\vmu_k\\
\vmu_{-k}
\end{bmatrix},
\qquad
\mSigma=
\begin{bmatrix}
\mSigma_{k,k} & \mSigma_{k,-k}\\
\mSigma_{-k,k} & \mSigma_{-k,-k}
\end{bmatrix}.
$$
Let
$\mSigma_{-k|k}= \mSigma_{-k,-k}
-
\mSigma_{-k,k}\mSigma_{k,k}^{-1}\mSigma_{k,-k}$.
Then the conditional distribution under the current Gaussian approximation is
\begin{equation}\label{eq12}
q(\vtheta_{-k}|\vtheta_k)
=
\phi\left(
\vtheta_{-k}
-
\vmu_{-k}
-
\mSigma_{-k,k}\mSigma_{k,k}^{-1}(\vtheta_k-\vmu_k);
\,
\mSigma_{-k|k}
\right).
%\tag{12}
\end{equation}

The GMP update is motivated by the conditional identity from Section~\ref{sec:VB}, i.e., \eqref{eq:elbo2}.
% $$
% \log p(\sD,\vtheta_k)
% =
% \bE_{q(\vtheta_{-k}|\vtheta_k)}
% \left[
% \log p(\sD,\vtheta)
% -
% \log q(\vtheta_{-k}|\vtheta_k)
% \right]
% +
% \mathrm{KL}[q(\theta_{-k}|\vtheta_k),
% p(\vtheta_{-k}|\sD,\vtheta_k)].
% $$
For the Gaussian conditional distribution in \eqref{eq12}, the entropy term
$\bE_{q(\vtheta_{-k}|\vtheta_k)}
[-\log q(\vtheta_{-k}|\vtheta_k)]$
does not depend on $\vtheta_k$. If, in addition, the conditional KL term varies slowly as a function of $\vtheta_k$, then the marginal posterior density of $\vtheta_k$ is approximated, up to normalization, by
\begin{equation}\label{eq13}
r_k(\vtheta_k)
\propto
\exp\left[
\bE_{q(\vtheta_{-k}|\vtheta_k)}
\{\log p(\sD,\vtheta)\}
\right].
%\tag{13}
\end{equation}

\noindent 
We refer to $r_k(\vtheta_k)$ as the local GMP marginal approximation for block $k$. It has dimension $d_k$, and so can often be normalized and integrated using low-dimensional numerical integration.

Suppose that the first two moments of $r_k$ are available: $\widetilde\vmu_k
= \bE_{r_k}(\vtheta_k)$,
and 
$\widetilde\mSigma_k
=\bV_{r_k}(\vtheta_k)$.
These moments may be calculated using quadrature, Monte Carlo, Laplace approximation, or other suitable low-dimensional approximations. We then update the joint Gaussian approximation by replacing the marginal distribution of $\vtheta_k$ with
$\widetilde q_k(\vtheta_k)
=\phi(\vtheta_k-\widetilde\vmu_k;\widetilde\mSigma_k)$,
while retaining the current conditional distribution $q(\vtheta_{-k}|\vtheta_k)$. That is, the updated joint approximation is
$\widetilde q(\vtheta)
=
q(\vtheta_{-k}|\vtheta_k)\widetilde q_k(\vtheta_k)$.
This is again a multivariate Gaussian distribution, with %mean
\begin{equation}\label{eq14}
\widetilde \vmu
=
\begin{bmatrix}
\widetilde\vmu_k\\
\vmu_{-k}
+
\mSigma_{-k,k}\mSigma_{k,k}^{-1}(\widetilde\vmu_k-\vmu_k)
\end{bmatrix}, \qquad \mbox{and}
%\tag{14}
\end{equation}
%
%\noindent and covariance
\begin{equation}\label{eq15}
\widetilde\mSigma
=
\begin{bmatrix}
\widetilde\mSigma_k
&
\widetilde\mSigma_k\mSigma_{k,k}^{-1}\mSigma_{k,-k}
\\
\mSigma_{-k,k}\mSigma_{k,k}^{-1}\widetilde\mSigma_k
&
\mSigma_{-k\mid k}
+
\mSigma_{-k,k}\mSigma_{k,k}^{-1}
\widetilde\mSigma_k
\mSigma_{k,k}^{-1}\mSigma_{k,-k}
\end{bmatrix}.
%\tag{15}
\end{equation}

\noindent 
After replacing $(\vmu,\mSigma)$ by $(\widetilde\vmu,\widetilde\mSigma)$, the algorithm proceeds to update another block. Cycling through all blocks until convergence yields the GMP approximation.

The GMP algorithm can be summarized as follows.

\begin{enumerate}
\item Initialize $(\vmu,\mSigma)$ using a Gaussian approximation, such as a Laplace, GVB, MFVB-based Gaussian, or EP approximations.

\item For $k=1,\ldots,K$, 
\begin{itemize}
    \item Compute the Gaussian conditional distribution $q(\vtheta_{-k}\mid\vtheta_k)$ from \eqref{eq13}.

\item Form the local marginal approximation $r_k(\vtheta_k)$ in \eqref{eq13}.

\item Compute
$\widetilde\vmu_k=\bE_{r_k}(\vtheta_k)$, and
$\widetilde\mSigma_k=\bV_{r_k}(\vtheta_k)$.

\item Update the joint Gaussian approximation using \eqref{eq14} and \eqref{eq15}.
\end{itemize}

\item Repeat  until convergence.
\end{enumerate}

The local density $r_k(\vtheta_k)$ and the Gaussian density $q(\vtheta)$ play different roles. The Gaussian density $q(\vtheta)$ is used to propagate dependence information between blocks through its conditional distributions. The local density $r_k(\vtheta_k)$ can be used directly as the marginal posterior approximation for $\vtheta_k$, while its first two moments are used to update $q(\vtheta)$.

\begin{Theorem}\label{Th:Theorem2}
If $\mSigma$ and $\widetilde\mSigma_k$ are positive definite, then the updated covariance matrix $\widetilde\mSigma$ in \eqref{eq15} is positive definite.
\end{Theorem}

\noindent 
The proof is given in Supplementary Section~\ref{Suppl:proof_theorem2}. Theorem~\ref{Th:Theorem2} ensures that each GMP update preserves positive definiteness of the Gaussian propagation covariance matrix. Notice that the updated off-diagonal blocks are nonzero if and only if the corresponding off-diagonal blocks in the current $\mSigma$ are nonzero. Therefore, GMP should be initialized with a Gaussian approximation that already contains useful posterior dependence information. In practice, we have found that the quality of the final GMP approximation depends on the quality of this initialization.

\subsection{Hybrid and approximate moment propagation updates}

The preceding sections describe two idealized MP updates. MFMP is attractive when conditional posterior moments are available, while GMP  requires a reasonable joint Gaussian approximation. In practice, neither set of requirements may hold exactly. The required moments in \eqref{eq10}, the expectations in \eqref{eq13}, or the moment matching equations for $\veta_k$ may be analytically intractable. In such cases, MP can be implemented using hybrid or approximate updates.

One possibility is to approximate the required moments using Monte Carlo, stochastic approximation, quadrature, Laplace approximation, or delta-method expansions. Another possibility is to apply MP only to blocks for which the conditional moment calculations are tractable, and use standard VB-style updates for the remaining blocks.
For example, suppose $V\subset M(k)$ is a subset of neighbouring blocks over which exact conditional moment propagation is inconvenient. Define
$\widetilde p_k(\vtheta_k|\sD,\vtheta_{M(k)\setminus V})
\propto
\exp\left[
\mathbb E_{q(\vtheta_V)}
\{\log p(\vtheta_k|\sD,\vtheta_{M(k)})\}
\right]$,
%
%\noindent
%
where
$q(\vtheta_V)=\prod_{j\in V}q_j(\vtheta_j)$.
If $\widetilde p_k$ can be normalized and its moments computed, then the MP moment update can be replaced by
\begin{equation}\label{eq16}
\bE[\mT_k(\vtheta_k)|\sD]
\approx
\bE_{q(\vtheta_{M(k)\setminus V})}
\left[
\bE_{\widetilde p_k(\vtheta_k|\sD,\vtheta_{M(k)\setminus V})}
\{\mT_k(\vtheta_k)\}
\right].
%\tag{16}
\end{equation}

\noindent 
This is a hybrid MP/VB update: the blocks in $V$ are averaged in a variational manner, while the remaining neighbouring blocks are propagated through conditional moments.

A second hybrid strategy is useful when a full conditional approximation $q(\vtheta_{-k}|\vtheta_k)$ is unavailable, but a lower-dimensional structured approximation is available. Suppose, for example, that for some $k'\neq k$,
$q(\vtheta)
=
q(\vtheta_k,\vtheta_{k'})
\prod_{j\notin\{k,k'\}}q_j(\vtheta_j)$,
and that $q(\vtheta_{k'}\mid\vtheta_k)$ is available. Then a lower-dimensional replacement for the GMP update is
\begin{equation}\label{eq17}
r_k(\vtheta_k)
\propto
\exp\left[
\mathbb E_{q(\vtheta_{k'}|\vtheta_k)
\prod_{j\notin\{k,k'\}}q_j(\vtheta_j)}
\{\log p(\sD,\vtheta)\}
\right].
%\tag{17}
\end{equation}

\noindent 
This approximation propagates dependence between $\vtheta_k$ and $\vtheta_{k'}$, while treating the remaining blocks in a mean-field manner.
Hybrid updates such as \eqref{eq16} and \eqref{eq17} may be less accurate than the ideal MFMP or GMP updates, but they can substantially expand the range of models for which moment propagation is computationally feasible. In general, MFMP is most useful when conditional posterior moments are available analytically or through low-dimensional integration; GMP is most useful when a good joint Gaussian approximation is available; and hybrid updates are useful when only part of the model admits tractable MP updates.

\section{Examples}
\label{sec:examples}

In this section, we illustrate moment propagation through a sequence of examples. The examples are chosen to demonstrate different aspects of the methodology. Linear regression and the multivariate normal model provide benchmark settings where exact marginal posterior distributions are available, allowing us to verify that suitable MP updates can recover the exact answer. The multivariate normal model with missing data demonstrates how MP can propagate uncertainty from latent quantities. Probit regression illustrates the use of MP in an auxiliary-variable model, including a case where not all approximating densities need to be specified explicitly. Generalized linear models demonstrate Gaussian moment propagation (GMP) in non-conjugate settings. Finally, the Bayesian Lasso example shows how GMP can be adapted to a model with a non-smooth prior.
Detailed derivations, theory, and algorithmic summaries are provided in the Supplementary Material. %Here we focus on the form of the updates and on the role each example plays in illustrating the MP framework.

\subsection{Linear regression}
\label{sec:mp_linear_model}

We begin with a conjugate linear regression model. This example is useful because the exact marginal posterior distributions are available in closed form, making it possible to assess whether moment propagation can recover the exact posterior.

Consider the linear model
$\vy| \vbeta,\sigma^2 \sim N_n(\mX\vbeta,\sigma^2 \mI_n)$,
with priors
$\vbeta|\sigma^2 \sim N_p(0,g\sigma^2(\mX^T\mX)^{-1})$,
$\sigma^2 \sim \mathrm{IG}(A,B)$,
where $g>0$ is fixed. Let
$\widehat\vbeta=(\mX^T\mX)^{-1}\mX^T\vy$, and 
$u=g/(1+g)$.
The posterior distribution is available in closed form. We consider two MP approximations. The first uses a Gaussian density for $q(\vbeta)$ and shows how MP improves upon the usual MFVB variance update. The second uses a multivariate $t$ 
%approximation 
density for $q(\vbeta)$ and recovers the exact marginal posterior.

%\subsubsection{Gaussian approximation for $q(\vbeta)$}
\subsubsection{Gaussian approximation for \texorpdfstring{$q(\vbeta)$}{q(beta)}}
\label{sec:mp_linear_model_approach1}

First suppose
$q(\vbeta)=N_p(\widetilde\vmu,\widetilde\mSigma)$,
and $q(\sigma^2) = \mathrm{IG}(\widetilde A,\widetilde B)$.
To update $q(\vbeta)$ given $q(\sigma^2)$, we propagate the conditional moments of $\vbeta|\vy,\sigma^2$. This gives
$$
\begin{array}{c}
\bE_r(\vbeta)
%=
%\bE_q[\bE(\vbeta|\vy,\sigma^2)]
=
u\widehat\vbeta,
\quad 
\mbox{and} 
\quad 
%\\ 
\ds \bV_r(\vbeta)
%=
%\bE_q[\bV(\vbeta|\vy,\sigma^2)]
%+
%\bV_q[\bE(\vbeta|\vy,\sigma^2)]
=
\frac{\widetilde B}{\widetilde A-1}
u(\mX^T\mX)^{-1}.
\end{array}
$$

\noindent 
Matching these moments with those of $q(\vbeta)$ gives leads to the updates
$\widetilde\vmu \leftarrow u\widehat\vbeta$,
and
$\widetilde\mSigma
\leftarrow
(\widetilde B/(\widetilde A-1))
u(\mX^T\mX)^{-1}$.
The update for $q(\sigma^2)$, uses the conditional distribution
$\sigma^2|\vy,\vbeta
\sim
\mathrm{IG}(A_{n,p},\sB(\vbeta))$,
where $A_{n,p} = A+(n+p)/2$ and
$\sB(\vbeta) = B+(\|\vy-\mX\vbeta\|^2)/2
+ (2g)^{-1}\vbeta^T\mX^T\mX\vbeta$. 

Moment propagation gives
$$
\bE_r(\sigma^2) =
\frac{\bE_q[\sB(\vbeta)]}{A_{n,p}-1},
\quad \mbox{and} \quad 
\bV_r(\sigma^2)
=
\frac{\bE_q[\sB(\vbeta)]^2}
     {(A_{n,p}-1)^2(A_{n,p}-2)}
+
\frac{\bV_q[\sB(\vbeta)]}
     {(A_{n,p}-1)(A_{n,p}-2)}.
$$

\noindent 
Here $\bE_q[\sB(\vbeta)]
=
B+ \|\vy-\mX\widetilde\vmu\|^2/2
+
\widetilde\vmu^T\mX^T\mX\widetilde\vmu/(2g)
+
\mathrm{tr}(\mX^T\mX\widetilde\mSigma)/(2u),
$ and 
$\bV_q[\sB(\vbeta)]
=
\mathrm{tr}[(\mX^T\mX\widetilde\mSigma)^2]/(2u^2)
$. Note that the MFVB approximation leads to the same approximation of posterior mean, and equals $\bV_r(\sigma^2)$ without the additional second term.
Matching the mean and variance of $q(\sigma^2)$ 
to $\bE_r(\sigma^2)$ and $\bV_r(\sigma^2)$ leads to the updates
$$
\widetilde A
\leftarrow
\frac{\bE_r(\sigma^2)^2}{\bV_r(\sigma^2)}+2,
\qquad
\widetilde B
\leftarrow
(\widetilde A-1)\bE_r(\sigma^2).
$$

This approximation shows the key MP correction: the variance of $B(\vbeta)$ contributes an additional term to the variance update for $\sigma^2$. However,  $q(\vbeta)$ is still misspecified because the exact marginal posterior of $\vbeta$ is multivariate $t$. The next approach corrects this.

\subsubsection{Multivariate 
\texorpdfstring{$t$}{t}
approximation for \texorpdfstring{$q(\vbeta)$}{q(beta)}}

Now suppose
$q(\vbeta)=t_p(\widetilde\vmu,\widetilde\mSigma,\widetilde\nu)$, and 
$q(\sigma^2)=\mathrm{IG}(\widetilde A,\widetilde B)$.
For fixed $q(\sigma^2)$, we match the first two moments of $\vbeta$, and  
$Q(\vbeta)^2$,
where
$Q(\vbeta)
=
(\vbeta-\widetilde\vmu)^T
\widetilde\mSigma^{-1}
(\vbeta-\widetilde\vmu)$.
This gives
$$
\widetilde\vmu \leftarrow u\widehat\vbeta,
\qquad
\widetilde\mSigma
\leftarrow
\frac{\widetilde B}{\widetilde A}
u(\mX^T\mX)^{-1},
\qquad
\widetilde\nu
\leftarrow
2\widetilde A.
$$

The update for $q(\sigma^2)$ again matches 
$\bE_r(\sigma^2)$ and $\bV_r(\sigma^2)$, but now the moments of $B(\vbeta)$ are computed under the multivariate $t$ approximation:
$$
\begin{array}{c}
\ds \bE_q[B(\vbeta)]
=
B+\frac12\|\vy-\mX\widetilde\vmu\|^2
+
\frac{1}{2g}\widetilde\vmu^T\mX^T \mX\widetilde\vmu
+
\frac{\widetilde\nu}{\widetilde\nu-2}
\frac{\mathrm{tr}(\mX^T\mX\widetilde\mSigma)}{2u} 
\qquad \mbox{and}
\\ [2ex]
\ds \bV_q[B(\vbeta)]
=
\frac{1}{2u^2}
\left[
\frac{\widetilde\nu^2
\mathrm{tr}\{(\mX^T \mX\widetilde\mSigma)^2\}}
     {(\widetilde\nu-2)(\widetilde\nu-4)}
+
\frac{\widetilde\nu^2
\{\mathrm{tr}(\mX^T\mX\widetilde\mSigma)\}^2}
     {(\widetilde\nu-2)^2(\widetilde\nu-4)}
\right],
\end{array}
$$

\noindent 
provided the required moments exist. Matching the mean and variance of \(q(\sigma^2)\) gives the same inverse-gamma moment equations as above.
% \noindent 
% {\bf Proposition 1.} For the Gaussian linear model with the prior specified above,
% the MP updates in Section 4.1.2 have a fixed point equal to the exact marginal
% posteriors $p(\vbeta|\vy)$ and 
% $p(\sigma^2|\vy)$.
% \bigskip 
The Supplementary Material \ref{sec:lm_example_details} shows that this second MP approach recovers the exact marginal posterior distributions $p(\vbeta|\vy)$ 
and $p(\sigma^2|\vy)$. Thus, the example demonstrates that MP can recover exact posterior marginals when the propagated moments are exact and the approximating families are sufficiently rich.

\subsection{Multivariate normal model}
\label{sec:mvn_model}

The next example gives a second exact-recovery benchmark.
Let
$\vx_i| \vmu,\mSigma \sim N_p(\vmu,\mSigma)$,
$i=1,\ldots,n$,
where $\vmu\in\bR^p$ and $\mSigma\in\mathcal S_p^+$. We use the conditionally conjugate priors
$\vmu|\mSigma \sim N_p(0,\lambda_0^{-1}\mSigma)$,
and $\mSigma\sim \mathrm{IW}_p(\mPsi_0,\nu_0)$,  an inverse-Wishart distribution,
where $\lambda_0>0$, $\nu_0>p-1$, and 
$\mPsi_0\in\mathcal S_p^+$. Without loss of generality, the prior mean of $\vmu$ is taken to be zero, after translating the observations if necessary. This model has a closed-form posterior distribution.

We approximate the marginal posterior distributions by
$q(\vmu)=t_p(\widetilde\vmu,\widetilde\mSigma,\widetilde\nu)$, and 
$q(\mSigma)=\mathrm{IW}_p(\widetilde\mPsi,\widetilde d)$.
The MP update for $q(\vmu)$ matches 
$\bE_r(\vmu)$, $\bV_r(\vmu)$, and a fourth radial moment to give
$$
\widetilde\vmu \leftarrow \frac{n\overline \vx}{\lambda_n},
\qquad
\widetilde\mSigma
\leftarrow
\frac{\widetilde\mPsi}{\lambda_n(\widetilde d-p+1)},
\qquad
\widetilde\nu
\leftarrow
\widetilde d-p+1,
$$

\noindent 
where $\overline \vx=n^{-1}\sum_{i=1}^n \vx_i$ and $\lambda_n=\lambda_0+n$.

To update $q(\mSigma)$, we match 
$\bE_r(\mSigma)$ and the diagonal element-wise variances of $\mSigma$. Let
$\mPsi_n
=
\mPsi_0+\sum_{i=1}^n \vx_i\vx_i^T
-
n\overline \vx\,\bar \vx^T
+
(\lambda_0/\lambda_n)\overline \vx\,\overline \vx^T$, and 
$\nu_n=\nu_0+n$.
The required propagated moments can be written in the form
$$
\begin{array}{c}
\ds \bE_r(\mSigma)=\frac{\mA}{\nu_n-p},
\qquad
\mbox{and}
\qquad 
\mathrm{dg}[\bE\bW\bV_r(\mSigma)]
=
\frac{
2\mathrm{dg}(\mA)^{\odot 2}
+
(\nu_n-p)\mathrm{dg}(\mB)
}
{(\nu_n-p)^2(\nu_n-p-2)},
\\ [2ex]
\ds \mbox{where} \qquad 
\mA
=
\mPsi_n+
\frac{\lambda_n\widetilde\nu}{\widetilde\nu-2}
\widetilde\Sigma,
\qquad \mbox{and} \qquad 
\mathrm{dg}(\mB)
=
\frac{
2\lambda_n^2\widetilde\nu^2(\widetilde\nu-1)
}
{
(\widetilde\nu-2)^2(\widetilde\nu-4)
}
\mathrm{dg}(\widetilde\mSigma)^{\odot 2}.
\end{array}
$$

\noindent 
Here $\mathrm{dg}(\mA)$ denotes the vector of diagonal elements of a square matrix $\mA$, and $\bE\bW\bV(\mA)$ denotes the matrix of element-wise variances of $\mA$.
Matching these quantities to the corresponding moments of an inverse-Wishart distribution gives
$$
\widetilde d
\leftarrow
\frac{
2\mathbf 1_p^T[\mathrm{dg}(\bE_r(\mSigma))^{\odot 2}]
}
{
\mathbf 1_p^\top \mathrm{dg}[\bE\bW\bV_r(\mSigma)]
}
+
p+3,
\qquad
\widetilde\mPsi
\leftarrow
(\widetilde d-p-1)\bE_r(\mSigma).
$$

The Supplementary Material \ref{sec:mvn_derivations} gives the full derivation and shows that the MP algorithm has a fixed point corresponding to the exact marginal posteriors under assumptions stated there. 

% We also show that initialization can matter: an additional undesirable fixed point can arise for some parameter values, and this is avoided by initializing with the exact conjugate-posterior values $\widetilde d=\nu_n$ and 
% $\widetilde\mPsi=\mPsi_n$.

% CHECK: Please verify the denominator in \(E_r(\Sigma)\). Depending on the
% inverse-Wishart parameterisation, this may need to be \(\nu_n-p-1\).
% Also check the statement about the undesirable fixed point in the current draft,
% which appears to assign two different values to \(\widetilde d\).

\subsection{Multivariate normal model with missing data}
\label{sec:missing}

We now extend the multivariate normal example to the case where some components of the observations are missing. This example illustrates how MFMP can propagate uncertainty from latent variables into the posterior approximation for model parameters.

Assume the model from Section~\ref{sec:mvn_model}, but suppose that for each observation $\vx_i$, only a subset of its components is observed. We assume an ignorable missingness mechanism, so that the missingness process does not enter the posterior distribution of $\vmu$, 
$\mSigma$, or the missing data.

Let $O_i$ and $M_i$ denote the observed and missing index sets for observation $i$, respectively. Let $\vz_i=\vx_{i,M_i}$ be the vector of missing components. Define $\vc_i$ to be the vector obtained from $\vx_i$ by replacing missing entries with zeros, and let 
$\mR_i$ be the matrix that inserts $\vz_i$ into the missing positions. Then
$\vx_i=\vc_i+\mR_i \vz_i$.
Let $\mX_M$ denote the collection of missing components.

We use the MFMP factorization
$q(\vmu,\mSigma,\mX_M)
=
q(\vmu,\mSigma)
\prod_{i\in\mathcal M} q(\vz_i)$,
where $\mathcal M$ indexes observations with at least one missing component, and
$q(\vmu,\mSigma)
=
\mathrm{NIW}_p(\widetilde\vmu,\widetilde\lambda,\widetilde\mPsi,\widetilde\nu)$, the p-dimensional Normal–Inverse-Wishart distribution, and 
$q(\vz_i)
=
N_{|M_i|}(\widetilde \vm_i,\widetilde \mV_i)$, for $i\in\sM$.
The update for each missing vector
$\vz_i$ is obtained by averaging the Gaussian conditional distribution of the missing components given the observed components over the current approximation to 
$(\vmu,\mSigma)$. The update for 
$(\vmu,\mSigma)$ then treats the completed data matrix through its first two moments, thereby propagating missing-data uncertainty into the parameter approximation.

For each $i\in\mathcal M$, moment matching gives
$\widetilde \vm_i
\leftarrow
\bE_r(\vz_i)$,
and 
$\widetilde \mV_i
\leftarrow
\bV_r(\vz_i)$.
If $O_i=\varnothing$ then 
$\bE_r(\vz_i) = \widetilde\vmu$ and
$\bV_r(\vz_i)
=(1+\widetilde\lambda^{-1})
\widetilde\mPsi/(\widetilde\nu-p-1)$ otherwise
$\bE_r(\vz_i) =\mL_i(\vx_{i,O_i}-\widetilde\vmu_{O_i})$
$$
\begin{array}{c}
\mbox{and} \qquad \ds \bV_r(\vz_i)
=
\left[
1+\widetilde\lambda^{-1}
+
(\vx_{i,O_i}-\widetilde\vmu_{O_i})^T
\widetilde\mPsi_{O_iO_i}^{-1}
(\vx_{i,O_i}-\widetilde\vmu_{O_i})
\right]
\frac{\widetilde\mPsi_{M_i|O_i}}
     {\widetilde\nu-|M_i|-1},
\end{array}
$$

\noindent 
where
$\mL_i
=
\bE_q(\mSigma_{M_iO_i}\mSigma_{O_iO_i}^{-1})
=
\widetilde\mPsi_{M_iO_i}\widetilde\mPsi_{O_iO_i}^{-1}$.
and
$\widetilde\mPsi_{M_i| O_i}
=
\widetilde\mPsi_{M_iM_i}
-
\widetilde\mPsi_{M_iO_i}
\widetilde\mPsi_{O_iO_i}^{-1}
\widetilde\mPsi_{O_iM_i}$.

For the update of $q(\vmu,\mSigma)$, define
$\widehat \vx_i=\bE_q(\vx_i)=\vc_i+\mR_i\widetilde \vm_i$,
$\mB_i=\bV_q(\vx_i)=\mR_i\widetilde \mV_i\mR_i^T$,
$\overline{\widehat \vx}=\sum_{i=1}^n \widehat \vx_i/n$,
$\lambda_n=\lambda_0+n$, and 
$\mA=\sum_{i=1}^n \mB_i$.
Then
$$
\bE_r(\vmu)
=
\frac{n\overline{\widehat \vx}}{\lambda_n}
\equiv
\widehat\vmu_n,
\qquad 
\mbox{and}
\qquad 
\bV_r(\vmu)
=
\frac{\bE_q(\mSigma)}{\lambda_n}
+
\frac{\mA}{\lambda_n^2}.
$$

\noindent 
Similarly,
$\bE_r(\mSigma)
=
\mC/(\nu_n-p-1)$,
where
$\mC = \bE_q(\mPsi_n) 
=
\mPsi_0
+
\sum_{i=1}^n \widehat \vx_i\widehat \vx_i^T
+
(1-\lambda_n^{-1})\mA
-
\lambda_n\widehat\vmu_n\widehat\vmu_n^T$.

The diagonal element-wise variances can be computed from the moments of the completed data; the resulting formulae are given in   Supplementary Material \ref{sec:mvn_missing_derivations}. Matching
$\bE_r(\vmu)$,
$\mathrm{tr}[\bV_r(\vmu)]$,
$\bE_r(\mSigma)$, and 
$\vone_p^T\mathrm{dg}[\bE\bW\bV_r(\mSigma)]$
to the corresponding NIW moments gives
$$
\widetilde\vmu\leftarrow \bE_r(\vmu),
\quad
\widetilde\lambda
\leftarrow
\frac{\mathrm{tr}[\bE_r(\mSigma)]}
     {\mathrm{tr}[\bV_r(\vmu)]},
\quad 
\widetilde\nu
\leftarrow
\frac{
2\mathbf 1_p^T[\mathrm{dg}\{\bE_r(\mSigma)\}^{\odot 2}]
}
{
\mathbf 1_p^T \mathrm{dg}[\bE\bW\bV_r(\mSigma)]
}
+
p+3,
\quad
\widetilde\mPsi
\leftarrow
(\widetilde\nu-p-1)\bE_r(\mSigma).
$$

% This example shows how MP can incorporate uncertainty in latent variables without replacing them by single imputations. The missing values affect the parameter updates through both their posterior means and variances.

\subsection{Probit regression}
\label{sec:probit_example}

We next consider probit regression. This example illustrates a useful feature of MP: not every auxiliary quantity requires a fully specified approximating density. In this case, the update for $q(\vbeta)$ requires only the first two moments of the auxiliary variables, so MP can propagate those moments directly.
Let $y_i\in\{0,1\}$, and let $\vx_i\in\mathbb R^p$. The probit likelihood can be written as
\begin{equation}\label{eq:probit_likelihood}
p(\vy|\vbeta)
=
\prod_{i=1}^n
\Phi(\vx_i^T\vbeta)^{y_i}
[1-\Phi(\vx_i^T\vbeta)]^{1-y_i}
=
\prod_{i=1}^n
\Phi(\vz_i^T\vbeta),
\end{equation}

\noindent 
where $\vz_i=(2y_i-1)\vx_i$.
Assume the Gaussian prior
$\vbeta\sim N_p(\vzero,\mD^{-1})$,
where $\mD$ is positive definite.
Using the Albert--Chib auxiliary-variable representation \parencite{AlbertChib1993},
$p(y_i,a_i|\vbeta)
=
\phi(a_i-z_i^\top\beta)\bI(a_i>0)$,
$i=1,\ldots,n$.
Integrating over $a_i$ recovers
$p(y_i|\vbeta)=\Phi(\vz_i^T\vbeta)$.
Let $\va=(a_1,\ldots,a_n)$ and 
$\mZ=\mathrm{diag}(2\vy-\mathbf 1_n)\mX$.

We use
$q(\vbeta)=N_p(\widetilde\vmu,\widetilde\mSigma)$,
but do not specify a full parametric form for $q(\va)$. Instead, at least at a conceptual level, we maintain only $\bE_q(\va)$ and $\bV_q(\va)$ because these are the quantities needed to update $q(\vbeta)$. Later, we merge updates so that $\bV_q(\va)$ is not explicitly stored.
Matching the propagated moments of $\vbeta$ gives
$\widetilde\vmu
\leftarrow
\mS\mZ^T \bE_q(\va)$, and
$\widetilde\mSigma
\leftarrow
\mS+\mS\mZ^T \bV_q(\va)\mZ\mS$,
where
$\mS=(\mZ^T \mZ+\mD)^{-1}$.
The required moments of $\va$ are obtained from the truncated-normal conditional distribution $\va|\vy,\vbeta$. Let
$\zeta_k(t) = d^k\log \Phi(t)/dt^k$.
Then
$$
\bE_q(\va)
\leftarrow
\mZ\widetilde\vmu_\beta
+
\bE_q[\zeta_1(\mZ\vbeta)],
\quad 
\mbox{and}
\quad 
\bV_q(\va)
\leftarrow
\mI_n
+
\mathrm{diag}\left[
\bE_q\{\zeta_2(\mZ\vbeta)\}
\right]
+
\bV_q[\mZ\vbeta+\zeta_1(\mZ\vbeta)].
$$

The expectations $\bE_q[\zeta_1(\mZ\vbeta)]$ and $\bE_q[\zeta_2(\mZ\vbeta)]$ can be evaluated element-wise via one dimensional quadrature. The full covariance term 
$\mV_q[\mZ\vbeta + \zeta_1(\mZ\vbeta)]$ would require two-dimensional quadrature for each pair of observations. For computational efficiency, we instead use a first-order delta-method approximation:
$$
\bV_q(\va)
\approx
\mI_n
+
\mathrm{diag}\left[
\bE_q\{\zeta_2(\mZ\vbeta)\}
\right]
+
[
\mI_n+\mathrm{diag}\{\zeta_2(\mZ\widetilde\vmu)\}]
\mZ\widetilde\mSigma \mZ^T
[\mI_n+\mathrm{diag}\{\zeta_2(\mZ\widetilde\vmu)\}].
$$

Substituting this expression directly into the update for $\widetilde\mSigma$ avoids storing the full $n\times n$ covariance matrix $\bV_q(\va)$. The resulting combined updates are
$$
\begin{array}{rl}
\bE_q(\va) & \ds
\leftarrow
\mZ\widetilde\vmu
+
\bE_q[\zeta_1(\mZ\vbeta)],
\qquad 
\widetilde\vmu
\leftarrow
\mS\mZ^\top \bE_q(\va), \qquad \mbox{and}
\\ [2ex]
\widetilde\mSigma
\leftarrow\;&
\mS
+
\mS\mZ^T
[\mI_n+\mathrm{diag}\{\bE_q(\zeta_2(\mZ\vbeta))\}]
\mZ\mS
\\
&+
\mS\mZ^T
[\mI_n+\mathrm{diag}\{\zeta_2(\mZ\widetilde\vmu)\}]
\mZ\widetilde\mSigma \mZ^T
[\mI_n+\mathrm{diag}\{\zeta_2(\mZ\widetilde\vmu)\}]
\mZ\mS .
\end{array}
$$

\noindent 
A careful implementation costs \(O(np^2+p^3)\) per iteration.
The following result summarizes the large-sample behaviour of the fixed point. The proof is given in  Supplementary Material \ref{supp:probit}.

\bigskip 
\noindent 
\begin{Theorem}\label{Th:Theorem3}
Under the regularity conditions stated in the Supplementary Material Section \ref{sec:probit_proof}, the fixed point of the probit MP updates satisfies
$$
\|\widetilde\vmu_\beta^\star-\widehat\vbeta\|
=
O_p(n^{-1}),
\qquad
\|\widetilde\mSigma_\beta^\star-\mV\|_F
=
O_p(n^{-2}),
$$

\noindent 
where $\widehat\vbeta$ is the posterior mode and
$\mV
= [\mZ^T\mathrm{diag}\{-\zeta_2(\mZ\widehat\vbeta)\}\mZ+\mD]^{-1}$.
\end{Theorem}

\bigskip 
\noindent 
Thus, the MP covariance approximation is asymptotically equivalent to the usual curvature-based covariance approximation. The result does not follow directly from Theorem~\ref{ConvergenceToLimitSolution} because of the first-order delta-method approximation used above.

% CHECK: In the current manuscript, the theorem refers to ``Froebenius'';
% this should be ``Frobenius''. Also check whether the theorem should remain
% in the main text or be moved to the Supplementary Material.

\subsection{GMP for generalized linear models}
\label{sec:example_glm}

We now consider non-conjugate generalized linear models. This example illustrates the GMP construction from Section~\ref{sec:GMP}: a joint Gaussian approximation is used as a propagation device, while low-dimensional local marginal approximations are used for inference and updating.

Consider a one-parameter exponential-family likelihood with Gaussian prior \\
$p(\vy|\vtheta)
=
\exp[\mathbf 1_n^T f(\vy,\mX\vtheta)]$,
$\vtheta\sim N(\vzero,\sigma^2 \mI_p)$,
%
%\noindent 
where
$f(\vy,\mX\vtheta)
=
(f(y_1,\vx_1^T\vtheta),\ldots,
f(y_n,\vx_n^T\vtheta))$.
Examples include logistic regression, with
$f(y,x)=yx-\log(1+\exp x)$,
probit regression, with
$f(y,x)=\log \Phi[(2y-1)x]$,
and Poisson regression, with
$f(y,x)=yx-\exp(x)-\log(y!)$.

Suppose that an initial Gaussian approximation
$q(\vtheta)=\phi(\vtheta-\vmu;\mSigma)$
has been obtained, for example from a Laplace approximation, GVB, or EP. We use the component-wise partition
$\vtheta=\{ \theta_1,\ldots,\theta_p\}$.
For the update of $\theta_k$, the Gaussian conditional distribution gives
$q(\vtheta_{-k}|\theta_k)
=
\phi(\va_k\theta_k + \vb_k,\mSigma_{-k|k})$,
where
$\va_k=\mSigma_{-k,k}/\Sigma_{k,k}$,
$\vb_k=\vmu_{-k} - \va_k\mu_k$, 
and $\mSigma_{-k|k}
= \mSigma_{-k,-k}
- \mSigma_{-k,k}\Sigma_{k,k}^{-1}\mSigma_{k,-k}$.
It follows that
$\mX\,\bE_q(\vtheta|\theta_k)
= \vc_k\theta_k + \vd_k$,
where $\vc_k = \mX_k + \mX_{-k}\va_k$,
and  $\vd_k=\mX_{-k}\vb_k$.
Let
$\vs_{-k|k}^2
= \mathrm{dg}(\mX_{-k}\mSigma_{-k|k}\mX_{-k}^T)$, and define
$F(y,\mu,\sigma^2)
=
\int_{-\infty}^{\infty}
f(y,x)\phi(x-\mu;\sigma^2)\,dx$.
Then the GMP local marginal approximation from Section~\ref{sec:GMP} is
\begin{equation}\label{eq:r_function}
r_k(\theta_k)
\propto
\exp\left[
\mathbf 1_n^T
F(\vy,\vc_k\theta_k + \vd_k,\vs_{-k\mid k}^2)
-
\frac{\|\va_k\|^2+1}{2\sigma^2}\theta_k^2
-
\frac{\va_k^T \vb_k}{\sigma^2}\theta_k
\right].
\end{equation}

\noindent 
The only model-specific ingredient is the function \(F\), the Gaussian average of the scalar log-likelihood contribution. For Poisson regression,
$F(y,\mu,\sigma^2)
=
y\mu-\exp(\mu+\sigma^2/2)-\log(y!)$,
whereas for logistic and probit regression it is evaluated using numerical quadrature.

After normalizing $r_k$, we compute
$\widetilde\mu_k=\bE_{r_k}(\theta_k)$,
and 
$\widetilde\Sigma_k=\bV_{r_k}(\theta_k)$,
and update the Gaussian propagation approximation using the GMP update equations from Section~\ref{sec:GMP}. The density
$r_k$ itself can be retained as the marginal approximation for $\theta_k$.

For a single coordinate update, the deterministic algebraic part of the GMP update can be implemented in \(O(np+p^2)\) time (see Supplementary Material \ref{Suppl:post_hoc_local_global}), and a full sweep over \(k=1,\ldots,p\) has cost \(O(np^2+p^3)\). If \(N\) denotes the number of quadrature nodes then evaluating \eqref{eq:r_function} 
requires approximating $F$ per sample
costing
$O(Nn)$, and then normalizing $r_k$ costing an additional $O(N)$
per coordinate. Thus the cost of quadrature is $O(Nnp)$ per iteration.
For the Poisson case this is reduced to $O(Np)$.

%per dimension, 
%quadrature cost is \(O(Nn)\) per coordinate.
%for Poisson regression and \(O(N^2n)\) per coordinate 
%for probit and logistic regression when bivariate quadrature is used.

% CHECK: Replace all current ``(??)'' references in this subsection with the
% relevant equations from Section~3.4.

\subsection{GMP for Bayesian Lasso}
\label{sec:gmp_lasso}

The final example considers a Bayesian Lasso model. This example illustrates a hybrid GMP/VB update: dependence among regression coefficients is propagated through a Gaussian approximation to $q(\vbeta)$, while uncertainty in $\sigma^2$ is averaged using a variational factor $q(\sigma^2)$. It also represents an example where the prior is not smooth.

Consider the linear model
$\vy| \vbeta,\sigma^2
\sim
N_n(\mX\vbeta,\sigma^2 \mI_n)$,
with $\vbeta|\sigma^2\sim\mbox{Laplace}(0,\sigma/\lambda)$
%$p(\vbeta| \sigma^2)
%\propto 
%\left(\lambda/(2\sigma)\right)^p
%\exp\left(-\lambda\|\vbeta\|_1/\sigma\right)$, 
and 
$\sigma^2\sim \mathrm{IG}(A,B)$, 
where $\lambda>0$ is a fixed constant. We assume that $\vy$ and $\mX$ have been centered so that the intercept can be omitted. The dependence of the Laplace prior on 
$\sigma^2$ is useful because, for fixed \(\lambda\), the resulting joint posterior is unimodal.
The joint density is proportional to
$$
p(\sD,\vbeta,\sigma^2)
\propto
\exp\left[
-\left(A+\frac n2+p+1\right)\log\sigma^2
-\frac{B}{\sigma^2}
-\frac{1}{2\sigma^2}\|\vy-\mX\vbeta\|^2
-\frac{\lambda}{\sigma}\|\vbeta\|_1
\right].
$$

\noindent 
For fixed $\sigma^2$, the negative log posterior is proportional in $\vbeta$ to the usual Lasso objective, up to the prior scaling.
Suppose an initial approximation has been obtained of the form
$q(\vbeta,\sigma^2)=q(\vbeta)q(\sigma^2)$,
where
$q(\vbeta)=\phi(\vbeta-\widetilde\vmu,\widetilde\mSigma)$,
and 
$q(\sigma^2)=\mathrm{IG}(\widetilde A,\widetilde B)$.
For the update of $\vbeta_k$, define
$\va_k$, $\vb_k$, $\vc_k$, $\vd_k$,
$\mSigma_{-k|k}$ as in Section \ref{sec:example_glm}.
Using the hybrid GMP/VB approximation,
$$
\begin{array}{l}
r_k(\beta_k)
\propto
\exp\left[
\bE_{q(\vbeta_{-k}|\beta_k)q(\sigma^2)}
\{\log p(\sD,\vbeta,\sigma^2)\}
\right]
\\ [2ex]
 
\quad \propto
\exp\Bigg[
-\tfrac{\widetilde A}{2\widetilde B}
\left\{
\|\vc_k\|^2\beta_k^2
-
2\vc_k^T(\vy-\vd_k)\beta_k
\right\}
-
\bE_q\left(\tfrac{\lambda}{\sigma}\right)
\left\{
|\beta_k|
+
\mathbf 1^T
\Psi(\va_k\beta_k+\vb_k, \vsigma_{-k|k}^2)
\right\}
\Bigg] 
\end{array}
$$

\noindent 
where 
$\bE_q(\sigma^{-1}) = \Gamma(\widetilde A+1/2)/
     {\Gamma(\widetilde A)\sqrt{\widetilde B}}$,   
$\Psi(\mu,\sigma^2)
=
\mu\{\Phi(\mu/\sigma)-\Phi(-\mu/\sigma)\}
+
2\sigma\phi(\mu/\sigma)$
is the expectation of $|Z|$ for $Z\sim N(\mu,\sigma^2)$,
and $\vsigma_{-k|k}^2 = \mathrm{dg}(\widetilde\mSigma_{-k|k})$.

After normalizing $r_k$, we compute
$\widetilde\mu_k=\bE_{r_k}(\beta_k)$, and
$\widetilde\Sigma_k=\bV_{r_k}(\beta_k)$,
and update the Gaussian propagation approximation for $\vbeta$ using the GMP equations from Section~\ref{sec:GMP}. The local density $r_k$ can also be retained as the marginal approximation for $\beta_k$.
For a single coordinate $k$, the local density can be normalized in $O(np+p^2+Np)$ time, where $N$ is the number of quadrature nodes (see Supplementary Material \ref{Suppl:post_hoc_local_global}). A full sweep over all $p$ coefficients has cost 
$O(np^2 + p^3 + Np^2)$.

% \subsection{Summary}

% These examples illustrate the range of MP updates. In conjugate models, MFMP can recover exact marginal posteriors when the propagated conditional moments are exact and the approximating families are chosen appropriately. In latent-variable models, MP propagates uncertainty from unobserved quantities into parameter updates. In non-conjugate models, GMP reduces marginal refinement to a sequence of low-dimensional local approximations. The numerical consequences of these updates are studied in Section~5.
%\input{examples}

\section{Numerical experience}\label{sec:Numerical_Experience}

We do not consider the linear regression or multivariate normal model examples here, since our MP approaches for these models are exact. 
%A short comparison with MFVB for these models can be found in the Supplementary Material Section X
%and Section Y.
For the remaining models, we consider MCMC as a gold standard where we use HMC (Hamiltonian Monte Carlo, with the no-U-turn sampler)  via \texttt{stan}  \parencite{HoffmanGelman2014, rstan}. For each model, we obtained a large number of HMC samples ($5\times 10^5$
samples; 5000 warm-up) as the gold standard through the {\tt R} package \texttt{rstan}. Separately, we obtain a short run of HMC samples  (5,000 samples; 1,000 warm-up) representing a bare minimum ``out of the box'' MCMC approach for comparisons with various ABI methods. 
%For the Bayesian Lasso we developed a fast Gibbs sampler implemented in {\tt Rcpp} and used Rao-Blackwellization for accurate approximations of the marginal posterior distributions.
%\bigskip 
%\noindent {\bf Metrics:} 
We use several metrics to assess the quality of marginal posterior density approximations. We use the accuracy metric of \textcite{FaesOrmerodWand2011} 
for the $j$-th marginal posterior density,  
\begin{equation}\label{eq:accuracy}
\mbox{accuracy}_j = 1 - \tfrac{1}{2}\int |p(\theta_j|\sD) - q(\theta_j)| d\theta_j, \quad j=1,\ldots,p,
\end{equation}

\noindent where $p(\theta_j|\sD)$ is estimated from a kernel density estimate from the long run of HMC. The quantity $\mbox{accuracy}_j$ takes values between $0$ and $1$ and is expressed as a percentage. For each dataset and method, we record the minimum (MIN), first quartile (Q1), and median (Q2) of the $p$ accuracies. Accuracies above 95\% are usually sufficiently accurate for most univariate inferences.
In addition to the accuracy, we record  $\sqrt{[\sum_{j} \{\bE_q(\theta_j) - \bE(\theta_j|\sD)\}^2]/p}$
denoted RMSE$(\mu)$, $\sqrt{[\sum_{j} \{\mbox{sd}_q(\theta_j) - \mbox{sd}(\theta_j|\sD)\}^2]/p}$ denoted 
RMSE$(\sigma)$
for each coefficient, where again posterior quantities are estimated using {\tt stan} using a large number of samples.
Lastly, we record $\| \bV(\vtheta|\sD) - \bV_q(\vtheta)\|_F$ using the Frobenius norm
as a multivariate metric.

Where appropriate, we compare our methodology against several ABI methods, including MFVB, 
%Partially Factorized Variational Bayes (PFVB) \citep{FasanoEtAl2022}, 
Laplace approximation, Expectation Propagation (EP) \parencite{Minka2001}, and Gaussian Variational Bayes (GVB)
\parencite{OpperArchambeau2009} where the ELBO is approximated via $n$  univariate integrals using numerical quadrature and optimized via the limited memory BFGS method of {\tt R}'s {\tt optim} function. 
GMP results initialized from the MFVB, Laplace approximation, GVB, and EP are denoted GMP-M, GMP-L, GMP-G
and GMP-E respectively. 
All methods are implemented using {\tt Rcpp} \parencite{rcpp}, and
{\tt RcppArmadillo} \parencite{RcppArmadilloManual} to facilitate comparisons.
All computations were performed on 
a computer with a 2.8 GHz Intel Core i7 processor and 32 GB of RAM to perform the computations.

\subsection{Multivariate normal model with missing data}

For this model we perform a short simulation study to compare our MFMP approximation against unblocked MFVB approximations where
$q(\vmu,\mSigma,\mX_\sM) = q(\vmu)q(\mSigma)\prod_{i\in\sM}q(\vx_{i,\sM_i})$ 
and blocked MFVB approximations where
$q(\vmu,\mSigma,\mX_\sM) = q(\vmu,\mSigma)\prod_{i\in\sM}q(\vx_{i,\sM_i})$. In the simulation study we
conducted 25 replications
where $p \in \{ 4, 8, 16\}$,
$n = p \times \{ 3, 5, 10\}$,
the true covariance matrix
$\mSigma$ had an AR(1) correlation structure
with $\rho\in \{ 0, 0.5, 0.9\}$,
where elements of $\vx_i$ were missing
completely at random with probabilities
$\{ 0.05, 0.1, 0.15, 0.2, 0.25, 0.3\}$.
Marginal posterior accuracies of elements of $\vmu$, diagonal elements of $\mSigma$,
and off-diagonal elements of $\mSigma$ were recorded separately.

Table \ref{tab:missing_reg_coeff} summarizes the coefficients
based on linear regression fits with ABI method, value of $n$, value of $p$, correlation $\rho$, and missingness percentages as predictors. This suggests that the MFMP method is 5.6\% more accurate for the posterior mean parameter $\mu_j$ across all simulations, and is roughtly 3\% more accurate for the posterior for elements of $\mSigma$. Increasing $n$ and/or $\rho$ increases accuracy of all methods on average, 
increasing $p$ and or the percent of missing values 
decreases the accuracy of all methods on average.
The average times where for the short run Gibbs sampling ($0.754$s),
MFVB ($0.004$s), MFVB-blocked ($0.012$s) and MFMP ($0.0095$s). Thus, 
MFMP represents a good accuracy versus speed trade-off.

\begin{table}
\centering
\begin{tabular}{l|r r r}
Coefficient                  & $\mu_j$                    & Diagonal $\mSigma$ & Off-diagonal $\mSigma$ \\
\hline
MFVB                    & $93.85$  $(0.23)$ & $97.79$ $(0.29)$  & $95.59$  $(0.31)$ \\
MFVB-blocked            & $+3.82$ $(0.16)$ & $+0.69$ $(0.20)$ & $+0.63$ $(0.21)$ \\
MFMP                    & $+5.64$ $(0.16)$ & $+2.89$ $(0.20)$ & $+3.00$ $(0.22)$ \\
Gibbs                   & $+5.76$ $(0.16)$ & $+4.14$ $(0.22)$ & $+4.97$ $(0.25)$ \\
$n$                     & $+0.02$ $(0.00)$ & $+0.02$ $(0.00)$ & $+0.01$ $(0.00)$ \\
$p$                     & $-0.11$ $(0.02)$ & $-0.23$ $(0.02)$ & $-0.03$ $(0.02)$ \\ 
$\rho$                  & $+2.86$ $(0.18)$ & $+4.43$ $(0.23)$ & $+6.64$ $(0.24)$ \\
Miss. \%                & $-16.08$ $(0.77)$ & $-26.04$ $(0.97)$ & $-32.14$ $(0.10)$ \\
\hline
\end{tabular}
\caption{Summary of coefficients based on linear regression fits using accuracy as the response across all simulation settings grouped by posterior mean $\mu_i$, posterior of diagonal elements of $\mSigma$, and off-diagonal elements of $\mSigma$. First row represents baseline accuracy, subsequent rows indicate adjustments from baseline accuracy. Bracketed values represent standard errors.}
\label{tab:missing_reg_coeff}
\end{table}

\subsection{Generalized Linear Models} 
\label{sec:probit_examples}

In this section, we compare several different approaches to fitting the Bayesian binary regression models via probit and logistic regression on commonly used benchmark datasets 
%{\tt O-ring}, 
%{\tt Heart}, 
%{\tt Glass}, and 
%{\tt Ionosphere} 
%from the UCI Machine Learning Repository \citep{Dua2019}. 
{\tt Breastfeed} ($n=135, p=10$), 
{\tt Ionosphere}, ($n=351, p=33$) 
{\tt Spam} ($n=4601, p=58$),
and 
{\tt Caravan}  ($n=5822, p=86$).  
The dataset {\tt Breastfeed} is based on the study
of \textcite{moustaki1998uk} and is available from the website of the book of \textcite{heritier2009robust}. 
The datasets {\tt Ionosphere} and {\tt Spam}
can be found in the UCI Machine Learning Repository \parencite{Dua2019}. 
The dataset {\tt Caravan} can be found from the {\tt ISLR} package in {\tt R} \parencite{ISLRcite}.

\begin{table}[!ht]
\begin{center}
{\footnotesize
\centering
\begin{tabular}{l|l|ccccccr}
Dataset & Method  & RMSE($\mu$) & RMSE($\sigma$) & MIN & Q1. & Q2. & Frob. & Time (s) \\
\hline
Breastfeed 
& stan  & 0.0093 &  0.0044 & 97.6 & 97.8 & 98.2 & 0.0031 & 0.768 \\
& MFVB  & 0.0846 &  0.1910 & 65.3 & 68.0 & 73.5 & 0.0615 & 0.000 \\
%& PFVB  & 0.0745 &  0.1240 & 78.9 & 82.0 & 85.5 & 0.0455 & 0.207 \\
& Laplace   & 0.0846 &  0.0124 & 88.9 & 92.8 & 95.1 & 0.0056 & 0.000 \\
%& IL    & 0.0017 &  0.0007 & 99.4 & 99.6 & 99.6 &        & 0.045 \\
& EP    & 0.0011 &  0.0060 & 96.4 & 98.3 & 99.2 & 0.0030 & 0.459 \\
& GVB   & 0.0012 &  0.0073 & 96.4 & 98.1 & 99.1 & 0.0036 & 0.012 \\
%& DMVB  & 0.0051 &  0.0076 & 95.8 & 97.9 & 99.0 & 0.0039 & 0.003 \\
& MP    & 0.0012 &  {\bf 0.0037} & 96.5 & 98.3 & 99.2 & {\bf 0.0018} & 0.000 \\
%& PHA-L & 0.0846 &  0.0119 & 98.4 & 98.5 & 98.6 & 0.0056 & 0.001 \\
%& PHA-G & 0.0012 &  0.0038 & 99.3 & 99.5 & 99.6 & 0.0036 & 0.001 \\
%& PHA-E & 0.0011 &  0.0037 & 99.3 & 99.5 & 99.6 & 0.0030 & 0.001 \\
& GMP-L & 0.0155 &  0.0079 & 98.9 & 99.0 & 99.1 & 0.0035 & 0.004 \\
& GMP-G & 0.0011 &  0.0038 & {\bf 99.3} & {\bf 99.5} & {\bf 99.6} & 0.0023 & 0.002 \\
& GMP-E & {\bf 0.0009} &  0.0038 & {\bf 99.3} & {\bf 99.5} & {\bf 99.6} & 0.0021 & 0.001 \\
\hline 
Ion.
& stan  & 0.004 &   0.003 & {\bf 97.4} & 97.9 & 98.2 & 0.0012 &       2.78 \\
& MFVB  & 0.114 &   0.171 & 33.5 & 48.5 & 52.4 & 0.0170 &       0.02 \\
%& PFVB  & 0.123 &   0.131 & 40.1 & 60.2 & 66.4 & 0.0151 &       0.48 \\
& Laplace   & 0.114 &   0.011 & 60.4 & 79.1 & 89.1 & 0.0023 &       0.00 \\
%& IL    & 0.015 &   0.004 & 95.0 & 96.8 & 98.0 &        &       2.48 \\
& EP    & {\bf 0.002} &   {\bf 0.004} & 97.0 & {\bf 98.3} & {\bf 98.9} & {\bf 0.0010} &       0.55 \\
& GVB   & 0.003 &   0.013 & 95.7 & 96.9 & 97.3 & 0.0021 &       0.12 \\
%& DMVB  & 0.030 &   0.037 & 86.7 & 91.5 & 93.9 & 0.0077 &       0.08 \\
& MP    & 0.003 &   0.009 & 96.5 & 97.6 & 98.0 & 0.0017 &       0.00 \\
%& PHA-L & 0.114 &   0.047 & 86.0 & 88.9 & 90.4 & 0.0023 &       0.01 \\
%& PHA-G & 0.003 &   0.013 & 96.7 & 97.1 & 97.6 & 0.0021 &       0.01 \\
%& PHA-E & 0.002 &   0.014 & 95.4 & 96.7 & 97.0 & 0.0010 &       0.01 \\
& GMP-L & 0.015 &   0.019 & 94.3 & 95.6 & 95.9 & 0.0034 &       0.06 \\
& GMP-G & 0.003 &   0.013 & 96.7 & 97.1 & 97.6 & 0.0020 &       0.03 \\
& GMP-E & 0.004 &   0.014 & 95.9 & 96.9 & 97.3 & 0.0023 &       0.02 \\ 
\hline 
Spam
& stan  & 0.0019 & 0.0026 & 96.6 & 97.8 & 98.0 & 0.0004 &      40.1 \\
& MFVB  & 0.1180 & 0.0948 &  2.7 & 40.2 & 52.8 & 0.0051 &       0.6 \\
%& PFVB  & 0.1400 & 0.0881 &  2.0 & 52.4 & 65.7 & 0.0050 &      43.4 \\
& Laplace   & 0.0319 & 0.0065 & 75.8 & 93.1 & 96.4 & 0.0004 &       0.2 \\
%& IL    & 0.0024 & 0.0014 & 97.5 & 99.0 & 99.2 &         &     144.0 \\
& EP    & {\bf 0.0013} & 0.0056 & 82.4 & 96.8 & 98.4 & 0.0004 &       2.0 \\
& GVB   & 0.0031 & 0.0097 & 80.3 & 96.1 & 98.4 & 0.0007 &       2.5 \\
%& DMVB  & 0.0038 & 0.0133 & 75.0 & 96.4 & 98.2 & 0.0009 &       7.6 \\
& MP    & 0.0031 & 0.0093 & 80.8 & 96.1 & 98.4 & 0.0007 &       0.1 \\
%& PHA-L & 0.0319 & 0.0079 & 90.2 & 97.7 & 98.5 & 0.00045 &       0.2 \\
%& PHA-G & 0.0031 & 0.0031 & 97.3 & 99.1 & 99.2 & 0.0007 &       0.2 \\
%& PHA-E & 0.0013 & 0.0036 & 96.8 & 99.1 & 99.2 & 0.00036 &       0.2 \\
& GMP-L & 0.0050 & 0.0047 & 96.5 & 98.6 & 98.9 & 0.0005 &       1.1 \\
& GMP-G & 0.0023 & {\bf 0.0033} & {\bf 97.2} & {\bf 99.1} & {\bf 99.2} & 0.0005 &       0.6 \\
& GMP-E & 0.0016 & 0.0036 & 96.8 & 99.0 & {\bf 99.2} & {\bf 0.0003} &       0.3 \\
\hline 
Caravan
& stan   & {\bf 0.018} & {\bf 0.025} & {\bf 95.2} & 97.2 & 97.7 & {\bf 0.023} &     1653.5 \\
& MFVB   & 0.412 & 0.604 & 13.7 & 55.5 & 60.7 & 0.192 &        1.1 \\
%& PFVB   & 0.422 & 0.515 & 19.3 & 69.4 & 74.5 & 0.173 &      14.9 \\
& Laplace    & 0.407 & 0.109 & 53.2 & 94.4 & 96.7 & 0.044 &       0.6 \\
%& IL     & 0.171 & 0.050 & 80.3 & 97.8 & 98.1 &       &    1167.1 \\
& EP     & 0.025 & 0.074 & 81.1 & 96.9 & 98.0 & 0.034 &       3.5 \\
& GVB    & 0.267 & 0.396 & 21.2 & 27.7 & 40.2 & 0.119 &       6.4 \\
%& DMVB   & 0.285 & 0.234 & 64.9 & 95.5 & 98.0 & 0.131 &      27.9 \\
& MP     & 0.455 & 0.361 & 18.7 & 55.8 & 65.5 & 0.156 &       3.5 \\
% & PHA-L  0.407 & 0.293 & 11.4 & 86.5 & 92.3 & 0.044 &       0.3 \\
% & PHA-G  0.267 & 0.358 & 10.7 & 45.8 & 58.3 & 0.119 &       0.3 \\
% & PHA-E  & 0.025 & 0.092 & 82.9 & 97.5 & 98.1 & 0.034 &       0.3 \\
 & GMP-L & 0.098 & 0.111 & 85.4 & 96.1 & 97.7 & 0.058 &       8.7 \\
 & GMP-G & 1.100 & 0.521 &  0.5 & 31.2 & 44.2 & 0.161 &      21.4 \\
 & GMP-E  & 0.070 & 0.098 & 86.9 & {\bf 97.5} & {\bf 98.1} & 0.053 &       7.0 \\
\hline

\end{tabular}
}
\end{center}
\caption{Performance metrics for Probit regression. }
\label{tab:Probit_reg}
\end{table}

Table \ref{tab:Probit_reg} 
summarize the performance metrics described in Section \ref{sec:Numerical_Experience}
for probit models. 
Additional results for logistic regression models can be found in Supplementary Material
\ref{sec:additional}. However,
these results are similar to those for the probit model case.
The MP entry in Table \ref{tab:Probit_reg} corresponds to the method outlined in Section \ref{sec:probit_example} and generally performs similarly to other Gaussian-based ABI methods. For probit regression, MFVB is based on
\textcite{ConsonniMarin2007}.
% while for logistic regression, MFVB is based on \textcite{DuranteRigon2019}
For almost all models and datasets, either EP or one of the GMP variants are
the most accurate with competitive RMSE's and Frob. metrics while being tens
of times faster than a short-run of {\tt stan}. Most GMP-based approximations
improve significantly over the respective method they were initialized with.
Occasionally, our GMP method performed poorly due to numerical instability 
(see GMP-G for probit regression for the {\tt Caravan} dataset). GMP is comparable
in speed to other ABI methods.

\subsection{Bayesian Lasso regression} 
\label{sec:BayesLasso_examples}

Here, we compare MFVB and GVB (described in Supplementary Material  \ref{Suppl:Sec_Bayes_Lasso}) with GMP initialized with these methods. 
The parameter $\lambda$ is chosen using the Gibbs sampler described in \textcite{ParkCasella2008} using the prior $\lambda^2\sim\mbox{Gamma}(10^{-4},10^{-4})$ and taking the posterior mean of $\lambda$ based on 5000 samples after a burn-in of 1000 samples.
Table \ref{tab:Blasso} summarizes the metrics described in Section \ref{sec:Numerical_Experience} using the 
{\tt Hitters} ($n=263$, $p=19$) (ISLR package; \parencite{ISLRcite}), {\tt Crime} ($n=2215,p=98$) (UCI Machine Learning Repository; \parencite{communitiesandcrime183})
and {\tt Eye} ($n=120, p=200$) (flare package; \parencite{flare}) datasets.

Table \ref{tab:Blasso} shows that
for each dataset almost all marginal posterior density approximations
for both GMP based approximations are nearly visually indistinguishable from those based on {\tt stan} whilst being at least tens of times faster than the corresponding short run MCMC approach.

\begin{table}[!ht]
\begin{center}
{\footnotesize
\centering
\begin{tabular}{l|l|ccccccr}
Data & Method  & RMSE($\mu$) & RMSE($\sigma$) & MIN & Q1. & Q2. & Frob. & Time (s) \\
\hline
Hitters
& stan & 0.0926       & 0.0177       & 87.5       &   94.0     &  96.1       & 0.0342       &  28.94 \\ 
& MFVB & 0.0913       & 0.0144       & 31.7       &   69.8     &  85.6       & 0.0197       &   0.01 \\
& GVB  & {\bf 0.0019} & 0.0075       & 86.3       &   89.9     &  92.1       & 0.0099       &   0.01 \\
& GMP-M & 0.0024       & 0.0042       & 96.8       &   97.8     &  98.2       & 0.0052       &   0.03 \\
& GMP-G & 0.0093       & {\bf 0.0030} & {\bf 97.7} & {\bf 98.6} &  {\bf 99.2} & {\bf 0.0038} &   0.01 \\
\hline 
Crime
& stan & 0.0939       & 0.0041       & 43.9       &   83.3     &  89.4       & 0.0034 &  93.60 \\
& MFVB & 0.0140       & 0.0033       & 36.1       &   82.0     &  88.4       & 0.0038 &   0.03 \\
& GVB  & {\bf 0.0002} & 0.0013       & 79.1       &   94.0     &  95.6       & 0.0019 &   0.03 \\
& GMP-M & 0.0005       & {\bf 0.0007} & 95.1       &   98.8     &  99.0       & 0.0011 &   0.87 \\
& GMP-G & 0.0010       & {\bf 0.0007} & {\bf 95.9} & {\bf 99.6} &  {\bf 99.7} & {\bf 0.0008} &   0.60 \\
\hline 
Eye
& stan & 0.0183       & {\bf 0.0022} & 94.8       &   96.8     &  97.3      & 0.0090 &  34.34 \\
& MFVB & 0.0138       & 0.0056       & 51.5       &   79.8     &  83.3      & 0.0085 &   0.33 \\
& GVB  & 0.0013       & 0.0036       & 83.3       &   87.7     &  88.7      & 0.0048 &   0.08 \\
& GMP-M & {\bf 0.0012} & 0.0058       & 92.7       &   93.5     &  93.7      & 0.0064 &   1.95 \\
& GMP-G & 0.0069       & 0.0027       & {\bf 99.1} & {\bf 99.4} & {\bf 99.5} & {\bf 0.0042} &   3.16 \\
\hline
\end{tabular}
}
\end{center}
\caption{Performance metrics for Bayesian Lasso. }
\label{tab:Blasso}
\end{table}

\section{Conclusion}\label{sec:conclusion}

We have introduced moment propagation (MP), a framework for improving marginal posterior approximations by propagating conditional posterior moment information between parameter blocks. In conjugate settings, MP identifies variance terms omitted by MFVB and uses moment matching to construct corrected marginal updates. We developed both mean-field and Gaussian variants, showed exact marginal posterior recovery for linear regression and multivariate normal models, and established asymptotically correct posterior moments in two-component settings. Across the examples considered, MP-based methods, particularly GMP, often produced substantially more accurate marginal posterior approximations than MFVB and other approximate Bayesian inference methods while remaining much faster than MCMC.

Several challenges remain. GMP can depend on the quality of the initial Gaussian approximation, and numerical integration may introduce additional cost or instability in some models. Theoretical properties are currently best understood in conjugate two-component settings, leaving open questions about convergence, rates, and accuracy in more general multi-component and non-conjugate models. Future work will focus on these theoretical issues, on improved initialization,
stability, and acceleration strategies, and on extensions to richer hierarchical models, including random-effects and state-space settings.

%\input{Numerical_experience}
%\input{Conclusion}

%\section{Competing interests}
%The authors declare that they have no conflict of interest.

\section{Acknowledgments}
The following sources of funding are gratefully acknowledged: Australian Research Council Discovery Project grant (DP210100521) to JO. We also thank Prof. Matt P. Wand for making comments on an early draft of the paper. % and Dr. Minh-Ngoc Tran for providing MATLAB code.

%\clearpage
%\nocite{*}

\printbibliography[title={References}]

\newpage

\doublespacing

\begin{center}
{\Large
Supplementary material to ``Moment Propagation''}

\bigskip 

\if1\blind
{
by John T. Ormerod, 
Weichang Yu,
Mohammad Javad Davoudabadi, \&
Yuhao Li
} \fi

\bigskip 
%\date{\today}
\end{center}

\noindent 
This material will appear as an online supplement. 

\appendix

\begin{refsection}

\section{Moment Results}

This appendix collects moment identities used in the derivations of the MP updates. 
Sections \ref{sec:mvn_moments} and \ref{Suppl:Moments_for_the_Multivariate_t_Distribution} are used for quadratic forms appearing in the linear regression 
and multivariate normal examples. Section \ref{sec:moments_iw} gives inverse-Wishart moment identities 
used in the multivariate normal examples. Section \ref{sec:IW_results} records block inverse-Wishart 
identities needed for the missing-data updates.

Throughout this appendix, $\operatorname{dg}(\mA)$ denotes the vector of diagonal elements, 
$\bE\bW\bV(\mA)$ denotes the matrix of element-wise variances,
$\odot$ denotes the Hadamard product, and $\mK_{p,p}$ denotes the commutation matrix.

\subsection{Moments for the Multivariate Gaussian Distribution}
\label{sec:mvn_moments}

\noindent %
It is shown in \textcite{MathaiProvost1992}  that the $s$-th cumulant of $\vx^T\mA\vx$ where $\mA$ is a $p\times p$ symmetric matrix and 
$\vx \sim N_p(\vmu,\mSigma)$ is given by
$$
\kappa(s) = 2^{s-1} s!\left[ \frac{\mbox{tr}\{(\mA\mSigma)^s\}}{s} + \vmu^T(\mA\mSigma)^{s-1} \mA\vmu \right]. 
$$

\noindent 
The $h$-th moment of $\vx^T\mA\vx$ can be calculated recursively using these cumulants via
$$
\bE[(\vx^T\mA\vx)^h] = \sum_{i=0}^{h-1} \frac{ (h - 1)!}{(h - 1 - i)! i!} \kappa(h - i) \bE[(\vx^T\mA\vx)^i]
$$

\noindent with $\bE[(\vx^T\mA\vx)^0] = 1$.
We can use the above moments to derive
\begin{equation}\label{eq:moments_of_quadratic_normal}
\begin{array}{rl}
\bE\left[ (\vx^T\mA\vx)^2 \right] 
& \ds = 2\mbox{tr}(\mA\mSigma\mA\mSigma) + 4\vmu^T\mA\mSigma\mA\vmu + 
\left[\vmu^T\mA\vmu + \mbox{tr}(\mA\mSigma)  \right]^{2},   
\\ 
\bV(\vx^T\mA\vx)
& \ds =  2\operatorname{tr}(\mA \mSigma\mA \mSigma) 
+ 4\vmu^T\mA\mSigma\mA\vmu, \qquad \mbox{and}

\\ 
\mbox{Cov}(\vx^T\mA\vx,\vx^T\mB\vx)
& \ds = 2 \operatorname{tr}(\mA\mSigma\mB \mSigma)
+ 4\vmu^T\mA\mSigma\mB\vmu.

\end{array} 
\end{equation}

\subsection{Moments for the Multivariate 
\texorpdfstring{$t$}{t}  Distribution}\label{Suppl:Moments_for_the_Multivariate_t_Distribution}

\noindent The $p$-variate $t$-distribution, denoted as $\vx \sim t_p(\vmu, \mSigma, \nu)$, has the density function given by
$$
p(\vx) = 
\frac{\Gamma\left[(\nu+p)/2\right]}{\Gamma(\nu/2)\nu^{p/2}\pi^{p/2}\mbox{det}({\boldsymbol\Sigma})^{1/2}}\left[1+\frac{1}{\nu}({\mathbf x}-{\boldsymbol\mu})^{\rm T}{\boldsymbol\Sigma}^{-1}({\mathbf x}-{\boldsymbol\mu})\right]^{-(\nu+p)/2}
$$

\noindent where $\vmu$, $\mSigma$ and $\nu$ are the location, scale and
degrees of freedom parameters respectively. The mean and variance of $\vx$ are given by $\bE(\vx) = \vmu$ (provided $\nu>1$), covariance $\bV(\vx) = \nu\mSigma/(\nu - 2)$ (provided $\nu>2$).
Fourth moments require $\nu>4$.

\bigskip 
\noindent 
{\bf Theorem 2.1 of \textcite{RongEtAl2012}}
If $\vx \sim t_p(\vmu,a\mSigma,b)$
and $\mA$ is a symmetric matrix then
\begin{equation}\label{eq:Rong1}
\begin{array}{rl}
\bE\left[ \left(   \vx^T\mA\vx \right)^2 \right]
& \ds = \frac{a^2b^2}{(b - 2)(b - 4)}\left[ 2\,\mbox{tr}(\mA\mSigma\mA\mSigma) + \mbox{tr}(\mA\mSigma)^2 \right] 
\\ [2ex]
& \ds \qquad 
+ \frac{4ab}{b-2} \vmu^T\mA\mSigma\mA\vmu 
+ (\vmu^T\mA\vmu)^2
+ \frac{2ab}{b-2}  (\vmu^T\mA\vmu)\mbox{tr}(\mA\mSigma),
\end{array} 
\end{equation}

\begin{equation}\label{eq:Rong2}
\ds \bV(\vx^T\mA\vx) 
= \frac{2a^2 b^2 \mbox{tr}(\mA\mSigma\mA\mSigma)}{(b - 2)(b - 4)}
+ \frac{2a^2 b^2[\mbox{tr}(\mA\mSigma)]^2}{(b - 2)^2(b - 4)}
+ \frac{4ab}{b - 2}\vmu^T\mA\mSigma\mA\vmu,
\end{equation} 

\noindent and 
\begin{equation}\label{eq:Rong3}
\mbox{Cov}(\vx^T\mA\vx,\vx^T\mB\vx)
\ds 
= \frac{2 a^2b^2\,\mbox{tr}(\mA\mSigma\mB\mSigma)}{(b-2)(b-4)}
+ \frac{2a^2b^2\,\mbox{tr}(\mA\mSigma)\mbox{tr}(\mB\mSigma)}{(b-2)^2(b-4)}
+ \frac{4ab (\vmu^T\mA\mSigma\mB\vmu )}{b-2}
\end{equation}

\noindent provided $b>4$.

\subsection{Moments for the Inverse-Wishart distribution}
\label{sec:moments_iw}

We use the parameterization of the inverse-Wishart distribution to have the following density
\begin{equation}
\label{eqn::priorMVN}
\begin{array}{rl}
\ds p(\mSigma) & 
\ds =  
\frac{ \mbox{det}(\mPsi)^{d/2}}{
2^{d p/2} \Gamma_p(d/2)}   |\mSigma|^{-(d+p+1)/2}\exp\left[ -\tfrac{1}{2}\mbox{tr}(\mPsi\mSigma^{-1}) \right]
\cdot 
\bI(\mSigma \in \bS_p^+),
\end{array} 
\end{equation}

\noindent where
$$
\Gamma_p(a)
=
\pi^{p(p-1)/4}
\prod_{j=1}^p \Gamma\{a+(1-j)/2\}
$$

\noindent 
is the multivariate gamma function.. The following Theorem and Corollary can be found in \textcite{Rosen1988}.

\bigskip  
\noindent 
{\bf Theorem 3.1 of \textcite{Rosen1988}:} 
Let $\mSigma \sim \mbox{IW}_p(\mPsi,d)$ then
\begin{itemize}
	\item[i.] $\ds\bE(\mSigma) = \frac{\mPsi}{d - p - 1}$, $\qquad$ if $d - p - 1>0$; 
	
%	\item[ii.]	$\bE\left( \stackrel{2}{\otimes} \mW^{-1} \right) 
%	=   c_1 \stackrel{2}{\otimes} \mSigma^{-1}
\end{itemize}

\medskip 
\noindent 
{\bf Corollary 3.1 of \textcite{Rosen1988}:} 
Let $\mSigma \sim \mbox{IW}_p(\mPsi,d)$ with
$$
\begin{array}{c}
c_2^{-1} = (d - p)(d - p - 1)(d - p - 3),  \qquad
c_1  = (d - p - 2) c_2, \\ 
\ds c_3  = \frac{d - p - 3}{(d - p - 5)(d - p +1)}, \qquad \mbox{and} \qquad 
c_4  = \frac{2}{(d - p - 5)(d + p + 1)}.
\end{array}
$$

\noindent Then,  
\begin{itemize}
\item[(i)] if $d - p - 3 > 0$,
$$
\begin{array}{rl}
\ds \bE\left[\mSigma^2\right]  = (c_1 + c_2)\mPsi^2 + \textcolor{red}{c_2}\mPsi\mbox{tr}(\mPsi);
\end{array} 
$$

\item[(ii)] if $d - p - 5 > 0$,
$$
\begin{array}{rl}
\ds\bE\left[\mSigma^{3}\right] 
& \ds  
= (c_3c_1 + c_3c_2 + c_4c_1 + 5c_4 c_2) \mPsi^3 + (2c_3c_2 + c_4c_1 + c_4c_2) \mbox{tr}(\mPsi)\mPsi^2 \\  
& \ds \quad 
- (c_3c_2 + c_4c_2) \mbox{tr}(\mPsi^{-2})\mPsi   - c_4 c_2 (\mbox{tr}(\mPsi))^2\mPsi;
\end{array} 
$$

\item[(iii)] if $d - p - 5 > 0$,
$$
\begin{array}{rl}
\ds\bE\left[\mbox{tr}(\mSigma^2)\mSigma\right]
& \ds  
= c_4(c_1 + c_2) \mPsi^{3} 
+ (c_4c_1 + c_3c_2) [\mbox{tr}(\mPsi)]^2\mPsi
\\
& \ds \quad 
+ (c_3c_1 + c_3c_2 + c_4c_2)  \mbox{tr}(\mPsi^2)\mPsi
- 2c_4c_2 \mbox{tr}(\mPsi)\mPsi^2;
\end{array} 
$$

\item[(iv)] if $d - p - 5 > 0$,
$$
\begin{array}{rl}
\ds\bE\left[\mbox{tr}(\mSigma)^3\right]
& \ds  
= c_3c_1\left[\mbox{tr}(\mPsi)^3\right] 
+ 4c_4c_2 \mbox{tr}(\mPsi^3) 
+ 2(c_4c_1 + c_3c_2) \mbox{tr}(\mPsi^2) \mbox{tr}(\mPsi);
\end{array} 
$$

\item[(v)] if $d - p - 3 > 0$,
$$
\bE\left[ \mbox{tr}(\mSigma)\mSigma\right]  
= c_1 \mbox{tr}(\mPsi)\mPsi
+ 2c_2 \mPsi^2
$$

\item[(vi)] if $d - p - 1 > 0$,
$$
\bE\left[ \mSigma \otimes \mSigma^{-1} \right]  
= \frac{d}{d - p - 1} \mPsi \otimes \mPsi^{-1}
- \frac{1}{d - p - 1}\left[ \mbox{vec}(\mI_p)\mbox{vec}(\mI_p)^T + \mK_{p,p} \right]; \quad\mbox{and}
$$

\item[(vii)] if $d - p - 1 > 0$,
$$
\bE\left[ \mbox{tr}(\mSigma)   \mSigma^{-1} \right]  
= \frac{d}{d - p - 1} \mbox{tr}(\mPsi)\mPsi^{-1}
- \frac{2}{d - p - 1} \mI_p.
$$

\end{itemize}

\noindent 
%where $\mSigma^2=\mSigma\mSigma$ and 
%$\mSigma^3=\mSigma\mSigma\mSigma$.
In \textcite{Rosen1988}, the coefficient multiplying $\mPsi^2$ in part (i) appears to contain a typographical error. The expression displayed above is the version used in our derivations.
% Note that the red highlighted  $c_2$ coefficient in (i) above is written as $c_1$ in \textcite{Rosen1988}, presumably a typo. The above expression in (i) is correct.
% Note that the expression for $c_3$ is different from the Wikipedia
% entry `Inverse-Wishart distribution'
%\bigskip
%\noindent 
%The corresponding $q$-density moments are:
%$$
%\begin{array}{rl}
%\bE_q(\mSigma) & \ds = \frac{ 
%	\widetilde{\mPsi}
%}{
%	\widetilde{d} - p  - 1
%} 
%\\ [2ex]
%
%\bE\bW\bV_q(\mSigma)
%& \ds 
%=  \frac{(\widetilde{d} - p+1)(\widetilde{\mPsi}\circ\widetilde{\mPsi})}{(\widetilde{d} - p)(\widetilde{d} - p-1)^2(\widetilde{d} - p-3)}  + \frac{  \mbox{dg}(\widetilde{\mPsi}) \mbox{dg}(\widetilde{\mPsi})^T}{(\widetilde{d} - p)(\widetilde{d} - p-1)(\widetilde{d} - p-3)}
%
%\\ [2ex]
%\mbox{dg}[\bE\bW\bV_q(\mSigma)]
%& \ds =  \frac{2\mbox{dg}(\widetilde{\mPsi})^2}{ %(\widetilde{d} - p-1)^2(\widetilde{d} - p-3)}
%
%\\ [2ex]
%
%\bE_q\left[ \ln \mbox{det}(\mSigma) \right]
%& \ds = -\psi_{p}=(\widetilde{d}/2) 
%- p\,\ln(2) + \ln\mbox{det}(\widetilde{\mPsi})   \\ [2ex]
%
%\bE_q\left[ \ln \mbox{det}(\mOmega) \right]
%& \ds =  \psi_{p}(\widetilde{d}/2) 
%+ p\,\ln(2) + \ln\mbox{det}(\widetilde{\mV})
%
%\\ [2ex]
%
%\bE_q(\mOmega) 
%& \ds = \widetilde{d}\,\widetilde{\mV}  
%
%\\ [2ex]
%
%
%\bE\bW\bV_q(\mOmega) 
%
%& \ds = \widetilde{d} [\widetilde{\mV}\circ\widetilde{\mV} + \mbox{dg}(\widetilde{\mV})\mbox{dg}(\widetilde{\mV})^T]
%
%\\ [2ex]
%
%\mbox{dg}[\bE\bW\bV_q(\mOmega) ] 
%& = 2\widetilde{d}\,\mbox{dg}(\widetilde{\mV})^2
%
%\end{array} 
%$$
%\noindent where $\widetilde{\mV} = \widetilde{\mPhi}^{-1}$.
The variances and covariances for elements 
of $\mSigma$
are given by 
\begin{equation}\label{eq:var_cov_inv_wishart}
\begin{array}{rl}
\bV(\Sigma_{ij}) 

& \ds = \frac{(d-p+1)\psi_{ij}^2 + (d-p-1)\psi_{ii}\psi_{jj}}
{(d-p)(d-p-1)^2(d-p-3)},

\\ [1ex]

\bV(\Sigma_{ii}) 
& \ds = \frac{2\psi_{ii}^2}{(d-p-1)^2(d-p-3)},
\qquad \mbox{and}

\\ [1ex]

\mbox{Cov}(\Sigma_{ij}, \Sigma_{k\ell}) 
& \ds = \frac{2\psi_{ij}\psi_{k\ell} + (d-p-1) (\psi_{ik}\psi_{j\ell} + \psi_{i\ell} \psi_{kj})}{(d-p)(d-p-1)^2(d-p-3)}.

\end{array} 
\end{equation} 

\noindent provided $d-p-3>0$ \parencite[see][]{press2005applied}. Note that the element-wise variance matrix for $\mSigma$ is
$$
\bE\bW\bV(\mSigma) = \frac{(d-p+1)(\mPsi\odot\mPsi)}{(d-p)(d-p-1)^2(d-p-3)}  + \frac{(d-p-1) \mbox{dg}(\mPsi) \mbox{dg}(\mPsi)^T}{(d-p)(d-p-1)^2(d-p-3)}
$$

\noindent 
%where $\odot$ denotes the Hadamard product, i.e., if $\mA$ and $\mB$ are of conforming dimensions then $(\mA\odot\mB)_{ij} = a_{ij}b_{ij}$, 
whose diagonal elements are
$$
\mbox{dg}[\bE\bW\bV(\mSigma)] = \frac{2\mbox{dg}(\mPsi)^{\odot 2}}{(d - p - 1)^2(d - p - 3)}.
$$

\subsection{Block Inverse-Wishart identities}
\label{sec:IW_results}

In Section \ref{sec:missing} some
identities concerning block inverse-Wishart distributions are
needed in the derivation  $\bE_r(\vx_{i,M_i})$ and  $\bV_r(\vx_{i,M_i})$
needed to update $q(\vx_{i,M_i})$.  
The update for $q(\vx_{i,M_i})$ requires expectations involving the
Gaussian conditional regression matrix
$\mSigma_{M_iO_i}\mSigma_{O_iO_i}^{-1}$
and the Schur complement
$\mSigma_{M_i|O_i}
  =
  \mSigma_{M_iM_i}
  -
  \mSigma_{M_iO_i}\mSigma_{O_iO_i}^{-1}\mSigma_{O_iM_i}$.
The following proposition collects the Inverse-Wishart identities used in Section~\ref{sec:missing} and   Section~\ref{sec:mvn_missing_derivations}.

\paragraph{Proposition.}
Let $\mSigma\sim IW_p(\mPsi,d)$,
and let $M$ and $O$ be non-empty disjoint index sets forming a partition
of $\{1,\ldots,p\}$, with $m=|M|$ and $o=|O|$. Define
$\mSigma_{M|O} =
  \mSigma_{MM}
  -
  \mSigma_{MO}\mSigma_{OO}^{-1}\mSigma_{OM},
$
and
$\mPsi_{M|O} = \mPsi_{MM} -
  \mPsi_{MO}\mPsi_{OO}^{-1}\mPsi_{OM}$.
The following proposition assumes 
$O\neq\varnothing$. The case $O=\varnothing$ is handled separately in the main text, where 
$\mSigma_{M| O}$ reduces to $\mSigma_{MM}$.
Then, 
$$
\mSigma_{M|O}\sim IW_m(\mPsi_{M|O},d),
\quad \mbox{and} \quad 
\mSigma_{MO}\mSigma_{OO}^{-1} \mid \mSigma_{M|O}
  \sim
  MN_{m\times o}
  \left(
    \mPsi_{MO}\mPsi_{OO}^{-1},
    \mSigma_{M|O},
    \mPsi_{OO}^{-1}
  \right)
$$

\noindent where $MN_{m\times o}(\mA,\mB,\mC)$ denotes the matrix-variate normal distribution.
Consequently,
$$
  \bE\left(\mSigma_{MO}\mSigma_{OO}^{-1}\right)
  =
  \mPsi_{MO}\mPsi_{OO}^{-1}, \quad \mbox{and} \qquad
  \bE(\mSigma_{M|O})
  =
  \frac{\mPsi_{M|O}}{d-m-1}
$$

\noindent provided $d>m+1$.
Moreover, for any fixed $o$-vector $\vx$,
$$
  \bV\left(
    \mSigma_{MO}\mSigma_{OO}^{-1}\vx
  \right)
  =
  \frac{\vx^T\mPsi_{OO}^{-1}\vx}{d-m-1}
  \mPsi_{M|O}.
$$

\paragraph{Proof sketch.}
The distributional statements follow from Theorem~3 of \textcite{BodnarOkhrin2008}, after relabelling the blocks. The first two moment identities
follow directly. For the final identity, condition on 
$\mSigma_{M|O}$:
$$
  \mSigma_{MO}\mSigma_{OO}^{-1}\vx
  \mid \mSigma_{M|O}
  \sim
  N_m
  \left(
    \mPsi_{MO}\mPsi_{OO}^{-1}\vx,
    (\vx^T\mPsi_{OO}^{-1}\vx)\mSigma_{M|O}
  \right).
$$

\noindent 
The conditional mean is non-random, so the law of total variance gives
$$
  \bV\left(
    \mSigma_{MO}\mSigma_{OO}^{-1}\vx
  \right)
  =
  (\vx^T\mPsi_{OO}^{-1}\vx)\,
  \bE(\mSigma_{M|O})
  =
  \frac{\vx^T\mPsi_{OO}^{-1}\vx}{d-m-1}
  \mPsi_{M|O}.
$$

\newpage 
\section{Proof of Theorems  \ref{ConvergenceToLimitSolution} to \ref{asympNormality} }\label{Suppl:profTheo1}

\subsection*{Recap of MFMP setup}

We approximate each marginal posterior of each subvector $p(\vtheta_k \mid \sD)$ with $q_k (\vtheta_k; \widehat{\vlambda}_k) \in \sQ_k = \{ q_k( \vtheta_k; \vlambda_k) \, : \, \vlambda_k \in \mLambda_k \}$, where each $\sQ_k$ is a convenient parametric family of distributions. Here, we specify $\{ \widehat{\vlambda}_k \}_{k=1}^K$ to solve the system of integral equations for $k = 1, \ldots, K$:
\begin{align*}
&\int \vtheta_k q_k (\vtheta_k; \vlambda_k) \, \text{d} \vtheta_k 
    = \int \vtheta_k \int^{\otimes^{K-1}} p( \vtheta_k \mid \sD, \vtheta_{-k} ) \prod_{j \neq k} q(\vtheta_j; \vlambda_j) \, \text{d} \vtheta_{j} \; \text{d} \vtheta_k  
\\
&\int \vtheta_k \vtheta_k^\top q_k (\vtheta_k; \vlambda_k) \, \text{d} \vtheta_k 
    = \int \vtheta_k \vtheta_k^\top \int^{\otimes^{K-1}} p( \vtheta_k \mid \sD, \vtheta_{-k} ) \prod_{j \neq k} q(\vtheta_j; \vlambda_j) \, \text{d} \vtheta_{j} \; \text{d} \vtheta_k 
\\
&\int \mathbf{s}_k (\vtheta_k) q_k (\vtheta_k; \vlambda_k) \, \text{d} \vtheta_k 
    = \int \mathbf{s}_k (\vtheta_k) \int^{\otimes^{K-1}} p( \vtheta_k \mid \sD, \vtheta_{-k} ) \prod_{j \neq k} q(\vtheta_j; \vlambda_j) \, \text{d} \vtheta_{j} \; \text{d} \vtheta_k
\end{align*}

\noindent 
Essentially, we are choosing the $q_k$'s such that their respective moments $\bE(\vtheta_k)$, $\bE( \vtheta_k \vtheta_k^\top)$, and $\bE[ \mathbf{s}_k (\vtheta_k) ]$ match those based on 
$\widetilde{r}_k (\vtheta_k; \vlambda_{-k}) = \int^{\otimes^{K-1}} p( \vtheta_k \mid \sD, \vtheta_{-k} ) \prod_{j \neq k} q(\vtheta_j; \vlambda_j) \, \text{d} \vtheta_{j}$. 
We mandate the matching of the first and second order moment of each $\vtheta_k$, while allowing the users flexibility to match other suitable moments through $\mathbf{s}_k$. If no additional moments are required to be matched, then $\mathbf{s}_k (\vtheta_k) = 1$. To obtain the solution $\{ \widehat{\vlambda}_k \}_{k=1}^K$, we iterate through a fixed point algorithm. More specifically, at iteration $\tau = 1, 2, \ldots$, we update $\lambda_k^{(\tau)}$ for $k = 1, \ldots, K$ such that
\begin{align*}
&\int \vtheta_k q_k (\vtheta_k; \vlambda_k^{(\tau)}) \, \text{d} \vtheta_k 
    = \int \vtheta_k \int^{\otimes^{K-1}} p( \vtheta_k \mid \sD, \vtheta_{-k} ) \prod_{j \neq k} q(\vtheta_j; \vlambda_j^{(\tau-1)}) \, \text{d} \vtheta_{j} \; \text{d} \vtheta_k  
\\
&\int \vtheta_k \vtheta_k^\top q_k (\vtheta_k; \vlambda_k^{(\tau)}) \, \text{d} \vtheta_k 
    = \int \vtheta_k \vtheta_k^\top \int^{\otimes^{K-1}} p( \vtheta_k \mid \sD, \vtheta_{-k} ) \prod_{j \neq k} q(\vtheta_j; \vlambda_j^{(\tau-1)}) \, \text{d} \vtheta_{j} \; \text{d} \vtheta_k 
\\
&\int \mathbf{s}_k (\vtheta_k) q_k (\vtheta_k; \vlambda_k^{(\tau)}) \, \text{d} \vtheta_k 
    = \int \mathbf{s}_k (\vtheta_k) \int^{\otimes^{K-1}} p( \vtheta_k \mid \sD, \vtheta_{-k} ) \prod_{j \neq k} q(\vtheta_j; \vlambda_j^{(\tau-1)}) \, \text{d} \vtheta_{j} \; \text{d} \vtheta_k
\end{align*}

\subsection*{Auxiliary notations and assumptions}
We define notations:
	$$
	\mathbf{T}_k (\vtheta_k)  = \begin{pmatrix}
		\vtheta_k \\
		\text{vech} ( \vtheta_k \vtheta_k^\top ) \\
		\mathbf{s}_k (\vtheta_k)
	\end{pmatrix}
	$$
	and
	$$
	\mPsi_n ( \vlambda ) = \vm (\vlambda) - \mG_n (\vlambda), \;\;\;\; \mPsi (\vlambda) = \vm(\vlambda) - \mG(\vlambda),
	$$
	where
	$$
	\vm (\vlambda)  = \begin{pmatrix}
		\int \mathbf{T}_1 (\vtheta_1) q_1 (\vtheta_1; \vlambda_1) \, \text{d} \vtheta_1 \\
		\vdots \\
		\int \mathbf{T}_K (\vtheta_K) q_K (\vtheta_K; \vlambda_K) \, \text{d} \vtheta_K 
	\end{pmatrix},
	$$
	and
	$$
	\mG_n (\vlambda)  = \begin{pmatrix}
		\int \mathbf{T}_1 (\vtheta_1) \int^{\otimes^{K-1}} p( \vtheta_1 \mid \sD, \vtheta_{-1} ) \prod_{j \neq 1} q(\vtheta_j; \vlambda_j) \, \text{d} \vtheta_{j} \; \text{d} \vtheta_1\\
		\vdots \\
		\int \mathbf{T}_K (\vtheta_K) \int^{\otimes^{K-1}} p( \vtheta_K \mid \sD, \vtheta_{-K} ) \prod_{j \neq K} q(\vtheta_j; \vlambda_j) \, \text{d} \vtheta_{j} \; \text{d} \vtheta_K
	\end{pmatrix}.
	$$
	Denote the solution set by $\sS_n = \{ \vlambda_{1:K} \, : \, \mPsi_n(\vlambda_{1:K}) = \vzero  \}$.
	Also, we denote the full conditional expectations as
	$$
	\vh_{kn} ( \vtheta _{-k}) = \bE \{  \mathbf{T}_k ( \vtheta_k ) \mid \sD_n, \vtheta_{-k} \}.
	$$
    Denote the true value of the parameter subvectors by $\vtheta_{1:K}^\star$. Denote the induced expectation and scaled variance of any point $\vlambda_k \in \mLambda_k$ by $\vmu_k (\vlambda_k ) = \int \vtheta_k q_k(\vtheta_k; \vlambda_k) \; \text{d} \vtheta_k$, $\mGamma_k^{(n)} (\vlambda_k) = n  \int \{\vtheta_k - \vmu_k (\vlambda_k )\} \{\vtheta_k - \vmu_k (\vlambda_k )\}^\top q_k(\vtheta_k; \vlambda_k) \; \text{d} \vtheta_k$, and $\mU_k ( \vlambda_k  ) = \tfrac{1}{n} \mGamma_k^{(n)} (\vlambda_k) + \vmu_k (\vlambda_k ) \vmu_k (\vlambda_k )^\top$.
    The theorems presented in this section may require only a subset of the following list of assumptions. The required assumptions for each theorem is stated in their respective technical statements.
	\begin{itemize}
		\item[(A1)] The variational parameter space $\mLambda_{1:K} = \mLambda_1 \times \ldots \times \mLambda_K$ is convex and compact.
		\item[(A2)] There exists a unique $\vtheta_{1:K}^\star = ( \vtheta_1^\star, \ldots, \vtheta_K^\star ) \in \mTheta$ and sequence $\epsilon_n \rightarrow 0$ such that
		$$
		p (  \lVert \vtheta_{1:K} - \vtheta_{1:K}^\star \rVert  > \epsilon_n \mid \sD_n ) \xrightarrow{\bP^\star} 0.
		$$
		Moreover, there exists functions $\{ \vh_k^\star(\cdot) \}_{k=1}^K$ such that $\vh_k^\star( \vtheta_{-k}^\star ) = \mathbf{T}_k ( \vtheta_{k}^\star )$ and \\
		$\sup_{\vtheta_{-k} \joc{\in} \mTheta_{-k}}  \lVert  \vh_{kn} ( \vtheta _{-k}) - \vh_k^\star( \vtheta_{-k} )  \rVert = O_p(\epsilon_n)$, for all $k = 1, \ldots, K$.

        \item[(A3)] The sequence of maps $\{ \mPsi_n \}_{n \ge 1}$ is continuous in $\vlambda_{1:K} \in \mLambda_{1:K}$. Moreover, the sequence of solution set $\sS_n$ is non-empty almost surely.
		\item[(A4)] There exists a unique $\vlambda_{1:K}^\star \in \mLambda_{1:K}$ that satisfies
		\begin{align*}
			&\int \mathbf{T}_k (\vtheta_k) q_k (\vtheta_k; \vlambda_k^\star ) \, \text{d} \vtheta_k = \int^{\otimes^{K-1}} \vh_k^\star( \vtheta_{-k} ) \prod_{j \neq k} q(\vtheta_j; \vlambda_j^\star) \, \text{d} \vtheta_{j},
		\end{align*}
		for all $k = 1, \ldots, K$. The integral $\int \mathbf{T}_k (\vtheta_k) q_k (\vtheta_k; \cdot ) \, \text{d} \vtheta_k$ is continuous in $\vlambda_k \in \mLambda_{k}$ and $\int^{\otimes^{K-1}} \vh_1^\star( \vtheta_{-k} ) \prod_{j \neq k} q(\vtheta_j; \cdot) \, \text{d} \vtheta_{j}$ is continuous in $\vlambda_{-k} \in \mLambda_{-k}$.
		\item[(A5)] Denote
		$$
		\mPsi (\vlambda) = \vm(\vlambda) - \mG(\vlambda)
		$$
		where
		$$
		\mG (\vlambda)  = \begin{pmatrix}
			\int^{\otimes^{K-1}} \vh_1^\star( \vtheta_{-1} ) \prod_{j \neq 1} q(\vtheta_j; \vlambda_j) \, \text{d} \vtheta_{j} \\
			\vdots \\
			\int^{\otimes^{K-1}} \vh_K^\star( \vtheta_{-K} ) \prod_{j \neq K} q(\vtheta_j; \vlambda_j) \, \text{d} \vtheta_{j}
		\end{pmatrix}.
		$$
		For every sufficiently large $\omega > 0$, the following lower bound holds: $\inf_{\vlambda_{1:K} \in \sG_n(\omega) } \lVert \mPsi (\vlambda) \rVert \ge M_\omega \epsilon_n$, where  $\sG_{n} (\omega) = \{ \vlambda_{1:K} \in \mLambda_{1:K} \, : \,  \lVert \vlambda_{1:K} - \vlambda_{1:K}^\star \rVert \ge \omega \epsilon_n  \}$, $\vlambda_{1:K}^\star$ is a point in $\mLambda_{1:K}$ such that $\mPsi( \vlambda_{1:K}^\star ) = \vzero$, and $\lim_{\omega \rightarrow \infty} M_\omega = \infty$.
        \item[(A6)] $\vmu_k (\vlambda_k^\star) = \vtheta_k^\star$ and $\mU_k  ( \vlambda_k^\star  )  = \vtheta_k^\star \vtheta_k^{\star \, \top}$ for all $k = 1, \ldots, K$.
        \item[(A7)]  The marginal posterior distribution of the localized parameter $\vz_k = \sqrt{n}(\vtheta_k - \vtheta_k^*)$ converges in total variation distance to a centered Gaussian distribution:
		$$
		\text{DTV} \{ p(\vz_k \mid \sD_n) \Vert \text{MVN} ( \vzero, \mV_k^\star )  \} \xrightarrow{\bP^{\star}} 0
		$$
		where $\mV_k^\star$ is a strictly positive definite asymptotic covariance matrix. Furthermore, we assume the scaled posterior covariance converges: $ n\text{Var}(\vtheta_k \mid \sD_n) \xrightarrow{\bP^\star} \mV_k^\star.$
		\item[(A8)] The full conditional mean $\vm_{kn}$ is continuously differentiable with a bounded Jacobian matrix in $\sB_{-k} = \left \{ \vtheta_{-k} \, : \, \lVert \vtheta_{-k} - \vtheta_{-k}^\star \rVert \le R \right \}$. The full conditional variance is bounded $\lVert \vomega_{kn}(\vtheta_{-k}) \rVert \le M_{\text{prior}}$ for all $\vtheta_{-k} \in \mTheta_{-k}$. Moreover, for all $\vtheta_{-k} \in \sB_{-k}$, the following expansion holds: $\vomega_{kn}(\vtheta_{-k}) = n^{-1} \widetilde{\vomega}_{kn}(\vtheta_{-k}) + \vr_n(\vtheta_{-k})$. The function $\widetilde{\vomega}_{kn}$ is strictly positive definite and twice-continuously differentiable with a bounded Hessian matrix, and the remainder satisfies $\sup_{\vtheta_{-k} \in \mathcal{B}_{-k}} \lVert n \cdot \vr_n(\vtheta_{-k}) \rVert \to 0$ almost surely.
		\item[(A9)] For each $n$, the pair of functions $\vlambda_k \mapsto (\vmu_{k}(\vlambda_k), \mGamma_{k}^{(n)}(\vlambda_k))$ defines a unique mapping between the variational parameter space $\mLambda_k$ and the location-variance space. Furthermore, this mapping is a twice-continuously differentiable diffeomorphism in an open neighborhood of $\vlambda_k^\star$ with a nonsingular Jacobian. Moreover, the limit $\mGamma_k^\star := \text{lim}_{n \rightarrow \infty} \mGamma_{k}^{(n)}( \widehat{\vlambda}_{k})$ is strictly positive definite and has bounded eigenvalues.
	\end{itemize}
	Condition (A1) is necessary to guarantee the existence of a solution under most versions of fixed point theorems. This assumption required to establish uniform convergence of the finite $n$ fixed point conditions to their respective limits. Condition (A2) assumes the exact posterior is consistent at rate $\epsilon_n$ and the expectations of the full conditionals are uniformly convergent. Condition (A3) is guarantees the existence of a sequence solutions to the finite $n$ fixed point equations. Condition (A4) assumes the uniqueness of solution to the limiting fixed point problem. Condition (A5) is an identifiability condition on the unique solution to the limiting fixed point equations. Condition (A6) is required to ensure that the solution to the limiting MFMP system of integral equations corresponds to a dirac mass on the true value. Condition (A7) ensures that the individual $K$ exact marginal posteriors exhibit a Berstein-von Mises property and is required to establish asymptotic normality of the MFMP posterior. Condition (A8) is a smoothness assumption on the full conditional mean and variances. Condition (A9) is a smoothness assumption on the mapping between the variational parameters and the location-scale space. (A8) and (A9) are essential for proving asymptotic equivalence between the MFMP posterior variance and the exact posterior variance. Condition (A10) assumes that the choice of approximating distributions admit an asymptotically normal form as $n$ diverges.

\subsection{Statement and proof of Theorem 1}

		Assume (A1) to (A5) holds. Then, there exists a sequence of $\{  n_l \}_{l=1}^\infty$ such that $\sS_{n_l} \neq \emptyset$. Moreover, there exists a corresponding sequence of solutions $\{ \widehat{\vlambda}_{1:K}^{(n_\ell)} \}_{\ell = 1}$ such that
		$$
		\lVert \widehat{\vlambda}_{1:K}^{(n_\ell)} - \vlambda_{1:K}^\star \rVert = O_p(\epsilon_n).
		$$
		\begin{proof}
			The existence of an increasing sequence $\{  n_l \}_{l=1}^\infty$ such that $\mS_{n_l} \neq \emptyset$ follows from (A3). Consider the event $\{ \widehat{\vlambda}_{1:K}^{(n_l)} \in \sG_{n_l} (\omega) \}$. Then, since $ \mPsi_{n_l} ( \widehat{\vlambda}_{1:K}^{(n_l)}  ) = \vzero$, the event $\{ \widehat{\vlambda}_{1:K}^{(n_l)} \in \sG_{n_l} (\omega) \}$ implies the event
			\begin{align*}
				\{ \lVert \mPsi ( \widehat{\vlambda}_{1:K}^{(n_l)}  ) - \mPsi_{n_l} ( \widehat{\vlambda}_{1:K}^{(n_l)}  )  \rVert \ge M_\omega \epsilon_{n_l} \}
			\end{align*}
			Then,
			$$
            \begin{array}{rl}
			\bP^\star \left \{ \widehat{\vlambda}_{1:K}^{(n_l)} \in \sG_{n_l} (\omega)  \right \}  
                &  \ds \le \bP^\star \left \{ \lVert \mPsi ( \widehat{\vlambda}_{1:K}^{(n_l)}  ) - \mPsi_{n_l} ( \widehat{\vlambda}_{1:K}^{(n_l)}  )  \rVert \ge M_\omega \epsilon_{n_l} \right \} 
                \\
                & \le \bP^\star \left \{ \sup_{\vlambda_{1:K} \in \mLambda_{1:K} } \lVert \mPsi ( \vlambda_{1:K}  ) - \mPsi_{n_l} ( \vlambda_{1:K}  )  \rVert \ge M_\omega \epsilon_{n_l} \right \}.
			\end{array}
            $$
			From Lemma \ref{supNormConvergence}, for any $\delta > 0$, we can find a $(\omega, M_\omega)$ pairing such that  
			$$
			\bP^\star \left \{ \widehat{\vlambda}_{1:K}^{(n_l)} \in \sG_{n_l} (\omega)  \right \} \le \bP^\star \left \{ \sup_{\vlambda_{1:K} \in \mLambda_{1:K} } \lVert \mPsi ( \vlambda_{1:K}  ) - \mPsi_{n_l} ( \vlambda_{1:K}  )  \rVert \ge M_\omega \epsilon_{n_l} \right \} < \delta.
			$$
			Hence, $ \lVert \widehat{\vlambda}_{1:K} -  \vlambda_{1:K}^{\star} \rVert = O_p(\epsilon_n)$ as required.
		\end{proof}

\subsection{Statement and proof of Theorem 2}
Assume (A1) to (A6) holds. Then, for any $\epsilon > 0$, 
		$$\mathbb{P}_{q_k(\cdot; \widehat{\vlambda}_k)} \left \{ \lVert \vtheta_k - \vtheta_k^\star \rVert > \epsilon \right \} \xrightarrow{\bP^\star} 0.$$
		\begin{proof}
			By the continuity of $\int \vtheta_k q_k(\vtheta_k; \vlambda_k) \; \text{d} \vtheta_k$ and $\int \vtheta_k \vtheta_k^\top q_k(\vtheta_k; \vlambda_k) \; \text{d} \vtheta_k$ in $\vlambda_k \in \mLambda_k$ as assumed in (A4) and Theorem \ref{ConvergenceToLimitSolution}, we have (by CMT):
			$$
			\vmu_k ( \widehat{\vlambda}_k ) \xrightarrow{\bP^\star} \int \vtheta_k q_k(\vtheta_k; \vlambda_k^\star) \; \text{d} \vtheta_k = \vtheta_k^\star 
			$$
			and
			$$
			\mU_k  ( \widehat{\vlambda}_k ) \xrightarrow{\bP^\star} \int \vtheta_k \vtheta_k^\top q_k(\vtheta_k; \vlambda_k^\star) \; \text{d} \vtheta_k = \vtheta_k^\star \vtheta_k^{\star \, \top}. 
			$$
			Hence
			$$
			\tfrac{1}{n} \mGamma_k^{(n)} ( \widehat{\vlambda}_k) \xrightarrow{\bP^\star} \vzero.
			$$
			Then,
            $$
				\mathbb{P}_{q_k(\cdot; \widehat{\vlambda}_k)} \left \{ \lVert \vtheta_k - \vtheta_k^\star \rVert > \epsilon \right \} \\
				\le \mathbb{P}_{q_k(\cdot; \widehat{\vlambda}_k)} \left \{ \lVert \vtheta_k - \vmu_{k} ( \widehat{\vlambda}_k )  \rVert > \epsilon/2 \right \} + \mathbb{I} \left \{ \lVert \vmu_{k} ( \widehat{\vlambda}_k ) - \vtheta_k^\star \rVert > \epsilon/2 \right \}.
			$$
			Now, by Chebychev's inequality, we have
			$$
			\mathbb{P}_{q_k(\cdot; \widehat{\vlambda}_k)} \left \{ \lVert \vtheta_k - \vmu_{k} ( \widehat{\vlambda}_k ) \rVert > \epsilon/2 \right \} \le  \frac{ 4 \text{tr} ( \bV_{\vtheta_k \sim q_k(\cdot; \widehat{\vlambda}_k) } \left [ \vtheta_k  \right ] )  }{ \epsilon^2} \xrightarrow{\bP^\star} 0.
			$$
			Moreover, by (A6), we have
			$$
			\lVert  \vmu_{k} ( \widehat{\vlambda}_k ) - \vtheta_k^\star \rVert  =  \lVert \vmu_{k} ( \widehat{\vlambda}_k ) - \vmu_{k} ( \vlambda_k^\star ) \rVert \xrightarrow{\bP^\star} 0,
			$$
			where the last stochastic convergence holds by CMT. Hence, we have proved our required result.

        \end{proof}

\subsection{Statement and proof of Theorem 3}
Assume (A1) to (A9) holds. Then, for each $k = 1, 2$, we have
		$$
		\lVert \mSigma_{kn} - \mathbf{V}_{kn} \rVert = O_p(n^{-3/2}),
		$$
		where $\mV_{kn} = \bV_{p(\vtheta_k \mid \sD_n)} ( \vtheta_k \mid \sD_n )$.
		\begin{proof}
			Consider the sequence of MFMP solutions $\{ \widehat{\vlambda}_k^{(n)} \}_{n \ge 1}$ and their corresponding location-scale map 
			$$\{  (\vmu_{k} ( \widehat{\vlambda}_k^{(n)} ) , \mGamma_{k}^{(n)} ( \widehat{\vlambda}_k^{(n)} ) ) \}_{n \ge 1}.$$ For notational brevity, we will write shorthand $\vmu_{kn} := \vmu_{k} ( \widehat{\vlambda}_k^{(n)} )$,  $\mGamma_{kn} := \mGamma_{k}^{(n)} ( \widehat{\vlambda}_k^{(n)} )$, and $\mSigma_{kn} := \tfrac{1}{n} \mGamma_{k}^{(n)} ( \widehat{\vlambda}_k^{(n)} )$. Now, by definition of the MFMP solution
			\begin{align*}
				\mSigma_{kn} &= \text{Var}_{\widetilde{r}_k( \cdot; \widehat{\vlambda}_{-k}^{(n)} )} ( \vtheta_k ) \\
				&= \bE_{q_{-k} }[ \vomega_{kn} ( \vtheta_{-k} )  ] + \text{Var}_{q_{-k}} [  \vm_{kn} ( \vtheta_{-k} ) ].
			\end{align*}
			where we recall that $\widetilde{r}_k (\vtheta_{-k}; \widehat{\vlambda}_{-k}^{(n)}  ) = \int p( \vtheta_k \mid \vtheta_{-k}, \sD_{n} ) \, q(\vtheta_{-k}; \widehat{\vlambda}_{-k}^{(n)} ) \, \text{d} \vtheta_{-k}$. We decompose the integral $ \bE_{q_{-k} }[ \vomega_{kn} ( \vtheta_{-k} )  ] $ using neighbourhood $\sB_{-k}$:
			\begin{equation}
				\label{KeyBound}
				\bE_{q_{-k} }[ \vomega_{kn} ( \vtheta_{-k} )  ] = \int_{\mathcal{B}_{-k}}  \vomega_{kn} ( \vtheta_{-k} ) q(\vtheta_{-k}; \widehat{\vlambda}_{-k}^{(n)} ) \, d\vtheta_{-k} + \int_{\mathcal{B}_{-k}^c} \vomega_{kn} (\vtheta_{-k}) q(\vtheta_{-k}; \widehat{\vlambda}_{-k}^{(n)} ) \, d\vtheta_{-k}.
			\end{equation}
			Due to the upper bound of $\vomega_{kn}$ from (A8), we can write
			$$
            \begin{array}{rl}
			\ds \int_{\mathcal{B}_{-k}^c} \vomega_{kn}(\vtheta_{-k}) q_{-k} (\vtheta_{-k}) \, d\vtheta_{-k} 
            & \ds \le M_{\text{prior}} \int_{\mathcal{B}_{-k}^c}  q_{-k} (\vtheta_{-k}) \, d\vtheta_{-k} \\
            & \ds \le \frac{M_{\text{prior}} \bE \left [ \lVert   \vtheta_{k} - \vmu_{k} ( \widehat{\vlambda}_k^{(n)} )   \rVert^4 \right ]}{(R/2)^4 } 

            \\
            & = O_p(n^{-2}) 
            \\
            & = o_p(n^{-1}),
			\end{array}
            $$
			where we write $q_{-k} (\vtheta_{-k})$ in place of $q_{-k} (\vtheta_{-k}; \widehat{\vlambda}_k^{(n)})$ for notational brevity. Hence
			$$
			\bE_{q_{-k}}[\vomega_{kn}(\vtheta_{-k})] = \int_{\mathcal{B}_{-k}} \vomega_{kn} (\vtheta_{-k}) q_{-k} (\vtheta_{-k}) \, d\vtheta_{-k} + o_p(n^{-1}).
			$$
			To complete our analysis of $\bE_{q_{-k} }[ \vomega_{kn}( \vtheta_{-k} )  ]$, it remains for us to analyse the first term in the RHS of equation (\ref{KeyBound}). In fact, by (A8), 
			\begin{align*}
				&\int_{\mathcal{B}_{-k}} \vomega_{kn}(\vtheta_{-k}) q_{-k} (\vtheta_{-k}) \, d\vtheta_{-k} \\
				& \qquad \le n^{-1} \int_{\mathcal{B}_{-k}} \widetilde{\vomega}_{kn}(\vtheta_{-k}) q_{-k} (\vtheta_{-k}) \, d\vtheta_{-k} + \int_{\mathcal{B}_{-k}} \vr_k(\vtheta_{-k}) q_{-k} (\vtheta_{-k}) \, d\vtheta_{-k}
			\end{align*}
			Now,
			$$
			\left \lVert \int_{\mathcal{B}_{-k}} \vr_k(\vtheta_{-k}) q_{-k} (\vtheta_{-k}) \, d\vtheta_{-k} \right \rVert \le  \int_{\mathcal{B}_{-k}} \lVert \vr_k(\vtheta_{-k}) \rVert q_{-k} (\vtheta_{-k}) \, d\vtheta_{-k} = o_p(n^{-1}),
			$$
			where the last stochastic convergence equality follows from (A8). To understand the asymptotic behaviour of the term $n^{-1} \int_{\mathcal{B}_{-k}} \widetilde{\vomega}_k(\vtheta_{-k}) q_{-k} (\vtheta_{-k}) \, d\vtheta_{-k}$, we consider an elementwise second-order Taylor's expansion of $\widetilde{\vomega}_{kn}$. In particular,
			$$
            \begin{array}{l}
			\widetilde{\vomega}_{knj} (\vtheta_{-k}) = \widetilde{\vomega}_{knj} ( \vmu_{-kn} ) + \nabla_{\vtheta_{-k}}  \widetilde{\vomega}_{knj} ( \vmu_{-kn} ) ( \vtheta_{-k} - \vmu_{-kn} ) \\ 
            \qquad + \tfrac{1}{2} ( \vtheta_{-k} - \vmu_{-kn} )^\top \tfrac{\partial^2  \widetilde{\vomega}_{knj} }{ \partial \vtheta_{-k} \partial \vtheta_{-k}^\top  } (\overline{\vtheta}_{-k})   ( \vtheta_{-k} - \vmu_{-kn} ),

            \end{array}
            $$
			where we use the shorthand notation $\vmu_{-kn} := \int \vtheta_{-k}  \; q (\vtheta_{-k};  \widehat{\vlambda}_{-k}^{(n)} ) \; \text{d} \vtheta_{-k}$. Hence, for each entry $\widetilde{\vomega}_{knj} (\vtheta_{-k})$, we have
			\begin{align*}
				&\int_{\sB_{-k}} \widetilde{\vomega}_{knj} (\vtheta_{-k}) q_{-k} (\vtheta_{-k}) \, d\vtheta_{-k} \\
				& \qquad = \widetilde{\vomega}_{knj} ( \vmu_{-kn} ) + \nabla_{\vtheta_{-k}}  \widetilde{\vomega}_{knj} ( \vmu_{-kn} ) \int_{\sB_{-k}} ( \vtheta_{-k} - \vmu_{-kn} ) q_{-k} (\vtheta_{-k}) \, d\vtheta_{-k} \\
				& \qquad+ \tfrac{1}{2} \int_{\sB_{-k}} ( \vtheta_{-k} - \vmu_{-kn} )^\top \tfrac{\partial^2  \widetilde{\vomega}_{knj} }{ \partial \vtheta_{-k} \partial \vtheta_{-k}^\top  } (\overline{\vtheta}_{-k})   ( \vtheta_{-k} - \vmu_{-kn} ) q_{-k} (\vtheta_{-k}) \, d\vtheta_{-k}
			\end{align*}
			Now,
			$$
			\vzero =  \int_{\sB_{-k}} ( \vtheta_{-k} - \vmu_{-kn} ) q_{-k} (\vtheta_{-k}) \, d\vtheta_{-k} + \int_{\sB_{-k}^c} ( \vtheta_{-k} - \vmu_{-kn} ) q_{-k} (\vtheta_{-k}) \, d\vtheta_{-k}
			$$
			and hence
			$$
			\left \lVert \int_{\sB_{-k}} ( \vtheta_{-k} - \vmu_{-kn} ) q_{-k} (\vtheta_{-k}) \, d\vtheta_{-k}  \right \rVert = \left \lVert  \int_{\sB_{-k}^c} ( \vtheta_{-k} - \vmu_{-kn} ) q_{-k} (\vtheta_{-k}) \, d\vtheta_{-k} \right \rVert.
			$$
			Following Cauchy-Schwarz inequality, we show that
			\begin{align*}
				&\left \lVert \int_{\sB_{-k}^c} ( \vtheta_{-k} - \vmu_{-kn} ) q_{-k} (\vtheta_{-k}) \, d\vtheta_{-k} \right \rVert \\
				& \qquad \le \int_{\mTheta_{-k}} \lVert \vtheta_{-k} - \vmu_{-kn} \rVert \bI\{ \vtheta_{-k} \in \sB_{-k}^c \} q_{-k} (\vtheta_{-k}) \, d\vtheta_{-k} \\
				& \qquad \le \sqrt{\bE_{q_{-k}} \left [ \lVert \vtheta_{-k} - \vmu_{-kn} \rVert^2 \right ] } \times \underbrace{\sqrt{ \int_{\sB_{-k}^c} q_{-k} (\vtheta_{-k}) \, d\vtheta_{-k}  }}_{ \le \frac{ \bE[ \lVert \vtheta_{-k} - \vmu_{-kn} \rVert^4 ] }{(R/2)^4} } = o_p(n^{-2}).
			\end{align*}
			Following (A8), each $\widetilde{\vomega}_{knj}$ is twice continuously differentiable. Hence there exists a constant $M$ such that $ \left \lVert \tfrac{\partial^2  \widetilde{\vomega}_{knj} }{ \partial \vtheta_{-k} \partial \vtheta_{-k}^\top  } (\vtheta_{-k}) \right \rVert \le M$ for all $\vtheta_{-k} \in \mTheta_{-k}$. Thus,
			\begin{align*}
				&\int_{\sB_{-k}} ( \vtheta_{-k} - \vmu_{-kn} )^\top \tfrac{\partial^2  \widetilde{\vomega}_{knj} }{ \partial \vtheta_{-k} \partial \vtheta_{-k}^\top  } (\overline{\vtheta}_{-k})   ( \vtheta_{-k} - \vmu_{-kn} ) \, d\vtheta_{-k} \\
				& \qquad \le M \int_{\sB_{-k}} \lVert \vtheta_{-k} - \vmu_{-kn} \rVert^2 q_{-k} (\vtheta_{-k}) \, d\vtheta_{-k} \le M \mathrm{tr}( \mSigma_{-kn}  ) = O_p(n^{-1}),
			\end{align*}
			where the last stochastic convergence follows from an implication on (A9), i.e., $\lVert \mSigma_{-kn} \rVert = O_p(n^{-1})$. Hence, we have
			$$
			\int_{\mathcal{B}_{-k}} \vomega_{kn}(\vtheta_{-k}) q_{-k} (\vtheta_{-k}) \, d\vtheta_{-k} = n^{-1} \widetilde{\vomega}_{kn} ( \vmu_{-kn} ) + o_p(n^{-1}).
			$$
			Furthermore, by delta method, it is straightforward to show that
			$$
			\bV_{q_{-k}} [ \vm_{kn} (\vtheta_{-k}) ] = \nabla_{\vtheta_{-k}} \vm_{kn} (\vmu_{-kn})^\top \mSigma_{-kn} \nabla_{\vtheta_{-k}} \vm_{kn} (\vmu_{-kn}) + o_p(n^{-1}).
			$$
			Moreover, since $\mSigma_{-kn} = O_p(n^{-1})$, hence $\text{Var}_{q_{-k}} [ \vm_{kn} (\vtheta_{-k}) ] = O_p(n^{-1})$. Thus, we have shown that 
			\begin{equation}
				\label{eqn::expansionSigma}
			n \mSigma_{kn} = \widetilde{\vomega}_{kn} ( \vmu_{-kn} ) + \nabla_{\vtheta_{-k}} \vm_{kn} (\vmu_{-kn})^\top \mSigma_{-kn} \nabla_{\vtheta_{-k}} \vm_{kn} (\vmu_{-kn}) + o_p(1).
			\end{equation}
			Next, we analyse the asymptotic behaviour of the exact posterior variance $\mV_{kn}$. By similar application of Taylor's expansion and delta method in the preceding segment of this proof, we can also show that
			\begin{equation}
				\label{eqn::expansionV}
			n \mV_{kn} = \widetilde{\vomega}_{kn} ( \vtheta_{-k}^\star ) + \nabla_{\vtheta_{-k}} \vm_{kn} ( \vtheta_{-k}^\star )^\top \mV_{-kn} \nabla_{\vtheta_{-k}} \vm_{kn} ( \vtheta_{-k}^\star ) + o_p(1).
		\end{equation}
			Hence, by combining \eqref{eqn::expansionSigma} and \eqref{eqn::expansionV}, we have
			\begin{align}
				\label{eq::CombinedBound}
				&n \lVert \mSigma_{kn}  - \mV_{kn} \rVert \nonumber  \\
				&\le \lVert \widetilde{\vomega}_{kn} (\vmu_{-kn}) - \widetilde{\vomega}_{kn} (\vtheta_{-k}^\star ) \rVert \nonumber \\
				&+ \lvert \nabla_{\vtheta_{-k}} \vm_{kn} (\vmu_{-kn})^\top \mSigma_{-kn}  \nabla_{\vtheta_{-k}} \vm_{kn} (\vmu_{-kn}) - \nabla_{\vtheta_{-k}} \vm_{kn} (\vtheta_{-k}^\star )^\top \mSigma_{-kn}  \nabla_{\vtheta_{-k}} \vm_{kn} ( \vtheta_{-k}^\star ) \rvert. 
			\end{align}
			To bound the first term on the RHS of \eqref{eq::CombinedBound}, we note from (A8) that $\widetilde{\vomega}_{kn}$ is twice continuously-differentiable and Lemma \ref{SqrtNconsistency} that $\lVert \vmu_{-kn} - \vtheta_{-k}^\star \rVert = O_p(n^{-1/2})$. Hence
			\begin{align*}
				&\lVert \widetilde{\vomega}_{kn} (\vmu_{-kn}) - \widetilde{\vomega}_{kn} (\vtheta_{-k}^\star ) \rVert \\
				&\le \underbrace{\sup_{t \in [0,1]} \left \lVert \nabla_{\vtheta_{-k}} \widetilde{\vomega}_{kn} ( t \vmu_{-kn} + (1-t) \vtheta_{-k}^\star ) \right \rVert}_{=O_p(1)}  \lVert \vmu_{-kn} - \vtheta_{-k}^\star \rVert= O_p(n^{-1/2}).
			\end{align*}
			Moreover, we can bound the second term in \eqref{eq::CombinedBound} as
			\begin{align*}
				&\lvert \nabla_{\vtheta_{-k}} \vm_{kn} (\vmu_{-kn})^\top \mSigma_{-kn}  \nabla_{\vtheta_{-k}} \vm_{kn} (\vmu_{-kn}) - \nabla_{\vtheta_{-k}} \vm_{kn} (\vtheta_{-k}^\star )^\top \mSigma_{-kn}  \nabla_{\vtheta_{-k}} \vm_{kn} ( \vtheta_{-k}^\star ) \rvert \\
				&\le \lVert \nabla_{\vtheta_{-k}} \vm_{kn} (\vmu_{-kn}) - \nabla_{\vtheta_{-k}} \vm_{kn} (\vtheta_{-k}^\star) \rVert \lVert \mSigma_{-kn} \rVert  \underbrace{\lVert \nabla_{\vtheta_{-k}} \vm_{kn} (\vmu_{-kn}) + \nabla_{\vtheta_{-k}} \vm_{kn} (\vtheta_{-k}^\star) \rVert}_{=O_p(1)} \\
				&+ \lVert \nabla_{\vtheta_{-k}} \vm_{kn} (\vtheta_{-k}^\star) \rVert^2 \lVert \mSigma_{-kn}  - \mV_{-kn} \rVert
			\end{align*}
			Following the twice-continuous differentiability of $\vm_{kn}$ in (A8) and Lemma \ref{SqrtNconsistency}, we have
			$$
            \begin{array}{l}
			\lVert \nabla_{\vtheta_{-k}} \vm_{kn} (\vmu_{-kn}) - \nabla_{\vtheta_{-k}} \vm_{kn} (\vtheta_{-k}^\star) \rVert 
            \\
            \qquad \le \underbrace{\sup_{t \in [0,1]} \left \lVert \nabla_{\vtheta_{-k}} \vm_{kn} ( t \vmu_{-kn} + (1-t) \vtheta_{-k}^\star ) \right \rVert}_{=O_p(1)}  \lVert \vmu_{-kn} - \vtheta_{-k}^\star \rVert = O_p(n^{-1/2}).
			\end{array}
            $$
			Moreover, recall from Lemma \ref{SqrtNconsistency} that $\lVert \mSigma_{-kn} \rVert = O_p(n^{-1})$. Now, by following similar steps as above, we obtain
			\begin{align}
				\label{eq::SigmaMinusKnVersion}
				&n \lVert \mSigma_{-kn}  - \mV_{-kn} \rVert \nonumber \\
				&\le O_p(n^{-1/2}) + \lVert \nabla_{\vtheta_{k}} \vm_{-kn} (\vmu_{kn}) - \nabla_{\vtheta_{k}} \vm_{-kn} (\vtheta_{k}^\star) \rVert \lVert \mSigma_{kn} \rVert  \lVert \nabla_{\vtheta_{k}} \vm_{-kn} (\vmu_{kn}) + \nabla_{\vtheta_{k}} \vm_{-kn} (\vtheta_{k}^\star) \rVert \nonumber \\
				&+ \lVert \nabla_{\vtheta_{k}} \vm_{-kn} (\vtheta_{k}^\star) \rVert^2  \lVert \mSigma_{kn}  - \mV_{kn} \rVert
			\end{align}
			By (A7) and Lemma \ref{SqrtNconsistency}, we have $\lVert \mSigma_{kn} \rVert = O_p(n^{-1})$ and $\lVert \mV_{kn} \rVert = O_p(n^{-1})$. Hence, a crude stochastic convergence order for $\lVert \mSigma_{kn}  - \mV_{kn} \rVert$ is $O_p(n^{-1})$. Hence, $\lVert \mSigma_{-kn}  - \mV_{-kn} \rVert = O_p(n^{-3/2})$. Consequently,
			$$
			\lvert \nabla_{\vtheta_{-k}} \vm_{kn} (\vmu_{-kn})^\top \mSigma_{-kn}  \nabla_{\vtheta_{-k}} \vm_{kn} (\vmu_{-kn}) - \nabla_{\vtheta_{-k}} \vm_{kn} (\vtheta_{-k}^\star )^\top \mSigma_{-kn}  \nabla_{\vtheta_{-k}} \vm_{kn} ( \vtheta_{-k}^\star ) \rvert  = O_p(n^{-3/2}).
			$$
			Hence, substituting the previous equation back into \eqref{eq::CombinedBound}, we have
			$$
			\lVert \mSigma_{kn} - \mathbf{V}_{kn} \rVert = O_p(n^{-3/2})
			$$
			as required.
		\end{proof}

\subsection{Supporting Lemmas}

\begin{Lemma}
		\label{supNormConvergence}
		Assume (A1) and (A2) holds. Then,
		$$
		\sup_{\vlambda_{1:K} \in \mLambda_{1:K}} \lVert \mPsi_n ( \vlambda ) - \mPsi (\vlambda) \rVert = O_p( \epsilon_n ).
		$$
		\begin{proof} 
			Denote
			$$
			\mPsi (\vlambda) = \vm(\vlambda) - \mG(\vlambda)
			$$
			where
			$$
			\mG (\vlambda)  = \begin{pmatrix}
				\int^{\otimes^{K-1}} \vh_1^\star( \vtheta_{-1} ) \prod_{j \neq 1} q(\vtheta_j; \vlambda_j) \, \text{d} \vtheta_{j} \\
				\vdots \\
				\int^{\otimes^{K-1}} \vh_K^\star( \vtheta_{-K} ) \prod_{j \neq K} q(\vtheta_j; \vlambda_j) \, \text{d} \vtheta_{j}
			\end{pmatrix}.
			$$
			\noindent For any $\vlambda_{1:K} \in \mLambda_{1:K}$,
			\begin{align*}
				\lVert \mPsi_n ( \vlambda_{1:K} ) - \mPsi (\vlambda_{1:K}) \rVert &= 	\lVert \mG_n ( \vlambda ) - \mG ( \vlambda ) \rVert \\
				&\le  \sum_{k=1}^K \int \lVert \vh_{kn} (\vtheta_{-k}) - \vh_k^\star( \vtheta_{-k} )  \rVert \prod_{j \neq k} q(\vtheta_j; \vlambda_j) \, \text{d} \vtheta_{j} \\
				&\le \sum_{k=1}^K \sup_{\vtheta_{-k} \in \mTheta_{-k}} \lVert \vh_{kn} (\vtheta_{-k}) - \vh_k^\star( \vtheta_{-k} )  \rVert   \underbrace{ \int \prod_{j \neq k} q(\vtheta_j; \vlambda_j) \, \text{d} \vtheta_{j} }_{=1} \\
				&= \sum_{k=1}^K \sup_{\vtheta_{-k} \in \mTheta_{-k}} \lVert \vh_{kn} (\vtheta_{-k}) - \vh_k^\star( \vtheta_{-k} )  \rVert  ,
			\end{align*}
			where the first equality follows from the definition of $\mPsi_n$ and $\mPsi$, the second inequality follows from the definition of $\mG_n$ and $\mG$ and then an application of triangle inequality, the third inequality follows from taking $\sup$ over $\vtheta_{-k}$.
			By noting from (A1) that $\mLambda_{1:K}$ is compact and the above upper bound doesn't depend on $\vlambda_{1:K}$, then
			$$
			\sup_{\vlambda_{1:K} \in \mLambda_{1:K}} 	\lVert \mPsi_n ( \vlambda_{1:K} ) - \mPsi (\vlambda_{1:K}) \rVert \le    \sum_{k=1}^K \sup_{\vtheta_{-k} \in \mTheta_{-k}} \lVert \vh_{kn} (\vtheta_{-k}) - \vh_k^\star( \vtheta_{-k} )  \rVert = O_p(\epsilon_n),
			$$
			where the stochastic convergence order follows from (A2).
		\end{proof}
\end{Lemma}

\begin{Lemma}
\label{SqrtNconsistency}
		Assume (A1) to (A6) and (A9) holds. Then, for each $k = 1, \ldots, K$, we have
		$$
		\lVert \vmu_{k} ( \widehat{\vlambda}_k ) - \vtheta_k^\star \rVert = O_p(n^{-1/2})
		$$
		and
		$$
		\lVert \tfrac{1}{n} \mGamma_{k}^{(n)} ( \widehat{\vlambda}_k ) \rVert = O_p(n^{-1}).
		$$
		\begin{proof}
			Following Theorem \ref{ConvergenceToLimitSolution}, our sequence of solutions converges as $\lVert \widehat{\vlambda}_k - \vlambda_k^\star \rVert = O_p(n^{-1/2})$. Following a straightforward Taylor's expansion, and applying the convergence rate of $\lVert \widehat{\vlambda}_k - \vlambda_k^\star \rVert$, we have
			$$
			\lVert \vmu_{k} ( \widehat{\vlambda}_k ) - \vmu_{k} ( \vlambda_k^\star ) \rVert  \le \lVert \nabla_{\vlambda_k} \vmu_{k} ( \vlambda_k^\star ) \rVert \lVert \widehat{\vlambda}_k - \vlambda_k^\star \rVert + o_p(n^{-1/2})
			$$
			and
			$$
			\lVert \text{vec} ( \mGamma_{k}^{(n)} (\widehat{\vlambda}_k )  ) - \text{vec} ( \mGamma_{k}^{(n)} (\vlambda_k^\star )  ) \rVert \le \lVert \nabla_{ \vlambda_k } \text{vec} ( \mGamma_k^{(n)}  ( \vlambda_k^\star ) ) \rVert \lVert \widehat{\vlambda}_k - \vlambda_k^\star \rVert + o_p(n^{-1/2}).
			$$
			From (A9), the map  $\vlambda_k \mapsto (\vmu_{k}(\vlambda_k), \mGamma_{k}^{(n)}(\vlambda_k))$ is twice-continuous differentiability. Hence, the two matrix norms of the first-order derivative are bounded: $\lVert \nabla_{\vlambda_k} \vmu_{k} ( \vlambda_k^\star ) \rVert = O(1)$ and $\lVert \nabla_{ \vlambda_k } \text{vec} ( \mGamma_k^{(n)}  ( \vlambda_k^\star ) ) \rVert = O(1)$. Hence, by (A6), we have
			$$
			\lVert \vmu_{k} ( \widehat{\vlambda}_k ) - \vmu_{k} ( \vlambda_k^\star ) \rVert = \lVert \vmu_{k} ( \widehat{\vlambda}_k ) - \vtheta_k^\star \rVert = O_p(n^{-1/2})
			$$
			and
			$$
			\lVert \text{vec} ( \mGamma_{k}^{(n)} (\widehat{\vlambda}_k )  ) - \text{vec} ( \mGamma_{k}^{(n)} (\vlambda_k^\star )  ) \rVert = O_p(n^{-1/2}).
			$$
			Thus have
			$$
			\lVert \text{vec} ( \tfrac{1}{n} \mGamma_{k}^{(n)} (\widehat{\vlambda}_k )  ) \rVert \le \tfrac{1}{n} \lVert \text{vec} ( \mGamma_{k}^{(n)} (\widehat{\vlambda}_k )  ) - \text{vec} ( \mGamma_{k}^{(n)} (\vlambda_k^\star )  ) \rVert + \tfrac{1}{n}  \lVert  \text{vec} ( \mGamma_{k}^{(n)} (\vlambda_k^\star )  )  \rVert = O_p(n^{-1}).
			$$
		\end{proof}
	\end{Lemma}

\newpage 

\section{Implementation details for GMP}\label{Suppl:post_hoc_local_global}

This section gives implementation details for Gaussian moment propagation. 
Section \ref{sec:downdates} shows how the local quantities required for $r_k(\theta_k)$ can be 
computed efficiently from precomputed quantities $\veta = \mX\vmu\in\bR^n$, $\mM = \mX\mSigma\in\bR^{n\times p}$,
and $\vs^2 = \mbox{dg}(\mX\mSigma\mX^T) \in \bR^n$.
Section \ref{sec:global_update} derives the Gaussian 
joint update after replacing a marginal distribution. Section \ref{Suppl:proof_theorem2} proves positive 
definiteness of the resulting covariance matrix. Section \ref{Suppl:updates} shows how to update 
the precomputed quantities $\veta$, $\mM$, and 
$\vs^2$ after each coordinate update. This reduces the cost of the updates and downdates from $O(np^2)$ to $O(np+p^2)$ for each coordinate $k$.

For simplicity in Section \ref{sec:downdates}
and 
Section \ref{Suppl:updates}, because this is the case used in Sections \ref{sec:example_glm}
and \ref{sec:gmp_lasso}, 
we describe the implementation for scalar coordinate updates. Block updates 
are analogous but require block-matrix versions of the same formulae.

% GMP updates require care to ensure the efficiency of the implementation.
% In particular,  we assume that the quantities $\veta = \mX\vmu\in\bR^n$, $\mM = \mX\mSigma\in\bR^{n\times p}$,
% and $\vs^2 = \mbox{dg}(\mX\mSigma\mX^T) \in \bR^n$ are precalculated
% at the start of the main loop. We will assume that $\mX$ is dense.
% These quantities are needed to efficiently calculate
% \eqref{eq:r_function} for multiple values of $\theta_k$. By maintaining these quantities,
% we can reduce the cost of ``downdating'', i.e., calculating $r(\theta_k)$,
% and updating the joint Gaussian distribution from $O(np^2)$ to $O(np + p^2)$ for each index $k$.

\subsection{Efficient downdates}
\label{sec:downdates}

The calculation of \eqref{eq:r_function} involves the quantities
\begin{itemize}
    \item $\va = \mSigma_{-k,k}/\Sigma_{k,k}$,
    \item $\vb = \vmu_{-k} - \va\mu_k$,
    \item $\vc = \mX_{k} + \mX_{-k}\va$,
    \item $\vd = \mX_{-k}\vb_k$, and 
\item $\vs_{-k|k}^2 = \mbox{dg}( \mX_{-k}\mSigma_{-k|k}\mX_{-k}^T)$
where $\mSigma_{-k|k}  = \mSigma_{-k,-k} - \Sigma_{k,k}\va\va^T$
\end{itemize}

\noindent need to be calculated for each $k = 1,\ldots,p$ and each iteration.
If calculated naively, this would cost $O(np^2)$ per index $k$ and for each iteration,
making it the most expensive cost of the algorithm for a total cost of $O(np^3)$. The vectors $\va$
and $\vb$ can be calculated in $O(p)$ time and do not need to be considered further.

Suppose that $\veta$, $\mM$ and $\vs^2$ have been pre-calculated. We first note that
$$
\veta = \mX_{k} \mu_k + \mX_{-k}\vmu_{-k}.
$$

\noindent Hence, $\mX_{-k}\vmu_{-k} = \veta - \mX_{k} \mu_k$.
Next, we let
$$
\begin{array}{rl}
\mM 
    & \ds = \left[ \begin{array}{cc} 
        \mX_k & \mX_{-k} 
        \end{array} \right]
        \left[ \begin{array}{cc} 
        \Sigma_{k,k}  & \mSigma_{k,-k} \\
        \mSigma_{-k,k} & \mSigma_{-k,-k}
        \end{array}\right]
\\ 
    & \ds = \left[ \begin{array}{cc} 
    \mX_k\Sigma_{k,k} + \mX_{-k} \mSigma_{-k,k}  
        & \mX_k\mSigma_{k,-k} + \mX_{-k} \mSigma_{-k,-k}
    \end{array} \right]
\equiv \left[ \begin{array}{cc} \mM_k & \mM_{-k} \end{array} \right]
\end{array}
$$

\noindent where $\mM_k$ and $\mX_k$ are the $k$-th columns of $\mM$ and $\mX$, respectively.
Hence, $\mX_{-k} \mSigma_{-k,k} = \mM_k - \mX_k\Sigma_{k,k}$.
Next, we consider
$$
\mM\mX^T = \mX_k\Sigma_{k,k}\mX_k^T + \mX_{-k} \mSigma_{-k,k}\mX_k^T +   \mX_k\mSigma_{k,-k}\mX_{-k}^T + \mX_{-k} \mSigma_{-k,-k}\mX_{-k}^T
$$

\noindent and hence,
$$
\begin{array}{rl}
\vs^2 
    & \ds = \Sigma_{k,k}\mX_k^{\odot 2} + 2 (\mX_{-k} \mSigma_{-k,k})\odot\mX_k +  \mbox{dg}(\mX_{-k} \mSigma_{-k,-k}\mX_{-k}^T)
    \\
    & \ds = \Sigma_{k,k}\mX_k^{\odot 2} + 2 (\mM_k - \mX_k\Sigma_{k,k})\odot\mX_k +  \mbox{dg}(\mX_{-k} \mSigma_{-k,-k}\mX_{-k}^T)
    \\
    & \ds = 2 (\mM_k \odot \mX_k)
        - \Sigma_{k,k}\mX_k^{\odot 2} 
        + \mbox{dg}(\mX_{-k} \mSigma_{-k,-k}\mX_{-k}^T)
\end{array}
$$

\noindent so that $\mbox{dg}(\mX_{-k} \mSigma_{-k,-k}\mX_{-k}^T) = \vs^2 + \Sigma_{k,k}\mX_k^{\odot 2} - 2 (\mM_k \odot \mX_k)$.
We can then calculate the vector $\vc$ via
$$
\begin{array}{rl}
\vc & \ds = \mX_{k} + \mX_{-k}\mSigma_{-k,k}/\Sigma_{k,k} \\
    & \ds = \mX_{k} + (\mM_k - \mX_k\Sigma_{k,k})/\Sigma_{k,k} \\
    & \ds = \mM_k/\Sigma_{k,k}.
\end{array}
$$

\noindent Hence, when $\mM$ is available the vector  $\vc$ can be calculated in $O(n)$ time rather than $O(np)$ time.
Next, we consider the vector $\vd$. We have
$$
\begin{array}{rl}
\vd 
    & \ds = \mX_{-k}\vmu_{-k} - \mX_{-k}\mSigma_{-k,k}\mu_k/\Sigma_{k,k}
    \\
    & \ds = \veta - \mX_{k} \mu_k - (\mM_k - \mX_k\Sigma_{k,k})\mu_k/\Sigma_{k,k}
    \\
    & \ds = \veta - \mM_k(\mu_k/\Sigma_{k,k})
    \\
    & \ds = \veta - \vc\mu_k.
\end{array}
$$

\noindent Hence, when $\veta$ and $\mM$ are available we can calculate $\vd$
in $O(n)$ time rather than $O(np)$ time. Lastly, we have
$$
\begin{array}{rl}
\vs_{-k|k}^2 
    & \ds = \mbox{dg}( \mX_{-k}\mSigma_{-k|k}\mX_{-k})
    \\ 
    & \ds = \mbox{dg}( \mX_{-k}\mSigma_{-k,-k}\mX_{-k}) - 
    \mbox{dg}(\mX_{-k}\mSigma_{-k,k}\Sigma_{k,k}^{-1}\mSigma_{k,-k}\mX_{-k})
    \\
    & \ds = \vs^2 + \Sigma_{k,k}\mX_k^{\odot 2} - 2 (\mM_k \odot \mX_k) - 
    (\mM_k - \mX_k\Sigma_{k,k})^{\odot 2}    \Sigma_{k,k}^{-1}
    \\
    & \ds = \vs^2 
    - \mM_k^{\odot 2}\Sigma_{k,k}^{-1}  
    \\
    & \ds = \vs^2 - \mM_k \odot \vc 
\end{array}
$$

\noindent so that if $\mM$ and $\vs^2$ are available, then $\vs_{-k|k}^2$ can be computed in $O(n)$ rather than $O(np^2)$ time.

Thus, using the precomputed quantities 
$\veta=\mX\vmu$, $\mM=\mX\mSigma$, and 
$\vs^2=\operatorname{dg}(\mX\mSigma \mX^T)$, the local GMP quantities are
$$
\va=\frac{\mSigma_{-k,k}}{\Sigma_{k,k}}, 
\qquad
\vb=\vmu_{-k}-\va\mu_k,
\qquad
\vc=\frac{\mM_k}{\Sigma_{k,k}},
\qquad
\vd=\veta-\vc\mu_k,
$$

\noindent and
$\ds \vs^2_{-k|k}
=
\vs^2-\mM_k^{\odot 2}/\Sigma_{k,k}
=
\vs^2-\mM_k\odot \vc$. 
%In numerical implementations, entries of $\vs^2_{-k|k}$ that are slightly negative 
%because of round-off error can be truncated to zero.

\subsection{Updating joint Gaussian with new marginal distribution}\label{sec:global_update}

Without loss of generality, suppose that we have partitioned the parameter vector $\vtheta$ into $K=2$ components $(\vtheta_1,\vtheta_2)$ with $q(\vtheta) = \phi(\vtheta - \vmu,\mSigma)$, where
$\vmu$ and $\mSigma$ have been partitioned conformably with the partitioning of $\vtheta$, i.e.,
$$\vmu = \left[ \begin{array}{c}
	\vmu_1 \\
	\vmu_2 
\end{array} \right]
\qquad \mbox{and} \qquad 
\mSigma = 
\left[ \begin{array}{cc}
	\mSigma_{11} & \mSigma_{12} \\
	\mSigma_{21} & \mSigma_{22} 
\end{array} \right].
$$

\noindent Suppose that, through an MP update (or otherwise), $q(\vtheta_1) = \phi(\vtheta_1 - \vmu_1,\mSigma_{11})$ is updated and replaced with
$\widetilde{q}(\vtheta_1) = \phi(\vtheta_1 - \widetilde{\vmu}_1,\widetilde{\mSigma}_1)$, with new (marginal) mean and covariance $\widetilde{\vmu}_1$ and $\widetilde{\mSigma}_1$ respectively. 
Let $\widetilde q(\vtheta)=\phi(\vtheta-\widetilde\vmu;\widetilde\mSigma)$ denote the updated joint distribution obtained by replacing 
$q(\vtheta)$ with
$$
\widetilde q(\vtheta)=q(\vtheta_2|\vtheta_1)\widetilde q(\vtheta_1)
$$
% Let
% $\widetilde{q}(\vtheta) = \phi(\vtheta - \widetilde{\vmu},\widetilde{\mSigma})$ be the updated joint distribution for $\vtheta$ where
% $\widetilde{\vmu}$ and $\widetilde{\mSigma}$
% be the updated mean and covariance after 
% $q(\vtheta)$ is replaced with
% $\widetilde{q}(\vtheta) = q(\vtheta_2|\vtheta_1) \widetilde{q}(\vtheta_1).$
\noindent 

\noindent and the updated mean and covariance $\widetilde{\vmu}$ and $\widetilde{\mSigma}$ is partitioned as
$$
\widetilde{\vmu} = \left[ \begin{array}{c}
	\widetilde{\vmu}_1 \\
	\widetilde{\vmu}_2 
\end{array} \right]
\qquad \mbox{and} \qquad 
\widetilde{\mSigma} = 
\left[ \begin{array}{cc}
	\widetilde{\mSigma}_{11} & \widetilde{\mSigma}_{12} \\
	\widetilde{\mSigma}_{21} & \widetilde{\mSigma}_{22} 
\end{array} \right].
$$

\noindent Note that $q(\vtheta_2|\vtheta_1) 
    = \phi(\vtheta_2 - (\vmu_2 + \mSigma_{21}\mSigma_{11}^{-1}\left(\vtheta_1 - \vmu_1\right)), \mSigma_{22} - \mSigma_{21} \mSigma_{11}^{-1}\mSigma_{12})$.
Then 
$$
\begin{array}{rl}
\widetilde{\vmu}_2 
%    & \ds = \bE_{\widetilde{q}}(\vtheta_2) 
%\\
    & \ds = \bE_{\widetilde{q}(\vtheta_1)}[\bE_q(\vtheta_2\mid\vtheta_1)] 
\\ 
    &  = \bE_{\widetilde{q}(\vtheta_1)}[\vmu_2 + \mSigma_{21}\mSigma_{11}^{-1}\left(\vtheta_1 - \vmu_1\right)]
\\
    & = \vmu_2 + \mSigma_{21}\mSigma_{11}^{-1}\left(\widetilde{\vmu}_1 - \vmu_1\right).
\end{array} 
$$

\noindent Let $d_1$ and $d_2$ be the dimensions of $\vtheta_1$ and $\vtheta_2$, respectively.
This calculation can be performed in $O(d_1^3 + d_2d_1)$ time. 
Similarly,
$$
\begin{array}{rl}
\widetilde{\mSigma}_{22}
%    & \ds = \bV_{\widetilde{q}}(\vtheta_2)
%\\ 
    & \ds = \bE_{\widetilde{q}(\vtheta_1)}[\bV_q(\vtheta_2|\vtheta_1)] 
    + \bV_{\widetilde{q}(\vtheta_1)}[\bE_q(\vtheta_2|\vtheta_1)] 
\\  
    &  = \bE_{\widetilde{q}(\vtheta_1)}(\mSigma_{2|1} )
    + \bV_{\widetilde{q}(\vtheta_1)}[ \vmu_2 + \mSigma_{21}\mSigma_{11}^{-1}\left(\vtheta_1 - \vmu_1\right)] 
\\ 
    & = \mSigma_{2|1} 
	+ \mSigma_{21}\mSigma_{11}^{-1}  \bV_{\widetilde{q}(\vtheta_1)}[\vtheta_1] \mSigma_{11}^{-1}  \mSigma_{12}
    \\
    & =  \mSigma_{2|1} 

    +\mSigma_{21}\mSigma_{11}^{-1}\widetilde{\mSigma}_{1} 
    \mSigma_{11}^{-1} \mSigma_{12}.
\end{array} 
$$

\noindent The update of 
$\widetilde{\mSigma}_{22}$ costs $O(d_1^3 + d_2^2 d_1)$
time.
%
%\bigskip 
%\noindent 
Lastly,
$$
\begin{array}{rl}
\widetilde{\mSigma}_{12}
%    & \ds = \mbox{Cov}_{\widetilde{q}}(\vtheta_1,\vtheta_2) 
%\\
%    & \ds = \bE[\{ \vtheta_1 - \bE(\vtheta_1)\}\{\vtheta_2 - \bE(\vtheta_2)\}^T] 
%\\ 
    & \ds =  \bE_{\widetilde{q}(\vtheta_1)}\left[ 
        \bE_q \left\{ 
            (\vtheta_1 - \bE(\vtheta_1))(\vtheta_2 - \bE(\vtheta_2))^T \mid \vtheta_1 \right\} \right]    
\\
    & \ds = \bE_{\widetilde{q}(\vtheta_1)}\left[ \left(\vtheta_1 - \widetilde{\vmu}_1 \right)\left(  \mSigma_{21}\mSigma_{11}^{-1}(\vtheta_1 - \widetilde{\vmu}_1)    \right)^T\right]
\\ 
    & \ds = \bE_{\widetilde{q}(\vtheta_1)}\left[ \left(\vtheta_1 - \widetilde{\vmu}_1 \right)\left(  \vtheta_1 - \widetilde{\vmu}_1 \right)^T\right] \mSigma_{11}^{-1}\mSigma_{12}
\\ 
    & \ds = \widetilde{\mSigma}_{1} \mSigma_{11}^{-1}\mSigma_{12}.
\end{array} 
$$

\noindent The update of $\widetilde{\mSigma}_{12}$ costs $O(d_1^3 + d_1^2d_2)$ time.

\subsection{Proof of Theorem \ref{Th:Theorem2}}\label{Suppl:proof_theorem2}

\begin{proof}\label{Suppl:proof_Theorem2}

Let
$$
\widetilde\mSigma
=
\begin{bmatrix}
\mA & \mC^T\\
\mC & \mD
\end{bmatrix},
\quad \mbox{where} \quad  \mA=\widetilde\mSigma_1.
$$

\noindent 
Since $\mA$ is positive definite, 
$\widetilde\mSigma$ is positive definite if and only if its Schur complement
$\mD-\mC\mA^{-1}\mC^T$ 
is positive definite. In the GMP update,
$$
\mC=\mSigma_{21}\mSigma_{11}^{-1}\widetilde\mSigma_1
\qquad \mbox{and} \qquad 
\mD
=
\mSigma_{2|1}
+
\mSigma_{21}\mSigma_{11}^{-1}
\widetilde\mSigma_1
\mSigma_{11}^{-1}\mSigma_{12}.
$$

\noindent 
Therefore,
$\mD-\mC\mA^{-1}\mC^T
=
\mSigma_{2|1}$.
Since $\mSigma$ is positive definite, its Schur complement
$$
\mSigma_{2|1}
=
\mSigma_{22}-\mSigma_{21}\mSigma_{11}^{-1}\mSigma_{12}
$$

\noindent 
is positive definite. Hence, $\widetilde\mSigma$ is positive definite.

\end{proof}

\subsection{Efficient updates}
\label{Suppl:updates}

Here we concern ourselves with maintaining the quantities $\veta$, $\mM$ and $\vs^2$ after the updates in Section
\ref{sec:global_update} have been performed. Using earlier notation
\begin{itemize}
    \item $\widetilde{\vmu}_{-k} = \va \widetilde{\mu}_k + \vb$

    \item $\widetilde{\mSigma}_{k,-k} = \widetilde{\Sigma}_k \va^T$,
$\widetilde{\mSigma}_{-k,k} = \widetilde{\Sigma}_k \va$, and

    \item $\widetilde{\mSigma}_{-k,-k} = \mSigma_{-k|k} + \widetilde{\Sigma}_k\va\va^T = \mSigma_{-k,-k} +  (\widetilde{\Sigma}_k - \Sigma_{k,k})\va\va^T$.
\end{itemize}

\noindent Hence,
$$
\widetilde{\mSigma} = \left[ \begin{array}{cc}
\mSigma_{-k|k} & \vzero \\
\vzero & 0
\end{array} \right] + \widetilde{\Sigma}_k\left( \begin{array}{c}
\va \\
1
\end{array} \right)\left( \begin{array}{c}
\va \\
1
\end{array} \right)^T.
$$

\noindent 
Thus, for a scalar coordinate update, assuming
$\mSigma_{-k|k}$ is available, the joint mean and covariance can be updated in $O(p^2)$ time.

Define $\widetilde{\veta} = \mX\widetilde{\vmu}$, $\widetilde{\mM} = \mX\widetilde{\mSigma}$, and $\widetilde{\vs}^2 = \mbox{dg}(\mX\widetilde{\mSigma}\mX^T)$, where 
$\widetilde{\vmu}$ and $\widetilde{\mSigma}$ are the updated mean and covariance matrix after the updates in Section \ref{sec:global_update} have been applied.
To update $\widetilde{\veta}$, we have
$$
\begin{array}{rl}
\widetilde{\veta}
%    & \ds = \mX\widetilde{\vmu} 
%    \\
    & \ds = \mX_k\widetilde{\mu}_k + \mX_{-k}\widetilde{\vmu}_{-k}
    \\
    & \ds = \mX\widetilde{\mu}_k + \mX_{-k}[\vmu_{-k} + \mSigma_{-k,k}\Sigma_{k,k}^{-1}(\widetilde{\mu}_k - \mu_k)]
    \\ 
    & \ds = \veta + \mX_k(\widetilde{\mu}_k   - \mu_k)
     + (\mM_k - \mX_k\Sigma_{k,k})\Sigma_{k,k}^{-1}(\widetilde{\mu}_k - \mu_k)
    \\ 
    & \ds = \veta  
     + \mM_k\Sigma_{k,k}^{-1}(\widetilde{\mu}_k - \mu_k)      
    \\ 
    & \ds = \veta  
     +  (\widetilde{\mu}_k - \mu_k)  \vc    
 
\end{array}
$$

\noindent where the right-hand side costs $O(n)$ to calculate.
Next we have 
$$
\begin{array}{rl}
\widetilde{\mM}_k
     \ds = \mX_k\widetilde{\Sigma}_{k} + \mX_{-k} \widetilde{\mSigma}_{-k,k}, \qquad \mbox{and} \qquad 
\widetilde{\mM}_{-k}
    \ds = \mX_k\widetilde{\mSigma}_{k,-k} + \mX_{-k} \widetilde{\mSigma}_{-k,-k}.
\end{array}
$$

\noindent Note that using the above expressions directly costs $O(np)$ to calculate $\widetilde{\mM}_k$ and $O(np^2)$ to calculate $\widetilde{\mM}_{-k}$.
Dealing with $\widetilde{\mM}_k$, we have
$$
\begin{array}{rl}
\widetilde{\mM}_k
    & \ds = \mX_k\widetilde{\Sigma}_{k}
    + \mX_{-k} \mSigma_{-k,k}\Sigma_{k,k}^{-1}\widetilde{\Sigma}_k
    \\ 
    & \ds = \mX_k\widetilde{\Sigma}_{k} 
    +  (\mM_k - \mX_k \Sigma_{k,k})\Sigma_{k,k}^{-1}\widetilde{\Sigma}_k
    \\ 
    & \ds = \mM_k \Sigma_{k,k}^{-1}\widetilde{\Sigma}_k
    \\ 
    & \ds = \vc \widetilde{\Sigma}_k.
\end{array}
$$

\noindent This costs $O(n)$ time to calculate.
Next, we consider $\widetilde{\mM}_{-k}$ given by
$$
\begin{array}{rl}
\widetilde{\mM}_{-k}
    & \ds = \mX_k \widetilde{\Sigma}_k \Sigma_{k,k}^{-1}\mSigma_{k,-k} + \mX_{-k} [\mSigma_{-k,-k} 
	+ \mSigma_{-k,k}\Sigma_{k,k}^{-1}( \widetilde{\Sigma}_{k}  -\Sigma_{k,k}) \Sigma_{k,k}^{-1} \mSigma_{k,-k}]
    \\
    & \ds = (\widetilde{\Sigma}_k \Sigma_{k,k}^{-1})\mX_k \mSigma_{k,-k} 
    + \mX_{-k} \mSigma_{-k,-k} 
    + (\Sigma_{k,k}^{-1}( \widetilde{\Sigma}_{k}  -\Sigma_{k,k}) \Sigma_{k,k}^{-1})\mX_{-k}\mSigma_{-k,k} \mSigma_{k,-k}

    \\
    & \ds = (\widetilde{\Sigma}_k \Sigma_{k,k}^{-1})\mX_k \mSigma_{k,-k} 
    + \mM_{-k} - \mX_k\mSigma_{k,-k}
    + (\Sigma_{k,k}^{-1}( \widetilde{\Sigma}_{k}  -\Sigma_{k,k}) \Sigma_{k,k}^{-1})(\mM_k - \mX_k\Sigma_{k,k})\mSigma_{k,-k}

    \\
    & \ds = 
    \mM_{-k}  
    + (\Sigma_{k,k}^{-1} \widetilde{\Sigma}_{k}\Sigma_{k,k}^{-1}  - \Sigma_{k,k}^{-1}) \mM_k\mSigma_{k,-k}

    \\
    & \ds = 
    \mM_{-k}  
    + (\Sigma_{k,k}^{-1}    - \widetilde{\Sigma}_k^{-1}) \widetilde{\mM}_k\mSigma_{k,-k}

\end{array}
$$

\noindent which uses $\mX_{-k} \mSigma_{-k,-k} = \mM_{-k} - \mX_k\mSigma_{k,-k}$ and
$\mX_{-k} \mSigma_{-k,k} = \mM_k - \mX_k\Sigma_{k,k}$.
This term costs $O(np)$ to compute. This is optimal, as $\widetilde{\mM}_{-k}$ is an $n \times p$ matrix.
Lastly, we have
$$
\begin{array}{rl}
\widetilde{\vs}^2 
    & \ds = 2 (\widetilde{\mM}_k \odot \mX_k)
        - \widetilde{\Sigma}_{k}\mX_k^{\odot 2} 
        + \mbox{dg}(\mX_{-k} \widetilde{\mSigma}_{-k,-k}\mX_{-k}^T)
\\
    & \ds = 2 (\widetilde{\mM}_k \odot \mX_k)
        - \widetilde{\Sigma}_{k}\mX_k^{\odot 2} 
        + \vs_{-k|k}^2
        + \widetilde{\Sigma}_k (\mX_{-k}\mSigma_{-k,k}\Sigma_{k,k}^{-1})^{\odot 2}

\\
    & \ds = 2(\Sigma_{k,k}^{-1}\widetilde{\Sigma}_k) (\mM_k  \odot \mX_k)
        - \widetilde{\Sigma}_{k}\mX_k^{\odot 2} 
        + \vs_{-k|k}^2
        \\
    & \ds \qquad +  
            \Sigma_{k,k}^{-1} \widetilde{\Sigma}_k\Sigma_{k,k}^{-1} \mM_k^{\odot 2}
                - 2\widetilde{\Sigma}_k\Sigma_{k,k}^{-1}(\mM_k\odot\mX_k) 
                + \widetilde{\Sigma}_k\mX_k^{\odot 2}

\\
    & \ds =  \vs_{-k|k}^2
    
            + (\Sigma_{k,k}^{-1} \widetilde{\Sigma}_k\Sigma_{k,k}^{-1}) \mM_k^{\odot 2}
                
\\
    & \ds =  \vs_{-k|k}^2
            + \widetilde{\mM}_k^{\odot 2}/\widetilde{\Sigma}_k

\end{array}
$$

\noindent which can be calculated in $O(n)$ time.

Thus, we can update $\widetilde{\veta}$, $\widetilde{\mM}$ and $\widetilde{\vs}^2$ via
$$
\widetilde\veta
=
\veta
+
(\widetilde\mu_k-\mu_k)\vc,
\qquad 
\widetilde \mM_k
=
\vc\widetilde\Sigma_k,
\qquad 
\widetilde \mM_{-k}
=
\mM_{-k}
+
\frac{\widetilde\Sigma_k-\Sigma_{k,k}}
     {\Sigma_{k,k}^2}
\mM_k\mSigma_{k,-k},
$$

% \noindent or equivalently,
% $$
% \widetilde \mM_{-k} = \mM_{-k}+(\Sigma_{k,k}^{-1}-\widetilde\Sigma_k^{-1})\widetilde \mM_k\mSigma_{k,-k}
% $$

\noindent 
and
$\widetilde \vs^2
=
\vs^2_{-k\mid k}
+
\widetilde \mM_k^{\odot 2}/\widetilde\Sigma_k$.

Therefore, for each coordinate $k$, the local quantities required to form 
$r_k(\theta_k)$ can be obtained in $O(n)$ time once $\veta$, $\mM$, and $\vs^2$ 
are maintained. Updating the Gaussian mean and covariance costs $O(p^2)$, while 
updating $\veta$, $\mM$, and $\vs^2$ costs
$O(np)$. Thus the deterministic algebraic 
cost of a full sweep over $p$ coordinates is 
$O(np^2+p^3)$, excluding the cost of 
normalizing and integrating the one-dimensional local densities.

\newpage 
\section{Additional details for linear models example}
\label{sec:lm_example_details}

In this section, we provide detailed derivations for all of the material in Section \ref{sec:mp_linear_model}. Throughout this section, 
$\mathrm{IG}(a,b)$ denotes the inverse-gamma distribution with density
proportional to $x^{-a-1}\exp(-b/x)$, so that
$$
\bE(X)=\frac{b}{a-1}, \qquad\mbox{and} \qquad 
\bV(X)=\frac{b^2}{(a-1)^2(a-2)}.
$$

\subsection{Exact posterior distributions}\label{Suppl:Exact_posterior_distributions} 

The full conditional distributions are given by
\begin{equation}\label{eq:full_conditionals} 
\begin{array}{c}
\vbeta|\vy,\sigma^2
\ds \sim N_p\left(  
u \widehat{\vbeta},   u \sigma^2 (\mX^T\mX)^{-1} \right)
\quad \mbox{and} \quad 
\sigma^2|\vy,\vbeta
\ds \sim \mbox{IG}\left( A + \tfrac{1}{2}(n+p), \mathcal{B}(\vbeta) \right),

\end{array} 
\end{equation}

\noindent where 
$\mathcal{B}(\vbeta) \equiv B +  \tfrac{1}{2} \|\vy - \mX\vbeta\|^2  + \tfrac{1}{2g}\vbeta^T\mX^T\mX\vbeta$,
$\ds \widehat{\vbeta} =  (\mX^T\mX)^{-1}\mX^T\vy$,
and $u = g/(1+g)$.

The exact posterior distributions for $\vbeta$ and $\sigma^2$ are given by
\begin{equation}\label{eq:exact_posterior_linear_model} 
\begin{array}{rl}
\vbeta\given\vy
& \ds \sim t\left( 
u\widehat{\vbeta},  \left(\frac{B + \frac{n}{2}\widehat{\sigma}_u^2}{A + \tfrac{n}{2}}\right) u (\mX^T\mX)^{-1},
2A + n \right), \qquad \mbox{and} 

\\  

\sigma^2\given\vy
& \ds \sim \mbox{IG}\left( A + \frac{n}{2}, B +  \frac{n\widehat{\sigma}_u^2}{2} \right),
\end{array} 
\end{equation}

\noindent where
\begin{equation}\label{eq:constants_lienar}
\begin{array}{c}
 \widehat{\sigma}_u^2 = \frac{1}{n}[\,\|\vy\|^2
- u \, \vy^T\mX  (\mX^T\mX)^{-1}\mX^T\vy\,].
\end{array} 
\end{equation}

\noindent Note that as $g\to \infty$, $\widehat{\sigma}_u^2$ converges to the maximum likelihood estimator for $\sigma^2$.

\subsection{MFVB  for linear models}\label{Suppl:MFVB_for_linear_models} 

The MFVB approximation for the linear model  corresponding to the factorization
$q(\vbeta,\sigma^2) = q(\vbeta)q(\sigma^2)$
has $q$-densities of the forms:
$$
\begin{array}{c}
\ds q(\vbeta)
= N_p(\widetilde{\vmu},\widetilde{\mSigma})
\qquad \mbox{and} \qquad 
q(\sigma^2) = \mbox{IG}(\widetilde{A}, \widetilde{B}),
\end{array}
$$

\noindent where the updates are given by equations (\ref{eq:vb_linear_beta}) and (\ref{eq:vb_linear_sigma2}) in Algorithm \ref{alg:vb_linear} which summarizes the MFVB method for the linear model. Note that the distributional form for $q(\vbeta)$ comes directly from (\ref{eq:vb_update}),
while the true posterior follows a $t$-distribution. Hence, for any factorized $q$-density MFVB cannot be exact for this model.

%\subsection{Derivations for MFVB}\label{Suppl:E2} 
For the linear model described in Section \ref{sec:mp_linear_model} we use the $q$-density factorization
$q(\vbeta,\sigma^2) = q(\vbeta)q(\sigma^2)$, the updates for $q(\vbeta)$
and $q(\sigma^2)$ are derived via
$$
\begin{array}{rl}
q(\vbeta)
& \ds \propto \exp\left[ \bE_{q(\sigma^2)} \left\{  -\tfrac{1}{2\sigma^2}\|\vy - \mX\vbeta\|^2 - \tfrac{1}{2g\sigma^2}\vbeta^T\mX^T\mX\vbeta 
\right\} 
\right] 

\\  

& \ds \propto \exp\left[ \bE_{q(\sigma^2)} \left\{ 
  \tfrac{\vy^T\mX\vbeta }{\sigma^2}
- \tfrac{1+1/g}{2\sigma^2}\vbeta^T\mX^T\mX\vbeta 
\right\} 
\right]

\\  

& \ds = N_p\left(  \widetilde{\vmu} \equiv 
u \widehat{\vbeta}, 
\widetilde{\mSigma} \equiv u(\widetilde{B}/\widetilde{A})  (\mX^T\mX)^{-1}
\right) \quad\mbox{and}

\\  

q(\sigma^2)
& \ds \propto 
\exp\Big[ 
- \left( A + \tfrac{n+p}{2} + 1\right) \log(\sigma^2)
\\
& \ds \qquad -  
\bE_{q(\vbeta)}\left\{ 
B + \tfrac{1}{2} \|\vy - \mX\widetilde{\vmu}\|^2 
+ \tfrac{1}{2g}\widetilde{\vmu}^T\mX^T\mX\widetilde{\vmu}
+ \tfrac{1}{2u}\mbox{tr}\left(\mX^T\mX\widetilde{\mSigma} \right) 
\right\}  \sigma^{-2}
\Big] 

\\  

& \ds = \mbox{IG}\left( 
\widetilde{A} \equiv A + \tfrac{n+p}{2}, 
\widetilde{B} \equiv B + \tfrac{1}{2} \|\vy - \mX\widetilde{\vmu}\|^2 
+ \tfrac{1}{2g}\widetilde{\vmu}^T\mX^T\mX\widetilde{\vmu}
+ \tfrac{1}{2u}\mbox{tr}\left(\mX^T\mX\widetilde{\mSigma} \right) 
 \right)
\end{array} 
$$

\noindent  noting $1+1/g = 1/u$.

\begin{algorithm}[!ht]
{\footnotesize 
\caption[Algorithm 2]{MFVB for the linear model}
\label{alg:vb_linear}
\begin{algorithmic}[1]
    \smallskip 
    \REQUIRE{$\vy\in \bR^n$, $\mX\in \bR^{n\times p}$,  $g>0$, $A>0$, and $B>0$.}  
		
	\smallskip 
	\STATE Calculate algorithm constants via  (\ref{eq:constants_lienar}).
	
	\smallskip 
	\STATE Initialize variational parameter $\widetilde{A} = A + \tfrac{1}{2}(n+p)$ and $\widetilde{B} = B + \tfrac{1}{2} \|\vy\|^2 $

	\REPEAT  
		\smallskip 
		\STATE Update $q(\vbeta)$ parameters:
		\begin{equation}\label{eq:vb_linear_beta}
		\ds \widetilde{\vmu} \leftarrow u \widehat{\vbeta}; \qquad 
		\ds \widetilde{\mSigma} \leftarrow \frac{\widetilde{B}}{\widetilde{A}}\, u (\mX^T\mX)^{-1}
		\end{equation} 
		
		\smallskip 
		\STATE Update $q(\sigma^2)$ parameters: 
		\begin{equation}\label{eq:vb_linear_sigma2}
		\begin{array}{rl} 
		\ds \widetilde{A}  \ds \leftarrow A + \frac{n+p}{2}; \qquad  
%		\\ [2ex]  
		\ds \widetilde{B}  \ds \leftarrow B + \frac{ \|\vy - \mX\widetilde{\vmu}\|^2 }{2}
+ \frac{\widetilde{\vmu}^T\mX^T\mX\widetilde{\vmu}}{2g}
+ \frac{\mbox{tr}(\mX^T\mX\widetilde{\mSigma}) }{2u}
        \end{array}
		\end{equation} 
		
		\smallskip 
	\UNTIL convergence criterion are met.
		
\end{algorithmic}
}
\end{algorithm}

\subsection{Derivations for MP - Approach 1} 
\label{Suppl:LinearModApproach1Derivations}

We consider the moment propagation approximation where 
$q(\vbeta) = N(\widetilde{\vmu},\widetilde{\mSigma})$ and
$q(\sigma^2) = \mbox{IG}(\widetilde{A},\widetilde{B})$. These are precisely the same distributional forms as MFVB. However, the updates for the $q$-density parameters is different.  For the update of $q(\vbeta)$ 
using the fact that $\vbeta\given\vy,\vsigma^2 \sim N(u \widehat{\vbeta},\sigma^2 u (\mX^T\mX)^{-1})$, we equate
\begin{align}\label{Supp:equ34}
\ds \bE_r(\vbeta) 
    & \ds = \bE_q[\bE(\vbeta\given\vy,\sigma^2)]  = u \widehat{\vbeta}   
\\ \label{Supp:equ35}
\ds \bV_r(\vbeta) 
    & \ds = \bE_q[\bV(\vbeta\given\vy,\sigma^2)] + \bV_q[\bE(\vbeta\given\vy,\sigma^2)] 
= \frac{\widetilde{B}}{\widetilde{A} - 1} u(\mX^T\mX)^{-1} 
\end{align}

\noindent with $\bE_q(\vbeta)=\widetilde{\vmu}$, and $\bV_q(\vbeta) = \widetilde{\mSigma}$.
Hence, by matching  \eqref{Supp:equ34} and \eqref{Supp:equ35} with $\bE_q(\vbeta)$ and $\bV_q(\vbeta)$, respectively, and solving for $\widetilde{\vmu}$ and $\widetilde{\mSigma}$, we obtain the update
$$
\widetilde{\vmu} \leftarrow  u \widehat{\vbeta} \quad \mbox{and} \quad  \widetilde{\mSigma} \leftarrow \frac{\widetilde{B}}{\widetilde{A} - 1} u(\mX^T\mX)^{-1}.
$$

\medskip 
\noindent Similarly, for the update of $q(\sigma^2)$ since $\sigma^2\given\vy,\vbeta \sim \mbox{IG}( A + \tfrac{1}{2}(n+p), B(\vbeta))$ we have 
\begin{align}\label{Supp:equ36}
\bE_r(\sigma^2) 
    & = \bE_q[\bE(\sigma^2\given\vy,\vbeta)]  = \frac{\bE_q[\mathcal{B}(\vbeta)]}{A + \tfrac{n+p}{2} - 1}, \qquad \mbox{and}
\\ \label{Supp:equ37} 
\bV_r(\sigma^2) 
    &  = \bE_q[\bV(\sigma^2\given\vy,\vbeta)] + \bV_q[\bE(\sigma^2\given\vy,\vbeta)]  
\\ \nonumber
    &  = \bE_q\left[ \frac{\mathcal{B}(\vbeta)^2}{(A+\tfrac{n+p}{2} - 1)^2(A+\tfrac{n+p}{2} - 2)} \right]  
       + \bV_q\left[ \frac{\mathcal{B}(\vbeta)}{A+\tfrac{n+p}{2} - 1} \right]   
\\ \nonumber 
    &  = \frac{\bE_q\left[ \mathcal{B}(\vbeta)\right]^2  }{(A+\tfrac{n+p}{2} - 1)^2(A+\tfrac{n+p}{2} - 2)}  
\\ \nonumber 
    &  \qquad 
    + \left[  \frac{  1 }{(A+\tfrac{n+p}{2} -     1)^2(A+\tfrac{n+p}{2} - 2)} 
    + \frac{1}{(A+\tfrac{n+p}{2} - 1)^2} \right]    \bV_q\left[ \mathcal{B}(\vbeta) \right]
\\ \nonumber
    &  = 
    \frac{  \bE_q\left[ \mathcal{B}(\vbeta)\right]^2  }{(A+\tfrac{n+p}{2} - 1)^2(A+\tfrac{n+p}{2} - 2)}  
    + \frac{\bV_q\left[ \mathcal{B}(\vbeta) \right]}{(A+\tfrac{n+p}{2} - 1)(A+\tfrac{n+p}{2} - 2)}.
\end{align}

\noindent 
Note that $u$ and $\mathcal{B}(\boldsymbol{\beta})$ are defined in Section \ref{Suppl:Exact_posterior_distributions}. 
Next, 
$$
\begin{array}{rl} 
\bE_q[\mathcal{B}(\vbeta)] 
& \ds = B +  \frac{1}{2} \|\vy - \mX\widetilde{\vmu}\|^2 
	+ \frac{\widetilde{\vmu}^T\mX^T\mX\widetilde{\vmu}}{2g} 
	+ \frac{\mbox{tr}(\mX^T\mX\widetilde{\mSigma})}{2u}
\\  
\bV_q[\mathcal{B}(\vbeta)] 

& \ds = \bV_q\left[ B +  \frac{1}{2} \|\vy - \mX\vbeta\|^2  + \frac{\vbeta^T\mX^T\mX\vbeta}{2g} \right] \\ 

%& \ds = \bV_q\left[ \tfrac{1}{2}\vbeta^T\mX^T\mX\vbeta    - \vy^T\mX\vbeta  + \tfrac{1}{2g} \vbeta^T\mX^T\mX\vbeta   \right] \\ [2ex]

& \ds = \bV_q\left[ \tfrac{1}{2}\vbeta^T( u^{-1}\mX^T\mX ) \vbeta - \vy^T\mX\vbeta     \right] \\  
 
%& \ds = \bV_q\left[ \tfrac{1}{2}\vbeta^T\mA\vbeta  - \vb^T\vbeta     \right] \\ [2ex]

%& \ds = \bV_q\left[ \tfrac{1}{2}\vbeta^T\mA\vbeta  - \vb^T\vbeta + \tfrac{1}{2}\vb^T\mA^{-1}\mA\mA^{-1}\vb -  \tfrac{1}{2}\vb^T\mA^{-1}\mA\mA^{-1}\vb \right] \\ [2ex]

& \ds = \frac{1}{4u^2} \bV_q\left[ ( \vbeta - u\widehat{\vbeta})^T(\mX^T\mX)( \vbeta - u\widehat{\vbeta}) \right] \\  

& \ds = 
\frac{\operatorname{tr}[ (\mX^T\mX \widetilde{\mSigma})^2 ] }{2u^2}.
\end{array} 
$$

\noindent In the derivation of $\bV_q[\mathcal{B}(\vbeta)]$ above, the first line absorbs constants into the variance operator, the second line completes the square in $\vbeta$, the third line takes constants outside the variance operator, and the last line follows from the variance formula for quadratic forms of multivariate normal variables, i.e., equation~(\ref{eq:moments_of_quadratic_normal}).

\bigskip 
\noindent Hence, by matching \eqref{Supp:equ36} and \eqref{Supp:equ37} with $\bE_q(\sigma^2)$ and $\bV_q(\sigma^2)$, respectively, we obtain
the following equations
$$
\bE_r(\sigma^2) = \frac{\widetilde{B}}{\widetilde{A} - 1} = \bE_q(\sigma^2)
\qquad \mbox{and} \qquad 
\bV_r(\sigma^2) = \frac{\widetilde{B}^2}{(\widetilde{A} - 1)^2(\widetilde{A} - 2)} = \bV_q(\sigma^2).
$$

\noindent 
Solving for $\widetilde{A}$ and $\widetilde{B}$ leads to the update
\begin{equation}\label{eq:inverse_gamma_moment_matching}
\widetilde{A} 
\leftarrow \frac{[\bE_r(\sigma^2)  ]^2}{\bV_r(\sigma^2) } + 2; 
\qquad 
\ds \widetilde{B} 
\leftarrow ( \widetilde{A} - 1 )\,\bE_r(\sigma^2) .
\end{equation} 

\noindent The MP approximation can then be summarized via Algorithm \ref{alg:mp_linear1}.

{\footnotesize 
\begin{algorithm}[!ht]
\caption[Algorithm 3]{MP for the linear model: Approach 1}
\label{alg:mp_linear1}
\begin{algorithmic}[1]
    \REQUIRE{$\vy\in \bR^n$, $\mX\in \bR^{n\times p}$,  $g>0$, $A>0$, and $B>0$.}  
		
	\STATE Calculate algorithm constants via  (\ref{eq:constants_lienar}).
	
	\STATE Initialize variational parameters: $\widetilde{A} = A + \tfrac{1}{2}(n+p)$ and $\widetilde{B} = B + \tfrac{n}{2}\widehat{\sigma}_u^2$.

	\REPEAT  
	
	\STATE Approximate posterior moments of $\vbeta$ assuming $q(\sigma^2)=\mbox{IG}(\widetilde{A},\widetilde{B})$, where $u = \frac{g}{1+g}$:
	\begin{equation}\label{alg:mp_linear1_beta_moments}
	\bE_r(\vbeta) \leftarrow u\widehat{\vbeta}; \qquad 
	\bV_r(\vbeta) \leftarrow \frac{\widetilde{B}}{\widetilde{A} - 1} u(\mX^T\mX)^{-1}
	\end{equation} 
	
	\STATE Update $q(\vbeta)$ parameters via moment matching
	\begin{equation}\label{alg:mp_linear1_beta}
	\ds \widetilde{\vmu} \leftarrow \bE_r(\vbeta); \qquad 
	\ds \widetilde{\mSigma} \leftarrow \bV_r(\vbeta)
	\end{equation} 
	
    \STATE Approximate posterior moments of $\sigma^2$ assuming $q(\vbeta)=N(\widetilde{\vmu},\widetilde{\mSigma})$:
	\begin{align}
	\ds \bE_q[\mathcal{B}(\vbeta)] & 
	    \ds \leftarrow B +  \tfrac{1}{2} \|\vy - \mX\widetilde{\vmu}\|^2 
	    + \tfrac{1}{2g} \widetilde{\vmu}^T\mX^T\mX\widetilde{\vmu}
	    + \tfrac{1}{2u}\mbox{tr}(\mX^T\mX\widetilde{\mSigma}) 
    \\ 
    \bV_q[\mathcal{B}(\vbeta)] & 
        \ds \leftarrow  
        \frac{\operatorname{tr}[ (\mX^T\mX \widetilde{\mSigma})^2 ] }{2u^2}
	\\ 
	\ds \bE_r(\sigma^2) &
	    \leftarrow \frac{ \bE_q[\mathcal{B}(\vbeta)]  }{A + \frac{n+p}{2} - 1}
	\\  
	\ds \bV_r(\sigma^2) &
	\leftarrow \frac{  \bE_q[\mathcal{B}(\vbeta)]^2 
}{(A + \frac{n+p}{2} - 1)^2(A + \frac{n+p}{2} - 2)}   
+   \frac{ \bV_q[\mathcal{B}(\vbeta)]}{(A + \frac{n+p}{2} - 1)(A + \frac{n+p}{2} - 2)}
\label{eq:variance_mp_update}
	\end{align}

    \STATE Update $q(\sigma^2)$ parameters via moment matching
    \begin{equation}\label{alg:mp_linear1_sigma2}
    \ds \widetilde{A} \leftarrow \frac{[\bE_r(\sigma^2)  ]^2}{\bV_r(\sigma^2) } + 2; \qquad \ds \widetilde{B} \leftarrow \bE_r(\sigma^2) ( \widetilde{A} - 1 );
    \end{equation} 

	\UNTIL convergence criteria are met.
		
\end{algorithmic}
\end{algorithm}
}

\subsection{Fixed-point properties of MP Approach 1}

\begin{result}\label{result2}
Let \(a=A+n/2\) and
$b=B+n\widehat\sigma_u^2/2$.
Assume $a>2$. At any admissible fixed point of the moment conditions
%\((35)\)--\((41)\)
(\ref{alg:mp_linear1_beta_moments})--(\ref{alg:mp_linear1_sigma2})
of Algorithm~\ref{alg:mp_linear1}, the propagated posterior mean and variance of
$\sigma^2$ are
$$
\bE_r^\ast(\sigma^2)
=
\frac{b}{a-1}
=
\frac{B+n\widehat\sigma_u^2/2}{A+n/2-1},
\quad \mbox{and} \quad 
\bV_r^\ast(\sigma^2)
=
\frac{[\bE_r^\ast(\sigma^2)]^2}
     {A+(n+p)/2-2}
\left[
1+
\frac{p/2}{A+(n+p)/2-1}
\right]
$$

\noindent
respectively. Furthermore,
$\widetilde\vmu^\ast=u\widehat\vbeta$,
and 
$\widetilde\mSigma^\ast
=
\bE_r^\ast(\sigma^2)\,u(\mX^T \mX)^{-1}$.
Hence the MP approximation gives the exact posterior means of $\vbeta$ and
$\sigma^2$, and the exact posterior covariance matrix of $\vbeta$. However, the propagated variance \
$\bV_r^\ast(\sigma^2)$ is smaller than the
exact posterior variance
$
\bV(\sigma^2|\vy)
=
[\bE(\sigma^2|\vy)]^2/(A+n/2-2)$.
For fixed $p$, the ratio
$\bV_r^\ast(\sigma^2)/\bV(\sigma^2|\vy)$
converges to $1$ as 
$n\to\infty$. Thus, Algorithm~\ref{alg:mp_linear1} recovers the exact first
posterior moments and the exact covariance of $\vbeta$, but it still
underestimates the posterior variance of $\sigma^2$ by a term that vanishes
asymptotically for fixed $p$.
\end{result}

\newpage 
 
\noindent {\bf Proof:} 
At a fixed point of Algorithm \ref{alg:mp_linear1}, the left-hand sides of the assignments are equal to the corresponding right-hand sides of equations (\ref{alg:mp_linear1_beta_moments})--(\ref{alg:mp_linear1_sigma2}).
% Algorithm \ref{alg:mp_linear1} converges when the left hand side of assignments ($\leftarrow$) are equal to the right hand of assignments (at least closely). 
This is equivalent to the following system of equations where $\widetilde{\vmu} = u \widehat{\vbeta}$ and $\widetilde{\mSigma} 
     = \frac{\widetilde{B}}{\widetilde{A} - 1} u(\mX^T\mX)^{-1}$: 
\begin{align*}
\bE_q[\mathcal{B}(\vbeta)]
    & \ds = 
    B +  \tfrac{1}{2} \|\vy - \mX\widetilde{\vmu}\|^2 
    + \tfrac{1}{2g} \widetilde{\vmu}^T\mX^T\mX\widetilde{\vmu}
    + \tfrac{1}{2u}\mbox{tr}(\mX^T\mX\widetilde{\mSigma}), \\
    \bV_q[\mathcal{B}(\vbeta)]
     & \ds = \frac{1}{2u^2}  \operatorname{tr}\left[ (\mX^T\mX \widetilde{\mSigma})^2 \right], 
\\ 
\bE_r(\sigma^2)
    & \ds  = \frac{\bE_q[\mathcal{B}(\vbeta)]}{A+\tfrac{n+p}{2} - 1},
\\
\bV_r(\sigma^2)
    & \ds =  
    \frac{\bE_q[\mathcal{B}(\vbeta)]^2}{(A+\tfrac{n+p}{2}  - 1)^2(A+\tfrac{n+p}{2} - 2)}   
   +\frac{\bV_q[\mathcal{B}(\vbeta)]  }{(A+\tfrac{n+p}{2}  - 1)  (A+\tfrac{n+p}{2} - 2)},
\\
\widetilde{A} 
    & \ds = \frac{[\bE_r(\sigma^2)]^2}{\bV_r(\sigma^2)} + 2, 
\qquad \widetilde{B} 
     = (\widetilde{A}  - 1)\bE_r(\sigma^2).  
\end{align*} 

\noindent Substituting $\widetilde{\vmu}$ and $\widetilde{\mSigma}$ into the expressions for 
$\bE_q[\mathcal{B}(\vbeta)]$ and $\bV_q[\mathcal{B}(\vbeta)]$,
we obtain
$$
\begin{array}{c}
\bE_q[\mathcal{B}(\vbeta)]
 \ds = 
B +  \tfrac{1}{2}n\widehat{\sigma}_u^2
+ \tfrac{1}{2} p \, \bE_r(\sigma^2)
\qquad \mbox{and} \qquad 
\bV_q[\mathcal{B}(\vbeta)]
 \ds = \tfrac{1}{2}  p\, [\bE_r(\sigma^2)]^2
\end{array}
$$

\noindent which follows from the facts that 
$$
\begin{array}{rl}
\ds \|\vy - \mX\widetilde{\vmu}\|^2 + g^{-1}\widetilde{\vmu}^T\mX^T\mX\widetilde{\vmu} 
& \ds = n\widehat{\sigma}_u^2 
\\ [2ex]
u^{-1}\operatorname{tr}\left[ (\mX^T\mX \widetilde{\mSigma})\right]
& \ds = \left( \frac{\widetilde{B}}{\widetilde{A} - 1} \right) p = p\, \bE_r(\sigma^2) 
\\ [2ex]
u^{-2}\operatorname{tr}\left[ (\mX^T\mX \widetilde{\mSigma})^2\right]
& \ds = \left( \frac{\widetilde{B}}{\widetilde{A} - 1} \right)^2 p = p\, [\bE_r(\sigma^2)]^2

\\ [2ex]

\ds \left( \frac{\widetilde{B}}{\widetilde{A} - 1} \right) 
& \ds = \bE_r(\sigma^2).

\end{array} 
$$

\noindent Substituting the above expressions for $\bE_q[\mathcal{B}(\vbeta)]$ and $\bV_q[\mathcal{B}(\vbeta)]$
into the equations for
$\bE_r(\sigma^2)$ and $\bV_r(\sigma^2)$
we obtain
\begin{equation}\label{eq:mp1_mean_and_variance}
\begin{array}{rl}
\bE_r(\sigma^2) 
& \ds 
= \frac{ B +  \tfrac{n\widehat{\sigma}_u^2}{2}  
	+ \tfrac{1}{2}p\, \bE_r(\sigma^2) }{A + \tfrac{n+p}{2} - 1}
\\ [2ex]
\bV_r(\sigma^2)
& \ds =  
\frac{\left[ B + \tfrac{n\widehat{\sigma}_u^2}{2}  
	+ \tfrac{1}{2}p \,\bE_r(\sigma^2)  \right]^2}{(A + \tfrac{n+p}{2} - 1)^2(A + \tfrac{n+p}{2} - 2)}   
+ \frac{   \tfrac{p}{2}[\bE_r(\sigma^2)]^2}{(A + \tfrac{n+p}{2} - 1)(A + \tfrac{n+p}{2} - 2)}.
\end{array} 
\end{equation}

\noindent Solving (\ref{eq:mp1_mean_and_variance}) for $\bE_r(\sigma^2)$ we obtain
$$
\bE_r(\sigma^2) = \frac{ B +  \tfrac{n\widehat{\sigma}_u^2}{2}  }{A + \tfrac{n}{2} - 1},
$$

\noindent which is the true posterior mean of $\sigma^2$. Furthermore, since the
expression for $\bE_r(\sigma^2)$ is exact, the expression for $\bE_r(\vbeta)$
is also exact.
Substituting $\bE_r(\sigma^2)$ into the expression for 
$\bV_r(\sigma^2)$, after simplification, we obtain
% Substituting $\bE_r(\sigma^2)$ into the expression for $\bE_r(\sigma^2)$ 
% after simplification we obtain
$$
\begin{array}{rl}
\bV_r(\sigma^2)
& \ds  =  \frac{1}{A + \tfrac{n+p}{2} - 2} \left( 
1
+ \frac{   \tfrac{p}{2}    }{A + \tfrac{n+p}{2} - 1} \right) \left(  \frac{ B +  \frac{n\widehat{\sigma}_u^2}{2} }{A + \tfrac{n}{2} - 1}
\right)^2.

\end{array} 
$$
 
\noindent From this, we can obtain expressions for $\widetilde{A}$
and $\widetilde{B}$ at convergence of Algorithm \ref{alg:mp_linear1}. However, the expressions $\widetilde{A}$ and $\widetilde{B}$ do not seem to 
lend themselves to additional insights.

\bigskip 
\noindent 
The exact posterior variance is
$$
\bV(\sigma^2\given\vy)
= \frac{1}{A + \tfrac{n}{2} - 2} \left(  \frac{ B +  \frac{n\widehat{\sigma}_u^2}{2} }{A + \tfrac{n}{2} - 1}
\right)^2.
$$

\noindent 
To see the underestimation explicitly, write \(t=p/2\). Then
\[
\frac{\bV_r^\ast(\sigma^2)}
     {\bV(\sigma^2\mid y)}
=
\frac{(a-2)(a+2t-1)}
     {(a+t-1)(a+t-2)}.
\]
Since
\[
(a+t-1)(a+t-2)-(a+2t-1)(a-2)=t(t+1)>0,
\]
this ratio is less than one for \(p>0\), and tends to one as \(a\to\infty\) with
\(p\) fixed.

% \noindent so that the posterior variance for $\sigma^2$ is underestimated by VB when
% $\bV_r(\sigma^2) < \bV(\sigma^2\given\vy)$ or equivalently when
% $$
% \frac{1}{A + \tfrac{n+p}{2} - 2} \left( 
% 1
% + \frac{   \tfrac{p}{2}    }{A + \tfrac{n+p}{2} - 1} \right) 
% < \frac{1}{A + \tfrac{n}{2} - 2}.
% $$

% \noindent Letting $x=A+n/2$ and $y=p/2$ this is equivalent to
% $$
% \begin{array}{rl}
% \ds \frac{1}{x + y - 2} \left( 
% 1
% + \frac{y}{x+y - 1} \right) 
% < \frac{1}{x - 2}.
% \end{array} 
% $$

% \noindent Assuming that $x>2$ (that is when $n\ge 2$ and $A>0$) after rearranging we have
% $$
% \begin{array}{rl}
% \ds  y^2 + y > 0
% \end{array} 
% $$

% \noindent which always holds. Hence, provided $n\ge 2$ and $A>0$
% we have $\bV_r(\sigma^2) < \bV(\sigma^2\given\vy)$.

Thus, Approach 1 corrects the posterior mean of 
$\sigma^2$ and the posterior mean and variance of $\vbeta$, but it does not recover the exact marginal variance of $\sigma^2$. Approach 2 resolves this by using a multivariate $t$ approximation for $q(\vbeta)$, matching the heavy-tailed exact marginal form.

\subsection{Exact recovery for MP Approach 2}\label{Suppl:Derivations_for_MP_Approach_2}

Now we have $q(\vbeta) = t_p(\widetilde{\vmu},\widetilde{\mSigma},\widetilde{\nu})$
and $q(\sigma^2) = \mbox{IG}(\widetilde{A},\widetilde{B})$.
To derive the update for $\vbeta$ we first note that
$\vbeta\given\vy,\vsigma^2 \sim N(u \widehat{\vbeta},\sigma^2 u (\mX^T\mX)^{-1})$. The mean and covariance of $r(\vbeta)$
are $\bE_r(\vbeta) = u\widehat{\vbeta}$, and  
$\bV_r(\vbeta) = (\widetilde{B}/(\widetilde{A} - 1)) u(\mX^T\mX)^{-1}$
as before. The corresponding mean and covariance of $q(\vbeta)$
are $\bE_q(\vbeta) = \widetilde{\vmu}$ and $\bV(\vbeta) = 
(\widetilde{\nu}/(\widetilde{\nu} - 2))\widetilde{\mSigma}$. Hence,
$$
\widetilde{\vmu} = u\widehat{\vbeta} 
\qquad \mbox{and} \qquad 
\widetilde{\mSigma} = 
\left(\frac{\widetilde{\nu} - 2}{\widetilde{\nu}}\right) 
\left( \frac{\widetilde{B}}{\widetilde{A} - 1} \right) u(\mX^T\mX)^{-1}.
$$

\noindent 
Let 
$Q(\vbeta) = (\vbeta - \widetilde{\vmu})^T\widetilde{\mSigma}^{-1}(\vbeta - \widetilde{\vmu})$. Then \eqref{eq:Rong1} implies
$$
\bE_q[Q(\vbeta)^2] = \frac{\widetilde{\nu}^2(2p + p^2)}{(\widetilde{\nu} - 2)(\widetilde{\nu} - 4)}.
$$

\noindent 
The fourth moment exists provided $\widetilde\nu>4$.
We want to match this with $\bE_r[Q(\vbeta)^2]$ where
$$
Q(\vbeta) =  \left(\frac{\widetilde{\nu}}{\widetilde{\nu} - 2}\right)  \left( \frac{\widetilde{B}}{\widetilde{A} - 1} \right) (\vbeta - u\widehat{\vbeta})^T\left(
u(\mX^T\mX)^{-1}\right)^{-1} (\vbeta - u\widehat{\vbeta}).
$$

\noindent Then using \eqref{eq:moments_of_quadratic_normal} where we replace
$\mSigma$ with $\sigma^2 u(\mX^T\mX)^{-1}$, $\vmu$ with $\vzero$, and 
$\mA = [u(\mX^T\mX)^{-1}]^{-1}$
we obtain
$$
\begin{array}{rl}
\bE_r[Q(\vbeta)^2] 
    & \ds = \bE_q[ Q(\vbeta)^2 | \sD, \sigma^2 ]
\\ [2ex]
    & \ds = \left[ \left(\frac{\widetilde{\nu}}{\widetilde{\nu} - 2}\right) 
    \left( \frac{\widetilde{A} - 1}{\widetilde{B}} \right)\right]^2(2p + p^2)
    \bE_q\left( \sigma^4\right)
\\ [2ex]
    & \ds = \left[ \left(\frac{\widetilde{\nu}}{\widetilde{\nu} - 2}\right) 
    \left( \frac{\widetilde{A} - 1}{\widetilde{B}} \right)\right]^2(2p + p^2)
    \frac{\widetilde{B}^2}{(\widetilde{A} - 1)(\widetilde{A} - 2)}
\\ [2ex]
    & \ds = \left( \frac{\widetilde{\nu}^2}{(\widetilde{\nu} - 2)^2} \right) \left( \frac{\widetilde{A} - 1}{\widetilde{A} - 2} \right)(2p + p^2).
\end{array}
$$

\noindent Equating $\bE_r[Q(\vbeta)^2]$ with
$\bE_q[Q(\vbeta)^2] $, we have
$$
\frac{\widetilde{\nu}^2(2p + p^2)}{(\widetilde{\nu} - 2)(\widetilde{\nu} - 4)} = \left( \frac{\widetilde{\nu}^2}{(\widetilde{\nu} - 2)^2} \right) \left( \frac{\widetilde{A} - 1}{\widetilde{A} - 2} \right)(2p + p^2)
$$

\noindent which, after cancellation and rearranging, yields
$$
\frac{\widetilde{\nu} - 2}{\widetilde{\nu} - 4} =  \frac{\widetilde{A} - 1}{\widetilde{A} - 2}.
$$

\noindent Solving for $\widetilde{\nu}$, we obtain $\widetilde{\nu} = 2\widetilde{A}$.
 Hence, the update for $q(\vbeta)$ is given by
$$
\widetilde{\vmu} \leftarrow u \widehat{\vbeta}; \qquad 
\widetilde{\mSigma} \leftarrow \frac{\widetilde{B}}{\widetilde{A}} u(\mX^T\mX)^{-1};  \quad \mbox{and} \quad  
\widetilde{\nu} \leftarrow 2\widetilde{A}.
$$

\noindent Similarly, for the update of $q(\sigma^2)$ since $\sigma^2\given\vy,\vbeta \sim \mbox{IG}( A + \tfrac{1}{2}(n+p), \mathcal{B}(\vbeta))$, we have 
$$
\begin{array}{rl}
\ds \bE_r(\sigma^2) 
& \ds = \bE_q[\bE(\sigma^2\given\vy,\vbeta)]  = \frac{\bE_q[\mathcal{B}(\vbeta)]}{A + \tfrac{n+p}{2} - 1} 

\\  
\ds \bV_r(\sigma^2) 
& \ds = 
\frac{  \bE_q\left[ \mathcal{B}(\vbeta)\right]^2  }{(A+\tfrac{n+p}{2} - 1)^2(A+\tfrac{n+p}{2} - 2)}  + 
  \frac{\bV_q\left[ \mathcal{B}(\vbeta) \right]}{(A+\tfrac{n+p}{2} - 1)(A+\tfrac{n+p}{2} - 2)}.

\end{array} 
$$

\noindent Next, using (\ref{eq:Rong2}) we have
$$
\begin{array}{rl} 
\bE_q[\mathcal{B}(\vbeta)] 
& \ds = B +  \frac{1}{2} \|\vy - \mX\widetilde{\vmu}\|^2 
	+ \frac{\widetilde{\vmu}^T\mX^T\mX\widetilde{\vmu}}{2g} 
	+ \frac{\widetilde{\nu}}{\widetilde{\nu} - 2}\frac{\mbox{tr}(\mX^T\mX\widetilde{\mSigma})}{2u}
\\ [2ex]
\bV_q[\mathcal{B}(\vbeta)] 

%& \ds = \frac{1}{4u^2} \bV_q\left[ ( \vbeta - u\widehat{\vbeta})^T(\mX^T\mX)( \vbeta - u\widehat{\vbeta}) \right] \\ [2ex]
 
& \ds = \frac{1}{2u^2}\left[ 
\frac{\widetilde{\nu}^2 \mbox{tr}\{(\mX^T\mX\widetilde{\mSigma})^2\}}{(\widetilde{\nu} - 2)(\widetilde{\nu} - 4)}
+ \frac{\widetilde{\nu}^2\{\mbox{tr}(\mX^T\mX\widetilde{\mSigma})\}^2}{(\widetilde{\nu} - 2)^2(\widetilde{\nu} - 4)} \right].

\end{array} 
$$

\noindent We can perform the update for $q(\sigma^2)$ via moment matching with the inverse-gamma distribution via Equation (\ref{eq:inverse_gamma_moment_matching}).

\begin{algorithm}[!ht]
{\footnotesize 
\caption[Algorithm 4]{MP for the linear model: Approach 2}
\label{alg:mp_linear2}
\begin{algorithmic}[1]
    \REQUIRE{$\vy\in \bR^n$, $\mX\in \bR^{n\times p}$,  $g>0$, $A>0$, and $B>0$.}  
		
	\STATE Calculate algorithm constants via  (\ref{eq:constants_lienar}).
		
	\STATE Initialize variational parameters: $\widetilde{A} \leftarrow A + \tfrac{1}{2}(n+p)$ and $\widetilde{B} \leftarrow B + \tfrac{n}{2}\widehat{\sigma}_u^2$.
    
	\REPEAT

	\STATE Update $q(\vbeta)=t(\widetilde{\vmu},\widetilde{\mSigma},\widetilde{\nu})$   via moment matching using approximate posterior
	means and covariances of $\vbeta$, and moments of $Q(\vbeta)^2$
	assuming $q(\sigma^2)=\mbox{IG}(\widetilde{A},\widetilde{B})$:
	\begin{equation}\label{alg:mp_linear2_beta}
	\ds \widetilde{\vmu} \leftarrow  u \widehat{\vbeta}; \qquad 
	\ds \widetilde{\mSigma} \leftarrow\frac{\widetilde{B}}{\widetilde{A}} u(\mX^T\mX)^{-1};  \qquad
\widetilde{\nu} \leftarrow 2\widetilde{A}
    \end{equation}
	
	\STATE Approximate posterior moments of $\sigma^2$ assuming $q(\vbeta)=t(\widetilde{\vmu},\widetilde{\mSigma},\widetilde{\nu})$:
\begin{align}
\ds \bE_q[\mathcal{B}(\vbeta)] 
    &\leftarrow B +  \frac{1}{2} \|\vy - \mX\widetilde{\vmu}\|^2 
+ \frac{1}{2g} \widetilde{\vmu}^T\mX^T\mX\widetilde{\vmu}
+ \frac{\widetilde{\nu}\,\mbox{tr}(\mX^T\mX\widetilde{\mSigma})}{2u(\widetilde{\nu} - 2)}
\label{alg:mp_linear2_EB}
\\ 
\bV_q[\mathcal{B}(\vbeta)] 
    & \ds \leftarrow  
\frac{1}{2u^2}\left[ 
\frac{\widetilde{\nu}^2 \mbox{tr}\{(\mX^T\mX\widetilde{\mSigma})^2\}}{(\widetilde{\nu} - 2)(\widetilde{\nu} - 4)}
+ \frac{\widetilde{\nu}^2\mbox{tr}\{(\mX^T\mX\widetilde{\mSigma})\}^2}{(\widetilde{\nu} - 2)^2(\widetilde{\nu} - 4)} \right] 
\label{alg:mp_linear2_VB}
\\
\ds \bE_r(\sigma^2) 
    & \leftarrow \frac{ \bE_q[\mathcal{B}(\vbeta)]  }{A + \frac{n+p}{2} - 1};
\label{eq:Er_sigma2}
\\ 
\ds \bV_r(\sigma^2) 
& \leftarrow \frac{  \bE_q[\mathcal{B}(\vbeta)]^2 
}{(A + \frac{n+p}{2} - 1)^2(A + \frac{n+p}{2} - 2)}   
+   \frac{ \bV_q[\mathcal{B}(\vbeta)]}{(A + \frac{n+p}{2} - 1)(A + \frac{n+p}{2} - 2)}
\label{eq:Vr_sigma2}
\end{align}

    \STATE Update $q(\sigma^2)$ parameters via moment matching:
    \begin{equation}\label{alg:mp_linear2_sigma2}
    \ds \widetilde{A} \leftarrow \frac{[\bE_r(\sigma^2)  ]^2}{\bV_r(\sigma^2) } + 2; \qquad \ds \widetilde{B} \leftarrow \bE_r(\sigma^2) ( \widetilde{A} - 1 );
    \end{equation} 

	\UNTIL convergence criteria are met.
		
\end{algorithmic}
}
\end{algorithm}

\newpage 

\subsection{MP Approach 2 Theory}

\begin{result}\label{result3} 
Suppose $\widetilde A>2$ and $\widetilde B>0$. The consistency conditions
(\ref{alg:mp_linear2_beta})--(\ref{alg:mp_linear2_sigma2}) of Algorithm~\ref{alg:mp_linear2} have a unique admissible fixed point, given by

% The unique admissible fixed point of the consistency conditions (\ref{alg:mp_linear2_beta})--(\ref{alg:mp_linear2_sigma2}) of Algorithm \ref{alg:mp_linear2} is
$$
q^\ast(\vbeta)=t_p(\widetilde\vmu^\ast,\widetilde\mSigma^\ast,\widetilde\nu^\ast),
\qquad
q^\ast(\sigma^2)=\mathrm{IG}(\widetilde A^\ast,\widetilde B^\ast),
$$

\noindent 
where
\[
\widetilde A^\ast=A+\frac n2,\qquad
\widetilde B^\ast=B+\frac n2\widehat\sigma_u^2,
\]
\[
\widetilde\vmu^\ast=u\widehat\vbeta,\qquad
\widetilde\mSigma^\ast=
\left(
\frac{B+n\widehat\sigma_u^2/2}{A+n/2}
\right)
u(\mX^T\mX)^{-1},
\qquad
\widetilde\nu^\ast=2A+n.
\]
These are the exact marginal posterior distributions for 
$\vbeta$ and $\sigma^2$.
\end{result}

% \begin{result}\label{result3} 
% The unique fixed point of the 
% consistency conditions (\ref{alg:mp_linear2_beta})--(\ref{alg:mp_linear2_sigma2})
% of Algorithm \ref{alg:mp_linear2} leads to the following $q$-densities:
% $$
% q^*(\vbeta) = t_p(\widetilde{\vmu}^*,\widetilde{\mSigma}^*,\widetilde{\nu}^*) \qquad \mbox{and} \qquad 
% q^*(\sigma^2) = \mbox{IG}(\widetilde{A}^*,\widetilde{B}^*)
% $$

% \noindent where $\widetilde{A}^* \equiv A + \tfrac{n}{2}$,
% $\widetilde{B}^*
% \equiv  B + \tfrac{n}{2}\widehat{\sigma}_u^2$,
% $$
% \begin{array}{c}
% \ds \widetilde{\vmu}^* \equiv u\widehat{\vbeta},
% \qquad 
% \widetilde{\mSigma}^* \equiv \left( \frac{B + \frac{ n\widehat{\sigma}_u^2}{2}}{A + \frac{n}{2}}
% \right) u (\mX^T\mX)^{-1}, 
% \quad \mbox{and} \quad  
% \widetilde{\nu}^* = 2A + n.
% \end{array} 
% $$

% \noindent These are the exact marginal posterior distributions for 
% $\vbeta$ and $\sigma^2$, respectively.
% \end{result}

\newpage 
\noindent {\bf Proof:} 
At a fixed point of Algorithm \ref{alg:mp_linear2}, the left-hand sides of the assignments are equal to the corresponding right-hand sides of equations (\ref{alg:mp_linear2_beta})--(\ref{alg:mp_linear2_sigma2}).
%Algorithm \ref{alg:mp_linear2} converges when the left %hand side of assignments ($\leftarrow$) are equal to the %right hand of assignments (at least closely). 
This is equivalent to solving the following system of equations: 
\begin{align} 
\widetilde{\vmu} & \ds = u \widehat{\vbeta} 
\label{eq:approach2_matching_mutilde}
\\  
\widetilde{\mSigma} 
& \ds = \frac{\widetilde{B}}{\widetilde{A}} u(\mX^T\mX)^{-1}
\\  
\widetilde{\nu} & \ds = 2\widetilde{A}
\label{eq:approach2_matching_nu}
\\  
\bE_q[\mathcal{B}(\vbeta)]
& \ds = 
B +  \frac{1}{2} \|\vy - \mX\widetilde{\vmu}\|^2 
	+ \frac{\widetilde{\vmu}^T\mX^T\mX\widetilde{\vmu}}{2g} 
	+ \frac{\widetilde{\nu}}{\widetilde{\nu} - 2}\frac{\mbox{tr}(\mX^T\mX\widetilde{\mSigma})}{2u}
\\  
\bV_q[\mathcal{B}(\vbeta)]
& \ds = \frac{1}{2u^2}\left[ 
\frac{\widetilde{\nu}^2 \mbox{tr}[(\mX^T\mX\widetilde{\mSigma})^2]}{(\widetilde{\nu} - 2)(\widetilde{\nu} - 4)}
+ \frac{\widetilde{\nu}^2[\mbox{tr}(\mX^T\mX\widetilde{\mSigma})]^2}{(\widetilde{\nu} - 2)^2(\widetilde{\nu} - 4)} \right] 
\\  
\bE_r(\sigma^2)
& \ds  = \frac{\bE_q[\mathcal{B}(\vbeta)]}{A+\tfrac{n+p}{2} - 1}
\\  
\bV_r(\sigma^2)
& \ds =  
\frac{  \bE_q[\mathcal{B}(\vbeta)]^2 
}{(A+\tfrac{n+p}{2}  - 1)^2(A+\tfrac{n+p}{2}  - 2)}   
+   \frac{\bV_q[\mathcal{B}(\vbeta)] }{(A+\tfrac{n+p}{2}  - 1)(A+\tfrac{n+p}{2}  - 2)}
\\  
\widetilde{A} 
& \ds = \frac{[\bE_r(\sigma^2)]^2}{\bV_r(\sigma^2)} + 2 
\label{eq:approach2_matching_Atilde}
\\  
\ds \widetilde{B} 
& \ds = (\widetilde{A}  - 1)\bE_r(\sigma^2)  
\label{eq:approach2_matching_Btilde}
\end{align}

\noindent 
The moment-matching equations for $\vbeta$, together with
$\widetilde\nu=2\widetilde A$, imply
$$
\widetilde\mSigma
=
\frac{\widetilde\nu-2}{\widetilde\nu}
\bE_r(\sigma^2)u(\mX^T\mX)^{-1}
\qquad \mbox{and} \qquad 
\frac{[\bE_r(\sigma^2)]^2}{\widetilde\nu-4}
=
\frac{\bV_r(\sigma^2)}{2}.
$$

% \noindent Matching second moments of $\vbeta$, i.e., 
% \begin{equation}\label{eq:linear_model_mathcing}
% \begin{array}{c}
% \bE_q(\vbeta) \quad  \mbox{with} \quad  \bE_r(\vbeta),  \qquad 
% \bV_q(\vbeta) \quad  \mbox{with} \quad  \bV_r(\vbeta),  \qquad  \mbox{and}  \\ [2ex]
% \bE_r\left[  Q(\vbeta)^2 \right] 
% = \bE_q\left[ \bE\left\{  Q(\vbeta)^2  \given \vy,\sigma^2 \right \} \right] 
% \quad \mbox{with} \qquad \bE_q[ Q(\vbeta)^2 ]
% \end{array} 
% \end{equation}
 
% \noindent and combining (\ref{eq:approach2_matching_nu})-
% (\ref{eq:approach2_matching_Atilde})
% leads to
% $$
% %\frac{\widetilde{\nu}  }{\widetilde{\nu}-2} \widetilde{\mSigma} =  \frac{\widetilde{B}}{\widetilde{A} - 1} u(\mX^T\mX)^{-1}
% %\qquad \mbox{and hence} \qquad 
% \widetilde{\mSigma} = 
% \frac{\widetilde{\nu}-2}{\widetilde{\nu}}
% \bE_r(\sigma^2) u (\mX^T\mX)^{-1}
% \qquad \mbox{and} \qquad
% \frac{[\bE_r(\sigma^2)]^2}{\widetilde{\nu} - 4} =\frac{\bV_r(\sigma^2)}{2}
% $$

% \noindent respectively.
%
%\noindent Also, combining
%$$
%\widetilde{\nu} = 2\widetilde{A} \qquad \mbox{and} \qquad \widetilde{A} 
%  = \frac{[\bE_q^{MP}(\sigma^2)]^2}{\bV_q^{MP}(\sigma^2)} + 2 
%$$
%
%\noindent implies that 
%$$
%\widetilde{\nu} - 4 = 2\widetilde{A}  - 4 =  2\left( \frac{[\bE_q^{MP}(\sigma^2)]^2}{\bV_q^{MP}(\sigma^2)} + 2  \right) - 4
%= 2\frac{[\bE_q^{MP}(\sigma^2)]^2}{\bV_q^{MP}(\sigma^2)}
%$$
%
%\noindent and hence,
%$$
%\frac{[\bE_q^{MP}(\sigma^2)]^2}{\widetilde{\nu} - 4} =\frac{\bV_q^{MP}(\sigma^2)}{2}.
%$$
%
%
\noindent Substituting the above expressions and $\widetilde{\vmu} = u\widehat{\vbeta}$ 
into the equations for $\bE_q[\mathcal{B}(\vbeta)]$ and $\bV_q[\mathcal{B}(\vbeta)]$ we obtain
$$
\begin{array}{rl}
\bE_q[\mathcal{B}(\vbeta)]
& \ds 
= B +  \frac{n}{2}\widehat{\sigma}_u^2
	+  \frac{p}{2}\bE_r(\sigma^2) \qquad \mbox{and}
\\ 
[2ex]
\bV_q[\mathcal{B}(\vbeta)]
& \ds = 
  \frac{p}{2} [\bE_r(\sigma^2)]^2  +
  \frac{p^2+2p}{2}\frac{[\bE_r(\sigma^2)]^2}{\widetilde{\nu} - 4}  

=\frac{p}{2} [\bE_r(\sigma^2)]^2  +
  \frac{p^2+2p}{4}   \bV_r(\sigma^2).
  
\end{array} 
$$

\noindent Hence,
$$
\begin{array}{rl}
\bE_r(\sigma^2)
& \ds  = \frac{ B +  \frac{n}{2}\widehat{\sigma}_u^2
	+  \frac{p}{2}\bE_r(\sigma^2)  }{A+\tfrac{n+p}{2} - 1}
\\ [2ex]
\bV_r(\sigma^2)
& \ds =  
\frac{ \left[ B +  \frac{n}{2}\widehat{\sigma}_u^2
	+  \frac{p}{2}\bE_r(\sigma^2)  \right]^2 
}{(A+\tfrac{n+p}{2}  - 1)^2(A+\tfrac{n+p}{2}  - 2)}   
+   
\frac{\frac{p}{2} [\bE_r(\sigma^2)]^2}{(A+\tfrac{n+p}{2}  - 1)(A+\tfrac{n+p}{2}  - 2)}
\\ [2ex]
& \ds +   
\frac{\frac{p^2+2p}{4}   \bV_r(\sigma^2) }{(A+\tfrac{n+p}{2}  - 1)(A+\tfrac{n+p}{2}  - 2)}.
  
\end{array} 
$$

\noindent Solving for the first equation gives
$$
\bE_r(\sigma^2) = \frac{B +  \frac{n}{2}\widehat{\sigma}_u^2
	  }{A+\tfrac{n }{2} - 1}
$$

\noindent which is the exact posterior mean for $\sigma^2$.

\bigskip 
\noindent To solve for  $\bV_r(\sigma^2)$ we
next use  $(A+\tfrac{n+p}{2} - 1)\bE_r(\sigma^2)
= B +  \frac{n}{2}\widehat{\sigma}_u^2 +  \frac{p}{2}\bE_r(\sigma^2)$
to obtain
$$
\begin{array}{rl}
\bV_r(\sigma^2)
& \ds =  
  \frac{(A+\tfrac{n+p}{2} - 1 + \frac{p}{2})[\bE_r(\sigma^2)]^2}{(A+\tfrac{n+p}{2} - 1)(A+\tfrac{n+p}{2} - 2)}   
+ \frac{\frac{p^2+2p}{4}   \bV_r(\sigma^2) }{(A+\tfrac{n+p}{2}  - 1)(A+\tfrac{n+p}{2}  - 2)}.
  
\end{array} 
$$

%\noindent Then
%$$
%[(A+\tfrac{n+p}{2}  - 1)(A+\tfrac{n+p}{2}  - 2) - \tfrac{p^2+2p}{4} 
%]
%\bV_q^{MP}(\sigma^2)
%=  
%(A+\tfrac{n+p}{2}  - 1 + \tfrac{p}{2}) \left[ \bE_q^{MP}(\sigma^2)  \right]^2 
%$$

%\noindent Next we note that
%$$
%(A+\tfrac{n+p}{2} - 1 + %\frac{p}{2})(A+\tfrac{n+p}{2} - 2 %- \frac{p}{2})
%= (A+\tfrac{n+p}{2} - %1)(A+\tfrac{n+p}{2} - 2)
%- \tfrac{p^2+2p}{4} 
%$$

%\noindent Hence,
%$$
%(A+\tfrac{n+p}{2} - 1 + %\frac{p}{2})(A+\tfrac{n+p}{2} - 2 %- \frac{p}{2})
%\bV_q^{MP}(\sigma^2)
%=  
%(A+\tfrac{n+p}{2}  - 1 + %\tfrac{p}{2}) \left[ %\bE_q^{MP}(\sigma^2)  \right]^2 
%$$

%\noindent 
%Dividing through by
%$(A+\tfrac{n+p}{2} - 1 + %\frac{p}{2})(A+\tfrac{n+p}{2} - 2 %- \frac{p}{2})$
%we get

\noindent Solving for $\bV_r(\sigma^2)$ we have
$$
\bV_r(\sigma^2) 
= \frac{[\bE_r(\sigma^2)]^2}{A+\tfrac{n}{2}  - 2} 
= \frac{\left( B +  \frac{n}{2}\widehat{\sigma}_u^2 \right)^2}{(A+\tfrac{n}{2}  - 1)^2(A+\tfrac{n}{2}  - 2)},
$$

\noindent which is the exact posterior variance for $\sigma^2$. Hence, the solution to the system of equations 
(\ref{eq:approach2_matching_mutilde})-(\ref{eq:approach2_matching_Btilde})
is given by
$$
\begin{array}{c}
\ds \widetilde{\vmu}^* = u\widehat{\vbeta},  
\qquad  
\widetilde{\mSigma}^* 
    = \left(\frac{B + \frac{n}{2}\widehat{\sigma}_u^2}{A + \tfrac{n}{2}}\right) u (\mX^T\mX)^{-1}, 
\qquad  
\widetilde{\nu}^* = 2A + n, 
\\ [2ex]
\widetilde{A}^* = A + \frac{n}{2}
\qquad \mbox{and} \qquad  
\widetilde{B}^* = B +  \frac{n}{2}\widehat{\sigma}_u^2,

\end{array} 
$$

\noindent which correspond to the parameters of the exact posterior distribution where the starred values denote the values of the corresponding parameters of $q(\vbeta)$ and $q(\sigma^2)$
at convergence.

\newpage 

\paragraph{Uniqueness.}
Let $a=A+n/2$,  
$b=B+n\widehat\sigma_u^2/2$, and
$c=A+(n+p)/2=a + p/2$.
We show that the admissible fixed point satisfying
$\widetilde A>2$ and $\widetilde B>0$ is unique.
At any fixed point of Algorithm~\ref{alg:mp_linear2}, Equation~\eqref{alg:mp_linear2_beta} determines
values for $\widetilde\vmu$, $\widetilde\mSigma$,
and $\widetilde\nu$.
Therefore the only remaining free parameters are
$\widetilde A$ and $\widetilde B$.
Using these expressions in \eqref{alg:mp_linear2_EB} gives
$$
\bE_q[\sB(\vbeta)]
=
b+
\frac{p\widetilde B}{2(\widetilde A-1)}.
$$

\noindent 
Indeed,
$$
\frac{\widetilde\nu}{\widetilde\nu-2}
\frac{\operatorname{tr}(\mX^T\mX\widetilde\mSigma)}{2u}
=
\frac{2\widetilde A}{2\widetilde A-2}
\frac{p u\widetilde B/\widetilde A}{2u}
=
\frac{p\widetilde B}{2(\widetilde A-1)}.
$$

\noindent 
The fixed-point condition for $\widetilde B$ from \eqref{eq:Er_sigma2} and \eqref{alg:mp_linear2_sigma2} is
$$
\widetilde B
=
(\widetilde A-1)
\frac{\bE_q[\sB(\beta)]}{c-1} 
 =
\frac{\widetilde A-1}{c-1}
\left[
b+\frac{p\widetilde B}{2(\widetilde A-1)}
\right].
$$

\noindent Rearranging,
$$
\widetilde B
\left[1-\frac{p}{2(c-1)}\right]
=
\frac{(\widetilde A-1)b}{c-1}.
$$

\noindent 
Since $c=a+p/2$,
$1-p/[2(c-1)] = (a-1)/(c-1)$.
Therefore, every admissible fixed point must satisfy
\begin{equation}\label{U1}
\widetilde B
=
b\,\frac{\widetilde A-1}{a-1}.
\end{equation}

\noindent 
It remains to identify $\widetilde A$. From \eqref{alg:mp_linear2_VB}, using
$\widetilde\nu=2\widetilde A$ and
$\widetilde\mSigma=(\widetilde B/\widetilde A)u(\mX^T \mX)^{-1}$, we obtain
$$
\bV_q[\sB(\vbeta)]
=
\widetilde B^2
\frac{
p\{2(\widetilde A-1)+p\}
}
{
4(\widetilde A-1)^2(\widetilde A-2)
}.
$$

\noindent 
Also, equations \eqref{eq:Er_sigma2}--\eqref{alg:mp_linear2_sigma2} imply
\begin{equation}\label{U2}
\bE_r(\sigma^2)
=
\frac{\widetilde B}{\widetilde A-1},
\qquad
\bV_r(\sigma^2)
=
\frac{[\bE_r(\sigma^2)]^2}{\widetilde A-2}
=
\frac{\widetilde B^2}
     {(\widetilde A-1)^2(\widetilde A-2)}.
\end{equation}

\noindent 
On the other hand, substituting \eqref{alg:mp_linear2_EB}  and \eqref{eq:Vr_sigma2} gives
$$
\bV_r(\sigma^2)
=
\frac{\widetilde B^2}{(\widetilde A-1)^2(c-2)}
+
\frac{\bV_q[\sB(\vbeta)]}{(c-1)(c-2)}.
$$

\noindent 
Combining this expression with \eqref{U2}, and canceling the positive factor
$\widetilde B^2/(\widetilde A-1)^2$, yields
$$
\frac{1}{\widetilde A-2}
=
\frac{1}{c-2}
+
\frac{
p[2(\widetilde A-1)+p]
}
{
4(\widetilde A-2)(c-1)(c-2)
}.
$$

\noindent 
Multiplying by $(\widetilde A-2)(c-2)$ gives the linear equation
$$
\widetilde A-c
+
\frac{p[2(\widetilde A-1)+p]}{4(c-1)}
=
0.
$$

\noindent 
Equivalently,
$[4(c-1)+2p]\widetilde A
= 4c(c-1)-p(p-2) = 4a(a+p-1)$.
Since $c=a+p/2$, the right-hand side simplifies to
and the coefficient of $\widetilde A$ is
$4(c-1)+2p = 4(a+p-1)$.
Thus,
$$
\widetilde A=a=A+\frac n2.
$$

\noindent 
Substituting this into \eqref{U1} gives
$$
\widetilde B=b=B+\frac n2\widehat\sigma_u^2.
$$

\noindent 
Therefore $\widetilde A$ and $\widetilde B$ are uniquely determined. Equation~\eqref{alg:mp_linear2_beta} then uniquely determines $\widetilde\vmu$, $\widetilde\mSigma$,
and $\widetilde\nu$.
Hence the admissible fixed point is unique.

\newpage 

\section{MP Derivations for MVN model}
\label{sec:mvn_derivations}

In this appendix, we provide detailed derivations for all of the material in Section \ref{sec:mvn_model}.
The MP updates below involve fourth moments of a multivariate $t$ distribution
and second moments of inverse-Wishart random matrices. We therefore assume
$\nu_n>p+3$,
so that the required moments exist at the exact posterior fixed point.

% intentionally left blank

\subsection{Exact posterior distributions for the MVN model}\label{Suppl:Exact_posterior_distributions_for_the_MVN_model}

 We next state the full conditional distributions for $\vmu$ and $\mSigma$, which are useful for deriving both the MFVB and MP approximations:
$$
\begin{array}{rl}
\vmu\given\mX,\mSigma
%& \ds \propto \exp\left[
%- \tfrac{1}{2} \sum_{i=1}^n (\vx_i - \vmu)^T\mSigma^{-1}(\vx_i - \vmu)
%- \tfrac{1}{2 } \lambda_0 \vmu^T\mSigma^{-1}\vmu 
%\right]
%\\ [3ex]
%& \ds \propto \exp\left[
%- \tfrac{\lambda_n}{2} \vmu^T\mSigma^{-1}\vmu
%+ \vmu^T\mSigma^{-1} n\overline{\vx} 
%\right]
%\\ [2ex]

\sim N_p\left( \vmu_n, \mSigma/\lambda_n \right)

%\ds \sim N\left( \frac{x_\bullet}{n + 1/g},\frac{\sigma^2}{n  + 1/g}\right)

\qquad \mbox{and} \qquad 

\mSigma\given\mX,\vmu
%& \ds \propto 
%\exp\left[ 
%-\tfrac{n+\nu+d+1}{2}\log\given\mSigma\given 
%-\tfrac{1}{2}\mbox{tr}\left\{  \mSigma^{-1}
%\left( \mPsi_0 + 
%\sum_{i=1}^n(\vx_i - \vmu)(\vx_i - \vmu)^T
%\right) 
%\right\}
%\right]
%\\ [2ex]
\sim \mbox{IW}_p\left( \mPsi(\vmu), \nu_n+1
\right) 

\end{array} 
$$

%$$
%\begin{array}{rl}
%\mSigma\given\mX,\vmu
%& \ds \propto 
% 
%\given\mSigma\given^{n/2}
%\exp\left[
%-\tfrac{1}{2}\mbox{tr}\left\{ \mSigma^{-1}
%\sum_{i=1}^n(\vx_i - \vmu)(\vx_i - \vmu)^T 
%\right\}
%\right]
%\\ [2ex]
%& \ds \qquad \times 
%\given\mSigma\given^{1/2}
%\exp\left[
%-\tfrac{1}{2}\mbox{tr}\left\{
% \mSigma^{-1} (\lambda_0\vmu\vmu^T) 
%\right\}
%\right]
%\\ [2ex]
%& \ds \qquad \times 
% \left\given\mSigma\right\given^{-(\nu_0+p+1)/2}\exp\left[ -\tfrac{1}{2}\mbox{tr}(\mPsi_0\mSigma^{-1}) \right]
%\\ [2ex]
%& \ds = \exp\left[ 
%-\tfrac{n+1+\nu_0+p+1}{2}\log\given\mSigma\given 
%-\tfrac{1}{2}\mbox{tr}\left\{  \mSigma^{-1}
%\left( \mPsi_0 + 
%\sum_{i=1}^n(\vx_i - \vmu)(\vx_i - \vmu)^T
%+ \lambda_0 \vmu\vmu^T
%\right) 
%\right\}
%\right]
%\\ [2ex]
%& \ds \sim \mbox{IW}_p\left( \mPsi(\vmu), \nu_n+1
%\right) 
%
%\end{array} 
%$$

\noindent where $\mPsi(\vmu)  = \mPsi_0 + \mS + n(\overline{\vx} - \vmu)(\overline{\vx} - \vmu)^T + \lambda_0 \vmu \vmu^T$,
\begin{equation}\label{eq:mvn_constants}
\begin{array}{c} 
\ds \overline{\vx} = \mX^T\vone_n/n,
\quad 
\lambda_n = \lambda_0 + n; \quad \nu_n  = \nu_0 + n,
\\ [2ex]
\vmu_n = n\overline{\vx}/\lambda_n,
\qquad   		
\mS = \mX^T\mX - n\overline{\vx} \ \overline{\vx}^T, 
\quad \mbox{and} \quad	  
\mPsi_n  
= \mPsi_0 
+ \mS 
+  n\lambda_0/\lambda_n \,\overline{\vx} \ \overline{\vx}^T.
\end{array}
\end{equation}	

\noindent For this example, we are able to calculate the exact posterior distribution and its corresponding moments. This can be used as a gold standard for assessing the quality of MFVB and MP algorithms.
The marginal posterior distribution of $\mSigma$ is 
$\ds \mSigma\given\mX \sim \mbox{IW}_p\left( \mPsi_n, \nu_n \right)$.

The marginal posterior distribution for $\vmu$ can be found via %by noting that if 
%$\vmu\given\mX,\mSigma \sim N_p\left( \vmu_n, \mSigma/\lambda_n \right)$ and
%$\mSigma\given\mX \sim \mbox{IW}_p\left( \mPsi_n, \nu_n \right)$ then
$$
\begin{array}{rl} 
\vmu\given\mX
%& \ds = \int p(\vmu\given\mX,\mSigma)p(\mSigma\given\mX) d\mSigma 
%\\
%& \ds = \int \exp\left[
%- \frac{n+\nu + d+ 2}{2}\log\given\mSigma\given
%- \frac{1}{2}\mbox{tr}\left(
%\mSigma^{-1}\left( \mPsi + \mS_g + (n+1/g)(\vmu - %\widehat{\vmu}_g)(\vmu - \widehat{\vmu}_g)^T \right)
%\right)
%\right] d\mSigma
%\\ 

& \ds \sim t_p\left( \vmu_n, \frac{\mPsi_n}{\lambda_n(\nu_n - p + 1)}, \nu_n - p + 1 \right),
\end{array} 
$$

\noindent where we have used the usual parametrization of the multivariate $t$ distribution
whose density and various expectations are summarised in section \ref{Suppl:Moments_for_the_Multivariate_t_Distribution} for reference.

\subsection{MFVB Derivations for the MVN model} 

For the MVN model in Section \ref{Suppl:Exact_posterior_distributions_for_the_MVN_model} with $q$-density factorization
$q(\vmu,\mSigma) = q(\vmu)q(\mSigma)$, the updates for $q(\vmu)$
and $q(\mSigma)$ are derived via
$$
\begin{array}{rl}
q(\vmu)
& \ds \propto \exp\left[
- \tfrac{1}{2} \sum_{i=1}^n (\vx_i - \vmu)^T\mSigma^{-1}(\vx_i - \vmu)
- \tfrac{1}{2}\lambda_0\vmu^T\mSigma^{-1}\vmu 
\right]
\\ [2ex]
& \ds \propto \exp\left[
- \tfrac{\lambda_n}{2} \vmu^T\bE_q(\mSigma^{-1})\vmu
+ \vmu^T\bE_q(\mSigma^{-1})n\overline{\vx}
\right]
\\ 

& \ds \propto \exp\left[
- \tfrac{\lambda_n\widetilde{d}}{2} \vmu^T\widetilde{\mPsi}^{-1}\vmu
+ \vmu^T\widetilde{\mPsi}^{-1} \widetilde{d} n\overline{\vx}
\right]
\\ 

& \ds 
\sim N_p( 
\widetilde{\vmu}, %\equiv \vmu_n, 
\widetilde{\mSigma} %\equiv \frac{\widetilde{\mPsi}}{\lambda_n\nu_n} 
)
\\ [2ex]
 
q(\mSigma )
& \ds \propto 
\exp\left[ 
-\tfrac{n+1+\nu_0+p+1}{2}\log\given\mSigma\given 
-\tfrac{1}{2}\mbox{tr}\left\{  \mSigma^{-1} \bE_q(\mPsi(\vmu))
\right\}
\right]
\\
& \ds \sim \mbox{IW}_p( 
\widetilde{\mPsi}, %\equiv \mPsi_* + n\widetilde{\mSigma}, 
\widetilde{d} %\equiv \nu_n
).

\end{array} 
$$

\noindent where the update for $q(\vmu)$ is given by
$\widetilde{\vmu} \leftarrow \vmu_n$
and $\widetilde{\mSigma} \leftarrow \widetilde{\mPsi}/(\lambda_n\widetilde{d})$,
and the update for $q(\mSigma)$ is given by
$\widetilde{\mPsi} \leftarrow \mPsi_n
 + \lambda_n \widetilde{\mSigma}$
and 
$\widetilde{d} \leftarrow \nu_n + 1$. Algorithm \ref{alg:vb_mvn} summarises the MFVB updates.

\begin{algorithm}[!ht]
\caption[]{MFVB method for the MVN model}
\label{alg:vb_mvn}
\begin{algorithmic}[1]
	\REQUIRE{$\mX\in \bR^{n\times p}$, $\mPsi_0 \in \bS_+^p$, $\lambda_0>0$, $\nu_0> p - 1$. }  

	\STATE Calculate algorithm constants via  (\ref{eq:mvn_constants}).
		
	\STATE Initialize: $\widetilde{d} \leftarrow \nu_n +1$ and  $\widetilde{\mPsi} \leftarrow \mPsi_n$.
 
	\REPEAT  
		\STATE Update $q(\vmu)$:
		\begin{equation}\label{alg:vb_mvn_update_mu}
		\ds \widetilde{\vmu} \leftarrow \vmu_n; \qquad 
		\widetilde{\mSigma} \leftarrow \frac{\widetilde{\mPsi}}{\lambda_n\widetilde{d}};
		\end{equation}

		\STATE Update $q(\mSigma)$: 
		\begin{equation}\label{alg:vb_mvn_update_Sigma}
		\widetilde{d} \leftarrow \nu_n +1; \qquad \widetilde{\mPsi}  \leftarrow \mPsi_n +  \lambda_n\widetilde{\mSigma};
		\end{equation}

	\UNTIL until convergence criteria are met.
		
	\end{algorithmic}
\end{algorithm}

\subsection{Derivations of MP update of 
\texorpdfstring{$q(\vmu)$}{q(mu)}
for the MVN model}\label{Suppl:Appendix_D5}

Equating $\bE_r(\vmu)$ with $\bE_q(\vmu)$, and $\bV_r(\vmu)$ with $\bV_q(\vmu)$  we have
\begin{equation}\label{eq:first_solve}
\widetilde{\vmu} = \vmu_n 
\qquad \mbox{and} \qquad 
\widetilde{\mSigma} 
= \frac{\widetilde{\nu}-2}{\widetilde{\nu}\lambda_n(\Delta - 1)} \widetilde{\mPsi}
\end{equation}
 
\noindent where $\Delta = \widetilde{d} - p$.
Using (\ref{eq:Rong1})  with $\vmu = \vzero$, $a=1$, $\mA = \mI$, $\mSigma = \widetilde{\mSigma}$ and $b = \widetilde{\nu}$ we obtain
\begin{equation}\label{eq:quadform_t}
\begin{array}{rl}
\bE_q\left[  \| \vmu - \widetilde{\vmu}\|^4   \right]
& \ds 
=  
\frac{\widetilde{\nu}^2[  2\,\mbox{tr}(\widetilde{\mSigma}^2)+\mbox{tr}(\widetilde{\mSigma})^2 ] }{(\widetilde{\nu}-2)(\widetilde{\nu}-4)}
\end{array} 
\end{equation}

\noindent and using (\ref{eq:first_solve}) we have 

\begin{equation}\label{eq:quadform_t_sub}
\begin{array}{rl}
\bE_q\left[  \| \vmu - \widetilde{\vmu}\|^4   \right]
& \ds 
=   \frac{(\widetilde{\nu}-2)[  2\,\mbox{tr}(\widetilde{\mPsi}^2)+\mbox{tr}(\widetilde{\mPsi})^2] }{\lambda_n^2(\Delta - 1)^2(\widetilde{\nu}-4)}.
\end{array} 
\end{equation}

\noindent Using Corollary 3.1 (i) and (iv) from \textcite{Rosen1988} summarized in Section
\ref{sec:moments_iw} we have
$$
\begin{array}{c} 
\bE\left[\mbox{tr}(\mSigma^2) \right] 
 \ds =  (c_1 + c_2)\mbox{tr}(\widetilde{\mPsi}^2) + c_2 \mbox{tr}(\widetilde{\mPsi})^2  \qquad \mbox{and} \qquad 

%\\ [1ex]

\bE\left[\mbox{tr}(\mSigma)^2\right] 
 \ds 
=  c_1 \mbox{tr}(\widetilde{\mPsi})^2
+  2c_2  \mbox{tr}(\widetilde{\mPsi}^2)
\end{array} 
$$

\noindent where $c_2^{-1} = \Delta(\Delta - 1)(\Delta - 3)$,  
$c_1 = (\Delta - 2) c_2$ and provided $\Delta>3$.
Hence, we obtain
\begin{equation}\label{eq:quadform_t_RHS}
\begin{array}{rl}
\ds \bE_r\left[  \|\vmu - \widetilde{\vmu}\|^4 \right]
    & \ds = \bE_q\left[  \bE\left\{   \|\vmu - \widetilde{\vmu}\|^4  \given \mX,\mSigma \right\} \right] 
\\  
    &  \ds = \lambda_n^{-2} \bE_q\left[ \mbox{tr}(\mSigma)^2 + 2\mbox{tr}(\mSigma^2) \right]
\\  
    & \ds =\lambda_n^{-2} \left[ 
c_1 \mbox{tr}(\widetilde{\mPsi})^2
+  2c_2  \mbox{tr}(\widetilde{\mPsi}^2) \right] 
+ \lambda_n^{-2}\left[  
(2c_1 + 2c_2)\mbox{tr}(\widetilde{\mPsi}^2) + 2c_2 \mbox{tr}(\widetilde{\mPsi})^2  
\right] 
\\  
    &  \ds = \lambda_n^{-2}\left(c_1   + 2c_2  \right)\left[ 2\mbox{tr}(\widetilde{\mPsi}^2) + \mbox{tr}(\widetilde{\mPsi})^2  
\right]
\\  
    &  \ds = \frac{2\mbox{tr}(\widetilde{\mPsi}^2) + \mbox{tr}(\widetilde{\mPsi})^2  }{\lambda_n^2(\Delta - 1)(\Delta - 3)}
\end{array} 
\end{equation}

\noindent noting that
$c_1 + 2c_2 = (\Delta - 2) c_2 + 2 c_2 = \Delta c_2 = 1/((\Delta - 1)(\Delta - 3))$.

 Equating (\ref{eq:quadform_t_sub}) with the final expression in (\ref{eq:quadform_t_RHS}), and substituting the expression for $\widetilde{\mSigma}$ in (\ref{eq:first_solve}), we obtain
$$
\frac{(\widetilde{\nu}-2) \left[  2\,\mbox{tr}(\widetilde{\mPsi}^2)+\mbox{tr}(\widetilde{\mPsi})^2 \right] }{\lambda_n^2(\Delta - 1)^2(\widetilde{\nu}-4)} = \frac{2\mbox{tr}(\widetilde{\mPsi}^2) + \mbox{tr}(\widetilde{\mPsi})^2  }{\lambda_n^2 (\Delta - 1)(\Delta - 3)}
$$

\noindent after  solving for $\widetilde{\nu}$ we find
$\widetilde{\nu} = \Delta + 1 = \widetilde{d} - p + 1$.

\subsection{MP posterior expectation of 
\texorpdfstring{$q(\mSigma)$}{q(Sigma)}} 

\noindent 
Since,
$\mSigma\given\mX,\vmu
\sim \mbox{IW}_p( \mPsi(\vmu), \nu_n+1)$ 
the corrected posterior expectation of $\mSigma$
is given by
$$
\begin{array}{rl}
\bE_r(\mSigma) 
    & \ds = \bE_q\left[
    \bE(\mSigma\given\mX,\vmu) 
    \right]  
\\  
    & \ds = \frac{1}{\nu_n + 1 - p - 1} \bE_q\left[ \mPsi(\vmu) \right]
\\ 
    & \ds = \frac{1}{\nu_n  - p} \left( \mPsi_n
    + \frac{\widetilde{\nu}}{\widetilde{\nu} - 2} \lambda_n\widetilde{\mSigma}
    \right)
\\ 
    & \ds = \frac{\mA}{\nu_n - p}  
\end{array} 
$$

\noindent where 
%$\nu^* = \nu_n + 1 - p$ and 
$
\ds \bE_q\left[  \mPsi(\vmu)  \right] = \mPsi_n  + \frac{ \lambda_n\widetilde{\nu}}{\widetilde{\nu} - 2}\widetilde{\mSigma} \equiv \mA. 
$

\subsection{MP posterior variance of elements of
\texorpdfstring{$\mSigma$}{Sigma}}

\noindent In the following derivations
the vector $\ve_i$ is a vector of zeros, except for the $i$th element which is equal to 1. The matrix $\mE_{ij}$ is a matrix whose elements are 0 except for the $(i,j)$th element which is equal to 1.

\bigskip
\noindent The MP variance of the diagonal elements of $\mSigma$ are given by 
$$
\begin{array}{rl}
\bV_r(\Sigma_{ii})
& \ds 
= \bE_q\left[ \bV(\Sigma_{ii}\given\mX,\vmu) \right] 
+ \bV_q\left[ \bE(\Sigma_{ii}\given\mX,\vmu) \right]
\\ [2ex]

& \ds 
= \bE_q\left[ \frac{2\{ \mPsi(\vmu) \}_{ii}^2 }{ (\nu_n - p)^2(\nu_n - p - 2)} \right] 
+ \bV_q\left[ \frac{ \{   \mPsi(\vmu) \}_{ii}}{\nu_n - p} 
\right]

\\ [2ex]

& \ds 
= \frac{2(a_{ii}^2 + b_{ii}) }{(\nu_n - p)^2(\nu_n - p - 2)}  
+ \frac{b_{ii}}{ (\nu_n - p)^2} 

\\ [2ex]

& \ds 
= \frac{2\,a_{ii}^2 + (\nu_n - p)b_{ii}}{(\nu_n - p)^2(\nu_n - p - 2)}

\end{array} 
$$

\noindent where  
$
\{ \mPsi(\vmu) \}_{ij}
= \psi_{0,ij} + s_{ij} + n(\overline{x}_i - \mu_i)(\overline{x}_j - \mu_j) + \lambda_0\mu_i\mu_j$,
$$
a_{ij} \equiv \bE_q\left[ \{ \mPsi(\vmu) \}_{ij} \right] 
\qquad \mbox{and} \qquad 
b_{ij} \equiv \bV_q\left[ \{ \mPsi(\vmu) \}_{ij} \right].
%c_{ij} \equiv \bC\left[ \{ \mPsi(\vmu) \}_{ii}, \{ \mPsi(\vmu) \}_{jj} \right].
$$

\noindent Hence, we can write
$$
\begin{array}{rl}
\mbox{dg}[\bE\bW\bV_r(\mSigma)]

& \ds = \frac{2\mbox{dg}(\mA)^{\odot 2} + (\nu_n - p)\mbox{dg}(\mB)}{(\nu_n - p)^2(\nu_n - p - 2)} . 
  
\end{array} 
$$

\noindent We have shown that $\bE_q[ \{ \mPsi(\vmu) \}_{ij}] = a_{ij}$, but we have yet to derive an expression for $b_{ij}$. % and $c_{ij}$.
%
%\bigskip 
%\noindent 
Expanding and completing the square   for $\mPsi(\vmu)$
we obtain
$$
\begin{array}{rl} 
\mPsi(\vmu) 
& \ds = \mPsi_0 + \mX^T\mX 
- n\, \overline{\vx}\overline{\vx}^T 
+  n(\overline{\vx} - \vmu)(\overline{\vx} - \vmu)^T 
+ \lambda_0 \vmu \vmu^T

\\ 

& \ds =  \mPsi_n + \lambda_n
\left( \vmu - \vmu_n\right)\left( \vmu - \vmu_n\right)^T.

\end{array} 
$$

\noindent 
Next, we  use (\ref{eq:Rong2}) to calculate
$$
\begin{array}{rl}
\bV_q\left[ \ds \{ \mPsi(\vmu) \}_{ij} \right]
& \ds = \lambda_n^2 \bV_q\left[ (\mu_i - \widetilde{\mu}_i)(\mu_j - \widetilde{\mu}_j) \right] 

\\ 

&  \ds = \lambda_n^2 \bV_q\left[ 
(\vmu - \widetilde{\vmu})\ve_i \ve_j^T(\vmu - \widetilde{\vmu})\right] 

\\ 

&  \ds = \lambda_n^2 \bV_q\left[ 
(\vmu - \widetilde{\vmu})\mE_{ij}(\vmu - \widetilde{\vmu})\right]. 
\end{array} 
$$

\noindent The diagonal elements of $\mB$ are given by
$$
\begin{array}{rl}
\bV_q\left[ \ds \{ \mPsi(\vmu) \}_{ii} \right]

&  \ds = \lambda_n^2 \bV_q\left[ 
(\vmu - \widetilde{\vmu})\mE_{ii}(\vmu - \widetilde{\vmu})\right] 

\\ 
&  \ds = \lambda_n^2 \Bigg[  \frac{2 \widetilde{\nu}^2 \, \mbox{tr}(\mE_{ii}\widetilde{\mSigma}\mE_{ii}\widetilde{\mSigma})}{(\widetilde{\nu} - 2)(\widetilde{\nu} - 4)}
+ \frac{2 \widetilde{\nu}^2\,\mbox{tr}(\mE_{ii}\widetilde{\mSigma})^2}{(\widetilde{\nu} - 2)^2(\widetilde{\nu} - 4)}
 \Bigg] 

\\ [2ex]
&  \ds = \lambda_n^2 \left[  \frac{2 \widetilde{\nu}^2 \, \widetilde{\Sigma}_{ii}^2}{(\widetilde{\nu} - 2)(\widetilde{\nu} - 4)}
+ \frac{2 \widetilde{\nu}^2\,\widetilde{\Sigma}_{ii}^2}{(\widetilde{\nu} - 2)^2(\widetilde{\nu} - 4)} 
  \right] 
  
\\ [1ex]
&  \ds =    \frac{2\lambda_n^2\widetilde{\nu}^2(\widetilde{\nu}  - 1)  \, \widetilde{\Sigma}_{ii}^2}{(\widetilde{\nu} - 2)^2(\widetilde{\nu} - 4)}

\\  
&    \ds \equiv b_{ii}.

\end{array} 
$$

\noindent Hence,
$$
\ds \vb \equiv \mbox{dg}\left[ \bE\bW\bV_q\left\{  \mPsi(\vmu)  \right\} \right]
= \frac{2 \lambda_n^2\widetilde{\nu}^2(\widetilde{\nu} - 1)}{(\widetilde{\nu} - 2)^2(\widetilde{\nu} - 4)}
\mbox{dg}(\widetilde{\mSigma})^{\odot 2}.
$$

\noindent Algorithm \ref{alg:mp_mvn} summarizes the MP algorithm for the multivariate normal model.

{\footnotesize 
\begin{algorithm}[!ht]
\caption[Algorithm 1]{Moment Propagation algorithm for the multivariate normal model}
\label{alg:mp_mvn}
	
\begin{algorithmic}[1]
		
\REQUIRE{$\mX\in \bR^{n\times p}$, $\mPsi_0 \in \bS_+^p$, $\lambda_0>0$, $\nu_0> p - 1$. }

\STATE Calculate algorithm constants via  (\ref{eq:mvn_constants}).

\STATE Initialize $q(\mSigma)$: $\widetilde{d} \leftarrow \nu_n$,  and $\widetilde{\mPsi} \leftarrow \mPsi_n$.
 
	\REPEAT  
	
	\STATE Update $q(\vmu)=t_p(\widetilde{\vmu},\widetilde{\mSigma},\widetilde{\nu})$   via moment matching using approximate posterior
	means and covariances of $\vmu$, and moments of $\|\vmu - \widetilde{\vmu}\|^4$
	assuming $q(\mSigma) = \mbox{IW}_p(\widetilde{\mPsi},\widetilde{d})$:
	\begin{equation}\label{alg:mp_mvn_mu_update}
	\ds \widetilde{\vmu}  \leftarrow \vmu_n; \qquad 
	\ds \widetilde{\mSigma} \leftarrow \frac{\widetilde{\mPsi}}{\lambda_n(\widetilde{d} - p + 1)}; 
	\qquad 
	\widetilde{\nu} \leftarrow \widetilde{d} - p + 1;
	\end{equation}
	
	\STATE Approximate posterior moments of $\mSigma$ assuming $q(\vmu)=t_p(\widetilde{\vmu},\widetilde{\mSigma},\widetilde{\nu})$:
\begin{equation}\label{eq:mvn_corr_moments1}
		\begin{array}{rl}
		\bE_r(\mSigma) & \ds \leftarrow  \frac{\mA}{\nu_n - p}  \qquad \mbox{and} 
		\\ [2ex]
		\mbox{dg}[\bE\bW\bV_r(\mSigma)]
		& \ds \leftarrow   
		
		\frac{2 \mbox{dg}(\mA)^{\odot 2} + (\nu_n - p  )\vb }{ (\nu_n - p)^2(\nu_n - p   -2)}  
 
		\end{array} 
		\end{equation} 
		
		\noindent  where
		\begin{equation}\label{eq:mvn_corr_moments2}
		\begin{array}{l}
		\ds  \mA  = \mPsi_n  + \frac{\lambda_n\widetilde{\nu}}{\widetilde{\nu} - 2}\widetilde{\mSigma}, 
		\qquad \mbox{and} \qquad 

        \vb
          =
		\frac{2 \lambda_n^2\widetilde{\nu}^2(\widetilde{\nu} - 1)}{(\widetilde{\nu} - 2)^2(\widetilde{\nu} - 4)}
\mbox{dg}(\widetilde{\mSigma})^{\odot 2}.
		\end{array} 
		\end{equation}

	\STATE Perform the MP update for $q(\mSigma)$ via moment matching: 
	\begin{equation}\label{alg:mp_mvn_Sigma_update}
	\ds \widetilde{d}  \leftarrow  \frac{2\vone^T[\mbox{dg}(\bE_r(\mSigma))^{\odot 2}]}{\vone_p^T\mbox{dg}[\bE\bW\bV_r(\mSigma)] } + p + 3; \qquad 
	\widetilde{\mPsi} \leftarrow \ds ( \widetilde{d} - p  - 1) \, \bE_r(\mSigma);
	\end{equation}

\UNTIL{convergence criteria are met}

\end{algorithmic}
\end{algorithm}
}

\newpage 
\subsection{Result \ref{result5}}\label{Suppl:Appendix_D8}

%% Note: Previous version had $\widetilde{\nu}=4$
%% as a solution. This can't be the case because of
%% of the moment conditions requires $\widetilde{\nu}>4$

\begin{result}\label{result5}
Assume $\nu_n>p+3$, so that the fourth moments required by Algorithm~\ref{alg:mp_mvn} exist.
Within the admissible region $\widetilde d>p+3$, the moment equations \eqref{alg:mp_mvn_mu_update}-\eqref{alg:mp_mvn_Sigma_update} have a unique fixed point,
$$
\widetilde\mPsi^\ast=\mPsi_n,\qquad
\widetilde d^\ast=\nu_n,
\qquad
\widetilde\vmu^\ast=\vmu_n,
\qquad
\widetilde\mSigma^\ast=
\frac{\mPsi_n}{\lambda_n(\nu_n-p+1)},
\qquad
\widetilde\nu^\ast=\nu_n-p+1.
$$

\noindent 
This fixed point corresponds to the exact marginal posterior distributions of
$\vmu$ and $\mSigma$. The algebraic fixed-point equation also has a boundary
root $\widetilde d=p+3$, equivalently 
$\widetilde\nu=4$, but this root is not
admissible because the fourth moment used in the update for $q(\vmu)$ is then
undefined.
\end{result}

\newpage

% \begin{result}\label{result5}
% There are two fixed point solutions of the moment equations (\ref{alg:mp_mvn_mu_update})-(\ref{alg:mp_mvn_Sigma_update}) for the MP method for
% the MVN model described by Algorithm \ref{alg:mp_mvn}. 
% \begin{itemize}
%     \item One solution corresponds to $\widetilde{\mPsi}^* = \mPsi_n$, 
% $\widetilde{d}^* = \nu_n$,
% $$
% \begin{array}{c}
% \widetilde{\vmu}^* = \vmu_n,
% \qquad 
% \ds \widetilde{\mSigma}^* = \frac{\mPsi_n}{\lambda_n(\nu_n - p + 1)}  
% \quad  \mbox{and} \quad 
% \widetilde{\nu}^* =  \nu_n - p + 1.
% \end{array} 
% $$

% \noindent 
% These correspond to the parameters of the exact posterior distribution.

% \item The second solution corresponds to $\widetilde{\vmu}  \leftarrow \vmu_n$, $\widetilde{d} = p + 3$,
% $$
% \ds\widetilde{\mPsi} = \frac{2\mPsi_n}{\nu_n - p - 1}, \quad 
% \widetilde{\mSigma} = \frac{\mPsi_n}{2\lambda_n(\nu_n - p - 1)}
% \quad \mbox{and} \quad 
% \widetilde{\nu} = 4.
% $$

% \noindent leading to an inexact estimation of the posterior distribution.
% \end{itemize}
% \end{result}

\noindent {\bf Proof:} Consider the set of consistency conditions as the set of equations (\ref{alg:mp_mvn_mu_update})-(\ref{alg:mp_mvn_Sigma_update}).
Matching $\bV_r(\vmu)$ with $\bV_q(\vmu)$ leads to the equation
$$
\begin{array}{rl}
\ds \frac{\widetilde{\mPsi}}{\lambda_n(\widetilde{d} - p - 1)} 
& \ds = \frac{\widetilde{\nu}\widetilde{\mSigma}  }{\widetilde{\nu} - 2}.
\end{array} 
$$

\noindent Equating $\bE_r(\mSigma)$  with $\bE_q(\mSigma)$ and the above equation we get
$$
\begin{array}{rl}
\ds \frac{ \widetilde{\mPsi}}{\widetilde{d} - p - 1} 
    & \ds = \frac{1}{\nu_n - p}  \left[ \mPsi_n  +  
    \frac{ \widetilde{\mPsi}}{ \widetilde{d} - p - 1}
    \right].
\end{array}
$$

\noindent Solving for $\widetilde{\mPsi}$
we obtain
$$ 
\widetilde{\mPsi} = \frac{\widetilde{d} - p - 1}{\nu_n - p - 1}    \mPsi_n. 
$$

\noindent Hence, we can write the expressions for $\mA$ and $\mbox{dg}(\mB)$ as
$$
\begin{array}{c}
\ds \mA  
=  \mPsi_n  +  
\frac{ \widetilde{\mPsi}}{ (\widetilde{d} - p - 1)}
= \frac{\nu_n - p }{\widetilde{d} - p - 1}
  \widetilde{\mPsi}
\end{array} 
$$

\noindent and
$$
\begin{array}{rl}
\mbox{dg}(\mB)
    & \ds =
\frac{\lambda_n^2\widetilde{\nu}^2(\widetilde{\nu} - 1)}{(\widetilde{\nu} - 2)^2(\widetilde{\nu} - 4)}
2\mbox{dg}(\widetilde{\mSigma})^2
\\ [2ex]

    & \ds = \frac{1}{ (\widetilde{d} - p - 1)^2}
\frac{  (\widetilde{\nu} - 1)}{(\widetilde{\nu} - 4)}
2\mbox{dg}(\widetilde{\mPsi})^2

\\ [2ex]

    & \ds =
\frac{  (\widetilde{\nu} - 1)}{(\nu_n - p)^2(\widetilde{\nu} - 4)}
2\mbox{dg}(\mA )^2.

\end{array}
$$

\noindent Hence, the moment matching equation for $\widetilde{d}$ can be written as
$$
\begin{array}{rl}
\ds \widetilde{d} 
& \ds =  \frac{\ds2\vone^T[\mbox{dg}(\bE_r(\mSigma))^2]}{\mbox{tr}[\bE\bW\bV_r(\mSigma)] } + p + 3 
\\ [2ex]
& \ds = \frac{\ds \frac{2}{(\nu_n - p)^2}  \vone^T[\mbox{dg}(\mA)^2] }{\ds 
\frac{2 \vone^T[\mbox{dg}(\mA)^2] + (\nu_n - p  )\mbox{tr}(\mB) }{ (\nu_n - p)^2(\nu_n - p   -2)}  
} + p + 3 

\\ [2ex]

& \ds = (\nu_n - p   -2)  \frac{\ds   2\vone^T[\mbox{dg}(\mA)^2] }{\ds 
2 \vone^T[\mbox{dg}(\mA)^2] + (\nu_n - p  )\mbox{tr}(\mB)  
} + p + 3 .
\end{array} 
$$

\noindent Substituting the above expressions for $\mA$ in terms of $\nu_n$, $\widetilde{d}$, p and $\widetilde{\mPsi}$, we have 
$$
\begin{array}{rl}
\widetilde{d}
& \ds =   \frac{\ds    (\nu_n - p   -2)}{\ds 
 1 +   	\frac{  (\widetilde{\nu} - 1)}{(\nu_n - p) (\widetilde{\nu} - 4)}
 
} + p + 3 

\\ [2ex]
& \ds = (\nu_n - p   -2)  \frac{\ds (\nu_n - p) (\widetilde{\nu} - 4)   }{\ds 
 (\nu_n - p) (\widetilde{\nu} - 4) +   	  (\widetilde{\nu} - 1)
} + p + 3

\\ [2ex]

& \ds = (\nu_n - p   -2) \left[ 1 -  \frac{\ds (\widetilde{\nu} - 1)  }{\ds 
 (\nu_n - p) (\widetilde{\nu} - 4) +   	  (\widetilde{\nu} - 1)
}  \right] + p + 3

\\ [4ex]

& \ds = \nu_n  + 1 -   \frac{\ds (\widetilde{\nu} - 1)(\nu_n - p   -2)  }{\ds 
 (\nu_n - p) (\widetilde{\nu} - 4) +   	  (\widetilde{\nu} - 1)
}

\\ [2ex]

& \ds = \nu_n  + 1 -   \frac{\ds (\widetilde{d} - p)(\nu_n - p   -2)  }{\ds 
 (\nu_n - p) (\widetilde{d} - p   - 3) +   	  (\widetilde{d} - p)
}   

\end{array} 
$$
%\wyc{How did you get from line 1 to line 2 in the previous block of equations?}
\noindent where the last line follows from using
$\widetilde{\nu} = \widetilde{d} - p + 1$. Letting $x= \widetilde{d} - p$
and $y=\nu_n - p$
leads to the equation
$$
\begin{array}{rl}
x  & \ds = y + 1 - \frac{\ds x(y - 2)}{y(x - 3) + x} \\ 
  & = y + 1 - \frac{\ds x(y + 1)  - 3x }{\ds x(y+1)    - 3y}.
%   & \ds = y + 1 - \frac{\ds xy -2x}{\ds xy    - 2y +   x - y}
\end{array}
$$

\noindent 
Multiplying by $x(y+1)    - 3y$, expanding, simplifying, and grouping by powers of $x$, and dividing through by $y+1$ we have
$$
x^2 - (y+3)x + 3y = 0
$$

\noindent The solutions are:
$$
x = \frac{ y + 3  \pm (y - 3)}{2} = y \, \mbox{ or } \,  3.
$$

% \noindent and so $\widetilde{d} = \nu_n$ or $\widetilde{d} = p + 3$.

The solution $x=y$ gives
$\widetilde d=\nu_n$,
and hence the exact posterior parameters. The second algebraic root $x=3$
corresponds to $\widetilde d=p+3$ and 
$\widetilde\nu=4$. This root lies on
the boundary of the admissible parameter space, since the fourth moment of
$t_p(\widetilde\vmu,\widetilde\mSigma,\widetilde\nu)$ exists only for
$\widetilde\nu>4$. Therefore, the exact-posterior fixed point is the unique
admissible fixed point.

% \bigskip 
% \noindent If $\widetilde{d} = \nu_n$ then $\widetilde{\mPsi} = \mPsi_n$
% and the MP approximation matches the exact solution.

% \bigskip 
% \noindent However, If $\widetilde{d} = p + 3$ then 
% $$
% \widetilde{\mPsi} = \frac{2\mPsi_n}{\nu_n - p - 1},
% $$

% \noindent and so $\widetilde{\vmu} = \vmu_n$,  
% $$
% \ds
% \widetilde{\mSigma} = \frac{\mPsi_n}{2\lambda_n(\nu_n - p - 1)}
% \quad \mbox{and} \quad 
% \widetilde{\nu} = 4,
% $$

% \noindent 
% leading to inexact $q$-densities.

\subsection{Example}
\label{Suppl:sec:Example_mvn_model}

Suppose we draw $n=4$ samples from $\vx_i \stackrel{\text{\scriptsize iid}}{\sim} N_2(\vmu,\mSigma)$
with 
$$
\vmu = \left[ \begin{array}{r}
-1 \\
 1
\end{array} \right] 
\qquad \mbox{and} \qquad 
\mSigma =
\left[ \begin{array}{cc}
 1    &  0.75 \\
 0.75 &    1
\end{array} \right] 
$$

\noindent with summary statistics
$$
\overline{\vx} = \left[ \begin{array}{r}
-0.9724726  \\
1.3202681
\end{array} \right] 
\qquad \mbox{and} \qquad 
\mS =
\left[ \begin{array}{rr}
0.8144316 & 0.5688416 \\
0.5688416 & 1.9682059 \\
\end{array} \right].
$$

\noindent Again, we have chosen the sample size to be small in order to see the biggest differences between MFVB and Exact methods.

\begin{figure}[!ht]
	\centering
	\includegraphics[width=0.9\linewidth]{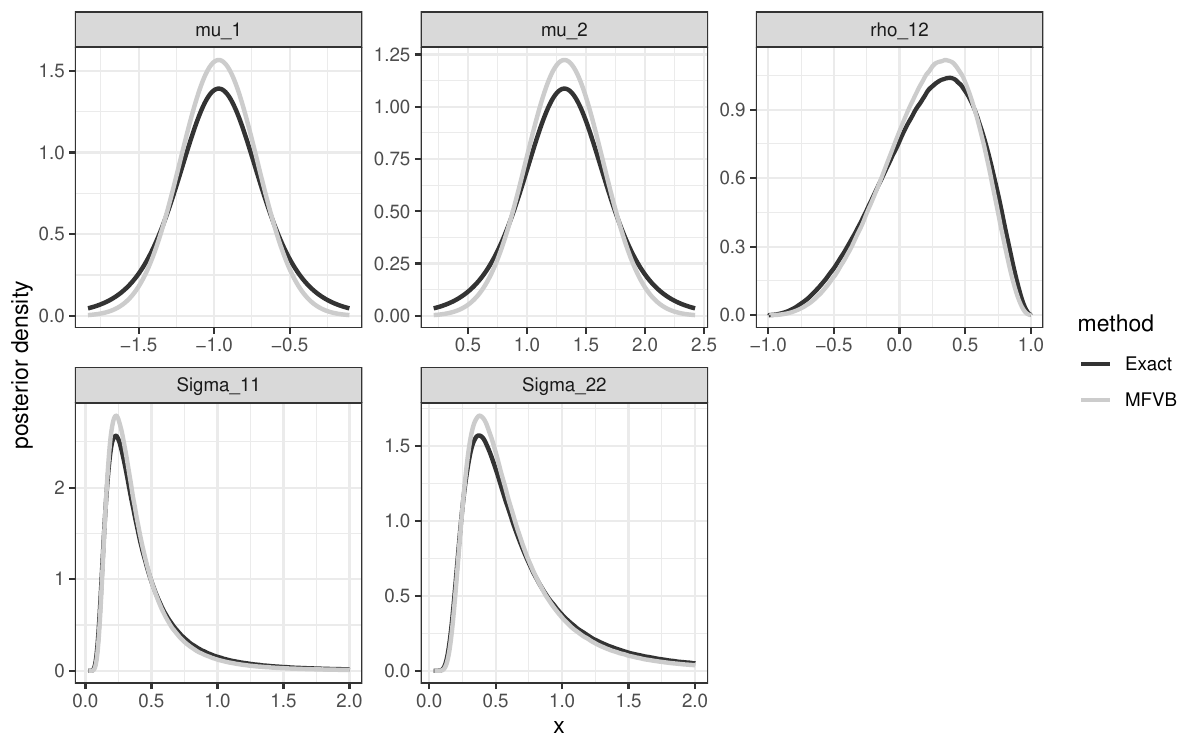}
	\caption{The fitted posteriors for the MVN model using described data. The posteriors densities corresponding to MP   are not plotted because they are exact. }
	\label{fig:mvn_example}
\end{figure}

The fitted values of the $q$-densities are summarised in Table \ref{tab:mvn_example}.
Figure \ref{fig:mvn_example} displays the posterior density estimates for MFVB against the exact posterior distributions. 
The posterior covariances for $\vmu$ are given by
$$
\bE_q^{VB}(\vmu) = \left[ \begin{array}{cc}
0.065 & 0.020 \\
0.020 & 0.106
\end{array} \right] 
\qquad \mbox{and} \qquad 
\bE(\vmu|\mX) = \left[ \begin{array}{cc}
0.114 & 0.035 \\ 
0.035 & 0.186
\end{array} \right],
$$

\noindent and the element-wise variances for $\mSigma$ are given by
$$
\bE_q^{VB}(\mSigma) = \left[ \begin{array}{cc}
0.116 & 0.085 \\
0.085 & 0.310
\end{array} \right] 
\qquad \mbox{and} \qquad 
\bE\bW\bV(\mSigma|\mX) = \left[ \begin{array}{cc}
0.208 & 0.148 \\
0.148 & 0.557
\end{array} \right],
$$

\noindent showing the exact posterior variances
for $\vmu$ are 1.75 times bigger than MFVB estimates,
and 
that the MFVB estimates, and the exact posterior element-wise variances for $\mSigma$ are 1.73-1.79 times bigger than
the corresponding MFVB estimates.

\begin{table}[H]
    \centering
    \begin{tabular}{l|c|c|c|c|c|c|c|c|c}
Method & $\widetilde{\nu}$ & $\widetilde{d}$ & $\widetilde{\Sigma}_{11}$ & $\widetilde{\Sigma}_{12}$ & $\widetilde{\Sigma}_{22}$ & $\widetilde{\Psi}_{11}$ & $\widetilde{\Psi}_{12}$ & $\widetilde{\Psi}_{22}$ & Iterations \\
\hline 
MFVB    &     &    8     &    0.065 & 0.0198 & 0.106 & 2.08 & 0.635 & 3.41 & 9 \\
MP      &    6  &     7  &    0.114 & 0.0347 & 0.186 & 1.82 & 0.556 & 2.99 & 22 \\
\hline
Exact   &    6  &     7  &    0.114 & 0.0347 & 0.186 & 1.82 & 0.556 & 2.99 & 
\end{tabular}
    \caption{Summary of fits for the MVN model using MFVB and MP methods  
for the described data. The values for $\widetilde{\vmu}$ are not shown because they are exact for all methods.}
    \label{tab:mvn_example}
\end{table}

\newpage 

\section{Derivations for MVN model with missing values}
\label{sec:mvn_missing_derivations}

We use the notation setup in Section 
\ref{sec:missing} which we spell out more explicitly here. Let $\mX$ be the $n\times p$ matrix whose $i$-th rows are equal to 
$\vx_i^T$ and $\mX_M$ denote the 
collection of missing values in the matrix $\mX$. 
For $i\in\sM$, write
$$
\vz_i=\vx_{i,M_i},
\qquad
\vx_i=\vc_i + \mR_i \vz_i,
$$

\noindent 
where $O_i$ and $M_i$ are index sets representing the elements of $\vx_i$ that are observed and missing respectively,
$\vc_i$ is the vector obtained by replacing the missing entries of $\vx_i$ by zero,
and $\mR_i$ is the matrix that inserts 
$\vz_i$ into the missing positions. 
For example, if $p=5$ and $M_i = \{ 2, 4, 5\}$ then 
$$
\vx_i = 
\left[ \begin{array}{c}
x_{i1} \\
x_{i2} \\
x_{i3} \\
x_{i4} \\
x_{i5} \\
\end{array} \right],
\quad 
\vc_i = 
\left[ \begin{array}{c}
x_{i1} \\
0 \\
x_{i3} \\
0 \\
0 \\
\end{array} \right],
\quad 
\mR_i = \left[ \begin{array}{ccc}
0 & 0 & 0 \\
1 & 0 & 0 \\
0 & 0 & 0 \\ 
0 & 1 & 0 \\
0 & 0 & 1 \\
\end{array} \right]
\quad \mbox{and} \quad 
\vz_i = \vx_{i,M_i} =
\left[ \begin{array}{c}
x_{i2} \\
x_{i4} \\
x_{i5} \\
\end{array} \right].
$$

\subsection{Fully factorized mean-field variational Bayes}

We first derive the MFVB updates corresponding to the factorization
$$
q(\vmu,\mSigma,\mX_M)
=
q(\vmu)q(\mSigma)\prod_{i\in\sM}q(\vx_{i,M_i}),
$$

\noindent 
where $\sM$ denotes the set of observations with at least one missing component.
We take
$$
q(\vmu) = N_p(\widetilde\vmu,\widetilde \mSigma),
\qquad
q(\mSigma)=\operatorname{IW}_p(\widetilde\mPsi,\widetilde d),
$$

\noindent 
and, for $i\in\sM$, $q(\vz_i)=q(\vx_{i,M_i})
=
N_{|M_i|}(\widetilde \vm_i,\widetilde \mV_i)$.

\paragraph{Update for $q(\vx_{i,M_i})$.}

For fixed $q(\vmu)$ and $q(\mSigma)$, the terms in the log joint density involving
$\vz_i=\vx_{i,M_i}$ are
$$
-\tfrac{1}{2}
(\vc_i + \mR_i \vz_i-\vmu)^T
\mSigma^{-1}
(\vc_i+\mR_i \vz_i-\vmu).
$$

\noindent 
Taking expectation with respect to 
$q(\vmu)q(\mSigma)$, and retaining only terms depending
on $\vz_i$, gives
$$
\log q^\star(\vz_i)
=
-\tfrac{1}{2}
\vz_i^T\mR_i^T \widetilde\mOmega \mR_i \vz_i
+
\vz_i^T \mR_i^T \widetilde\mOmega(\widetilde\vmu-\vc_i)
+
\operatorname{const.}
$$

\noindent where 
$\widetilde{\mOmega}=
\bE_q(\mSigma^{-1})
=
\widetilde d\,\widetilde\mPsi^{-1}$.
Therefore,
$$
q^\star(\vz_i)
=
N_{|M_i|}(\widetilde \vm_i,\widetilde \mV_i),
\quad 
\mbox{where}
\quad 
\widetilde \mV_i
=
(\mR_i^T \widetilde\mOmega \mR_i)^{-1},
\quad \mbox{and} \quad 
\widetilde \vm_i
=
\widetilde \mV_i \mR_i^T \widetilde\mOmega(\widetilde\vmu-\vc_i).
$$

\noindent 
Note that this can be equivalently written as follows. If $O_i\neq\varnothing$, this can be written in block form as
$\widetilde \mV_i
= \widetilde\mOmega_{M_iM_i}^{-1}$,
and
$\widetilde \vm_i
=
\widetilde\vmu_{M_i}
-
\widetilde\mOmega_{M_iM_i}^{-1}
\widetilde\mOmega_{M_iO_i}
(\vx_{i,O_i}-\widetilde\vmu_{O_i})$.
If $O_i=\varnothing$, then the update reduces to
$\widetilde \vm_i=\widetilde\vmu$, and 
$\widetilde \mV_i=\widetilde\mOmega^{-1}$.

It is useful to define the first two moments of the completed observation
$$
\widehat \vx_i
=
\bE_q(\vx_i)
=
\vc_i+\mR_i\widetilde \vm_i,
\quad \mbox{and}
\quad 
\mB_i
=
\bV_q(\vx_i)
=
\mR_i\widetilde \mV_i\mR_i^T.
$$

\noindent 
For observations
with no missing components, we set 
$\widehat \vx_i=\vx_i$ and $\mB_i=\vzero$.

\paragraph{Update for $q(\vmu)$.}

For fixed $q(\mSigma)$ and 
$\{ q(\vx_{i,M_i}):i\in\sM\}$, the terms in the log joint
density involving $\vmu$ are
$$
-\frac{1}{2}
\sum_{i=1}^n
(\vx_i-\vmu)^T\mSigma^{-1}(\vx_i-\vmu)
-
\frac{\lambda_0}{2}\vmu^T\mSigma^{-1}\vmu.
$$

\noindent 
Taking expectation with respect to 
$q(\mSigma)\prod_{i\in\sM}q(\vx_{i,M_i})$, and
retaining only terms depending on $\vmu$, gives
$$
\log q^\star(\vmu)
=
-\frac{1}{2}
\lambda_n
\vmu^T\widetilde\mOmega\vmu
+
\vmu^T\widetilde\mOmega
\sum_{i=1}^n \widehat \vx_i
+
\operatorname{const},
$$

\noindent 
where $\lambda_n=\lambda_0+n$.
Thus,
$q^\star(\vmu)
=
N_p(\widetilde\vmu,\widetilde \mSigma),
$
with
$$
\widetilde\vmu
=
\frac{1}{\lambda_n}
\sum_{i=1}^n \widehat \vx_i
\equiv
\widehat\vmu_n,
\qquad \mbox{and} \qquad 
\widetilde \mSigma
=
(\lambda_n\widetilde\mOmega)^{-1}.$$

\paragraph{Update for $q(\mSigma)$.}

For fixed $q(\vmu)$ and 
$\{ q(\vx_{i,M_i}) : i\in\sM\}$, the terms in the log joint
density involving $\mSigma$ are
$$
-\frac{\nu_0+p+n+2}{2}\log|\mSigma|
-\frac12\operatorname{tr}(\mPsi_0\mSigma^{-1})
%-\frac{n}{2}\log|\mSigma|
-\frac{1}{2}
\sum_{i=1}^n
(\vx_i-\vmu)^T\mSigma^{-1}(\vx_i-\vmu)
%-\frac{1}{2}\log|\mSigma|
-\frac{\lambda_0}{2}\vmu^T\mSigma^{-1}\vmu.
$$

\noindent 
%The additional $-\tfrac{1}{2}\log|\mSigma|$ term arises from the conditional prior
%$\vmu|\mSigma$. 
Hence,
$q^\star(\mSigma)
=
\operatorname{IW}_p(\widetilde\mPsi,\widetilde d)$, where
$$
\widetilde d
=
\nu_0+n+1
\quad \mbox{and} \quad  
\widetilde\mPsi
=
\mPsi_0
+
\sum_{i=1}^n
\bE_q\left[ (\vx_i-\vmu)(\vx_i-\vmu)^T \right]
+
\lambda_0 \bE_q(\vmu\vmu^T).
$$

\noindent 
Since $q(\vmu)$ is independent of 
$q(\vx_{i,M_i})$, we have
$$
\bE_q\left[(\vx_i-\vmu)(\vx_i-\vmu)^T\right]
=
\mB_i+\widetilde \mSigma
+
(\widehat \vx_i-\widetilde\vmu)(\widehat \vx_i-\widetilde\vmu)^T,
\quad \mbox{and} \quad 
\bE_q(\vmu\vmu^T)
=
\widetilde \mSigma+\widetilde\vmu\widetilde\vmu^T.
$$

\noindent 
Therefore,
$$
\widetilde\mPsi
=
\mPsi_0
+
\sum_{i=1}^n
\left[
\mB_i+
(\widehat \vx_i-\widetilde\vmu)(\widehat \vx_i-\widetilde\vmu)^T
\right]
+
\lambda_0\widetilde\vmu\widetilde\vmu^T
+
\lambda_n\widetilde \mSigma.
$$

\noindent 
At the update for $q(\vmu)$, 
we have $\widetilde\vmu=\widehat\vmu_n$, where
$\widehat\vmu_n
= \lambda_n^{-1}
\sum_{i=1}^n\widehat \vx_i$.
Thus the scale update can equivalently be written as
$$
\widetilde\mPsi
=
\mPsi_0
+
\sum_{i=1}^n
\widehat \vx_i\widehat \vx_i^T
+
\sum_{i=1}^n \mB_i
-
\lambda_n\widehat\vmu_n\widehat\vmu_n^T
+
\lambda_n\widetilde \mSigma.
$$

This expression highlights the difference between the fully factorized MFVB update and the
blocked update for $q(\vmu,\mSigma)$: the factorization $q(\vmu)q(\mSigma)$ contributes the extra
term $\lambda_n\widetilde\mSigma$ to the inverse-Wishart scale matrix and changes the degrees
of freedom from $\nu_0+n$ to 
$\nu_0+n+1$.

\newpage 

\subsection{Blocked mean-field variational Bayes}

We now derive the MFVB updates corresponding to the partially factorized approximation
$$
q(\vmu,\mSigma,\mX_M)
=
q(\vmu,\mSigma)\prod_{i\in\sM}q(\vx_{i,M_i}).
$$

\noindent  We take
$$
q(\vmu,\mSigma)
=
\operatorname{NIW}_p(\widetilde\vmu,\widetilde\lambda,\widetilde\mPsi,\widetilde\nu),
$$

\noindent  meaning that
$q(\vmu|\mSigma)
=
N_p(\widetilde\vmu,\widetilde\lambda^{-1}\Sigma)$, and 
$q(\mSigma)
=
\operatorname{IW}_p(\widetilde\mPsi,\widetilde\nu)$,
and, for $i\in\sM$,
$q(\vz_i)=q(\vx_{i,M_i})
=
N_{|M_i|}(\widetilde \vm_i,\widetilde \mV_i)$. Define
$
\widetilde\mOmega
=
\bE_q(\mSigma^{-1})
=
\widetilde\nu\,\widetilde\mPsi^{-1}$.

\paragraph{Update for $q(\vx_{i,M_i})$.}

For fixed $q(\vmu,\mSigma)$, the terms in the log joint density involving
$\vz_i=\vx_{i,M_i}$ are
$$
-\frac{1}{2}
(\vc_i+\mR_i \vz_i-\vmu)^T
\mSigma^{-1}
(\vc_i+\mR_i \vz_i-\vmu).
$$

\noindent 
Taking expectation with respect to 
$q(\vmu,\mSigma)$, and retaining only terms depending
on $\vz_i$, gives
$$
\log q^\star(\vz_i)
=
-\frac{1}{2}
\vz_i^T \mR_i^T \widetilde\mOmega \mR_i \vz_i
+
\vz_i^T \mR_i^T \widetilde\mOmega(\widetilde\vmu-\vc_i)
+
\operatorname{const}.
$$

\noindent 
Therefore,
$q^\star(\vz_i)
=
N_{|M_i|}(\widetilde \vm_i,\widetilde \mV_i)$,
where
$$
\widetilde \mV_i
=
(\mR_i^T \widetilde\mOmega \mR_i)^{-1},
\qquad\mbox{and} \qquad 
\widetilde \vm_i
=
\widetilde \mV_i \mR_i^T\widetilde\mOmega(\widetilde\vmu-\vc_i).
$$

\noindent 
Equivalently, if $O_i\neq\varnothing$, this update can be written in block form as
$$
\widetilde \mV_i
=
\widetilde\mOmega_{M_iM_i}^{-1},
\qquad \mbox{and \qquad }
\widetilde \vm_i
=
\widetilde\vmu_{M_i}
-
\widetilde\mOmega_{M_iM_i}^{-1}
\widetilde\mOmega_{M_iO_i}
(\vx_{i,O_i}-\widetilde\vmu_{O_i}).
$$

\noindent 
If $O_i=\varnothing$, this reduces to
$\widetilde \vm_i=\widetilde\vmu$, and 
$\widetilde \mV_i=\widetilde\mOmega^{-1}$.
After updating \(q(z_i)\), define
$$
\widehat \vx_i
=
\bE_q(x_i)
=
\vc_i+\mR_i\widetilde \vm_i,
\qquad \mbox{and} \qquad 
\mB_i
=
\bV_q(x_i)
=
\mR_i\widetilde \mV_i\mR_i^T.
$$

\noindent 
As previously, for observations
with no missing components, we set
$\widehat \vx_i=\vx_i$, and 
$\mB_i=\vzero$.

\paragraph{Update for $q(\vmu,\mSigma)$.}

For fixed $\{q(\vx_{i,M_i}) : i\in\sM\}$, the terms in the log joint density involving
$(\vmu,\mSigma)$ are
$\log p(\mSigma)
+
\log p(\vmu|\mSigma)
+
\sum_{i=1}^n \bE_q[\log p(\vx_i|\vmu,\mSigma)]$.
Ignoring constants not depending on 
$(\vmu,\mSigma)$, this is
$$
-\frac{\nu_0+n+p+2}{2}\log|\mSigma|
-\frac{1}{2}\operatorname{tr}(\mPsi_0\mSigma^{-1})
%-\frac12\log|\Sigma|
-\frac{\lambda_0}{2}\vmu^T\mSigma^{-1}\vmu
%-\frac n2\log|\Sigma|
-\frac{1}{2}
\sum_{i=1}^n
\bE_q\left[ (\vx_i-\vmu)^T\mSigma^{-1}(\vx_i-\vmu)\right].
$$

\noindent 
Since
$\bE_q(\vx_i)=\widehat \vx_i$, and 
$\bV_q(\vx_i)=\mB_i$, we have
$$
\bE_q\left[ (\vx_i-\vmu)^T\mSigma^{-1}(\vx_i-\vmu)\right]
=
\operatorname{tr}\left[
\mSigma^{-1}
\left\{
\mB_i+(\widehat \vx_i-\vmu)(\widehat \vx_i-\vmu)^T
\right\}
\right].
$$

\noindent 
Therefore, the coordinate update for 
$q(\vmu,\mSigma)$ is proportional to
$$
|\mSigma|^{-(\nu_0+n+p+2)/2}
\exp\left[
-\frac{1}{2}
\operatorname{tr}
\left\{
\mSigma^{-1}
\left(
\mPsi_0
+
\sum_{i=1}^n \mB_i
+
\sum_{i=1}^n
(\widehat \vx_i-\vmu)(\widehat \vx_i-\vmu)^T
+
\lambda_0\vmu\vmu^T
\right)
\right\}
\right].
$$

\noindent 
Let
$\lambda_n=\lambda_0+n$,
$\overline{\widehat \vx}
=
\sum_{i=1}^n\widehat \vx_i/n$,
$\widehat\mu_n
=
n\overline{\widehat \vx}/\lambda_n
=
\lambda_n^{-1}\sum_{i=1}^n\widehat \vx_i$.
Completing the square in $\vmu$ gives
$$
\sum_{i=1}^n
(\widehat \vx_i-\vmu)(\widehat \vx_i-\vmu)^\top
+
\lambda_0\vmu\vmu^T
=
\lambda_n(\vmu-\widehat\vmu_n)(\vmu-\widehat\vmu_n)^T
+
\sum_{i=1}^n \widehat \vx_i\widehat \vx_i^T
-
\lambda_n\widehat\vmu_n\widehat\vmu_n^T.
$$

\noindent 
Hence,
$q^\star(\vmu,\mSigma)
=
\operatorname{NIW}_p
(
\widetilde\vmu,
\widetilde\lambda,
\widetilde\mPsi,
\widetilde\nu
)$, where
$$
\widetilde\vmu
=
\widehat\vmu_n,
\qquad
\widetilde\lambda
=
\lambda_n,
\qquad
\widetilde\nu
=
\nu_0+n,
\quad \mbox{and} \quad 
\widetilde\mPsi
=
\mPsi_0
+
\sum_{i=1}^n \widehat \vx_i\widehat \vx_i^T
+
\sum_{i=1}^n \mB_i
-
\lambda_n\widehat\vmu_n\widehat\vmu_n^T.
$$

\noindent 
Equivalently, 
$\widetilde\mPsi
=
\mPsi_0
+
\sum_{i=1}^n
\left[
\mB_i+
(\widehat \vx_i-\widehat\vmu_n)(\widehat \vx_i-\widehat\vmu_n)^T
\right]
+
\lambda_0\widehat\vmu_n\widehat\vmu_n^T$.

The blocked approximation differs from the fully factorized MFVB approximation in two ways.
First, the variational distribution preserves the conditional dependence
$$q(\vmu|\mSigma)
=
N_p(\widetilde\vmu,\widetilde\lambda^{-1}\mSigma).$$

Second, the update for $q(\mSigma)$ has degrees of freedom $\nu_0+n$, rather than
$\nu_0+n+1$, because $\vmu$ is updated jointly with $\mSigma$ rather than factorized from it.

\subsection{Mean-field moment propagation}

We now derive the MFMP updates for the factorization
$$
q(\vmu,\mSigma,\mX_M)
=
q(\vmu,\mSigma)\prod_{i\in\sM}q(\vz_i),
$$

\noindent 
where $\vz_i=\vx_{i,M_i}$ denotes the missing components of observation $i$. We use
$$q(\vmu,\mSigma)
=
\operatorname{NIW}_p(\widetilde\vmu,\widetilde\lambda,\widetilde\mPsi,\widetilde\nu),
\qquad \mbox{and} \qquad 
q(\vz_i)
=
N_{|M_i|}(\widetilde \vm_i,\widetilde \mV_i), \qquad i\in\mathcal M .
$$

The update for $q(\vz_i)$ is obtained by propagating the conditional moments of
$\vx_{i,M_i}$ over the current approximation $q(\vmu,\mSigma)$. The update for
$q(\vmu,\mSigma)$ is then obtained by propagating the conditional posterior moments of
$(\vmu,\mSigma)$ over the current approximation to the missing values.

\paragraph{Update for $q(\vz_i)$.}

First suppose $O_i=\varnothing$. 
Then $\vz_i=\vx_i$, and conditional on
$(\vmu,\mSigma)$, we have 
$\vx_i|\vmu,\mSigma \sim N_p(\vmu,\mSigma)$.
Therefore, $\bE_r(\vz_i)
=
\bE_q(\vmu)
=
\widetilde\vmu$.
Also, by the law of total variance,
$\bV_r(\vz_i)
=
\bE_q(\mSigma)+\bV_q(\vmu)$.
Under the normal-inverse-Wishart approximation,
$\bV_q(\vmu)
=
\widetilde\lambda^{-1}\bE_q(\mSigma)$,
and hence
$$
\bV_r(\vz_i)
=
(1+\widetilde\lambda^{-1})\bE_q(\mSigma)
=
(1+\widetilde\lambda^{-1})
\frac{\widetilde\mPsi}{\widetilde\nu-p-1}.
$$

\noindent 
Thus, if $O_i=\varnothing$, the moment-matching update is
$$
\widetilde \vm_i
\leftarrow
\widetilde\vmu,
\qquad \mbox{and} \qquad 
\widetilde \mV_i
\leftarrow
(1+\widetilde\lambda^{-1})
\frac{\widetilde\mPsi}{\widetilde\nu-p-1}.
$$

Now suppose \(O_i\neq\varnothing\). Conditional on $(\vmu,\mSigma)$, the missing components
have Gaussian conditional distribution
$$
\vx_{i,M_i}| \vx_{i,O_i},\vmu,\mSigma
\sim
N_{|M_i|}
\left(
\vmu_{M_i}
+
\mSigma_{M_iO_i}\mSigma_{O_iO_i}^{-1}
(\vx_{i,O_i}-\vmu_{O_i}),
\,
\mSigma_{M_i|O_i}
\right),
$$

\noindent 
where
$\mSigma_{M_i|O_i}
=
\mSigma_{M_iM_i}
-
\mSigma_{M_iO_i}\mSigma_{O_iO_i}^{-1}\mSigma_{O_iM_i}$.
Define
$\mL_i
=
\bE_q(\mSigma_{M_iO_i}\mSigma_{O_iO_i}^{-1})
=
\widetilde\mPsi_{M_iO_i}\widetilde\mPsi_{O_iO_i}^{-1}$ (see Section \ref{sec:IW_results}),
and
$\widetilde\mPsi_{M_i|O_i}
=
\widetilde\mPsi_{M_iM_i}
-
\widetilde\mPsi_{M_iO_i}
\widetilde\mPsi_{O_iO_i}^{-1}
\widetilde\mPsi_{O_iM_i}$.
Then
$\bE_r(\vz_i)
=
\widetilde\vmu_{M_i}
+
\mL_i(\vx_{i,O_i}-\widetilde\vmu_{O_i})$.

The propagated variance is obtained using the law of total variance. We have
$$
\bV_r(\vz_i)
=
\bE_q(\mSigma_{M_i|O_i})
+
\bV_q\left[
\vmu_{M_i}
+
\mSigma_{M_iO_i}\mSigma_{O_iO_i}^{-1}
(\vx_{i,O_i}-\vmu_{O_i})
\right].
$$

\noindent 
Using the block inverse-Wishart identities in Section~\ref{sec:IW_results}, this gives
$$
\bV_r(\vz_i)
=
\left[
1+\widetilde\lambda^{-1}
+
(\vx_{i,O_i}-\widetilde\vmu_{O_i})^T
\widetilde\mPsi_{O_iO_i}^{-1}
(\vx_{i,O_i}-\widetilde\vmu_{O_i})
\right]
\frac{\widetilde\mPsi_{M_i|O_i}}
     {\widetilde\nu-|M_i|-1}.
$$

\noindent 
Hence, for $O_i\neq\varnothing$,
$$
\widetilde \vm_i
\leftarrow
\widetilde\vmu_{M_i}
+
\mL_i(\vx_{i,O_i}-\widetilde\vmu_{O_i}),
$$

\noindent and
$$
\widetilde \mV_i
\leftarrow
\left[
1+\widetilde\lambda^{-1}
+
(\vx_{i,O_i}-\widetilde\vmu_{O_i})^T
\widetilde\mPsi_{O_iO_i}^{-1}
(\vx_{i,O_i}-\widetilde\vmu_{O_i})
\right]
\frac{\widetilde\mPsi_{M_i|O_i}}
     {\widetilde\nu-|M_i|-1}.
$$

\noindent 
After updating $q(\vz_i)$, define
$\widehat \vx_i
=
\bE_q(\vx_i)
=
\vc_i+\mR_i\widetilde \vm_i$ 
and
$\mB_i
=
\bV_q(\vx_i)
=
\mR_i\widetilde \mV_i\mR_i^T$.
Again, for observations with no missing components, set $\widehat \vx_i=\vx_i$, and $\mB_i=\vzero$.

\paragraph{Update for $q(\vmu,\mSigma)$.}

For fixed $\{q(\vz_i):i\in\sM\}$, the conditional posterior of
$(\vmu,\mSigma)$ given the completed data matrix is normal-inverse-Wishart. Let
$\lambda_n=\lambda_0+n$,
$\nu_n=\nu_0+n$,
and define
$\widehat\vmu_n
=
\lambda_n^{-1}\sum_{i=1}^n \widehat \vx_i$,
and 
$\mA
=
\sum_{i=1}^n \mB_i$.
The propagated mean of $\vmu$ is
$\bE_r(\vmu)
=
\widehat\vmu_n$.
The propagated variance of $\vmu$ is
$$
\bV_r(\vmu)
=
\frac{\bE_r(\mSigma)}{\lambda_n}
+
\frac{\mA}{\lambda_n^2}.
$$

\noindent 
The second term is the contribution from uncertainty in the missing values.

The following moment calculations assume that the required inverse-Wishart moments exist; in particular, we require 
$\widetilde\nu>p+1$ for $\bE_q(\mSigma)$ and $\nu_n>p+3$ for the propagated element-wise variances of $\mSigma$.
The propagated mean of $\mSigma$ is
$$
\bE_r(\mSigma)
=
\frac{\mC}{\nu_n-p-1}
$$

\noindent where
$\mC
=
\mPsi_0
+
\sum_{i=1}^n \widehat \vx_i\widehat \vx_i^T
+
(1-\lambda_n^{-1})\mA
-
\lambda_n\widehat\vmu_n\widehat\vmu_n^T$.
To update the inverse-Wishart degrees of freedom, we also need the diagonal element-wise
variances of $\mSigma$. Let
$$
d_j
=
\bV_q\left(\{\mPsi_n(X)\}_{jj}\right),
\qquad j=1,\ldots,p,
$$

\noindent 
where
$$
\mPsi_n(\mX)
=
\mPsi_0
+
\sum_{i=1}^n \vx_i\vx_i^T
-
\lambda_n\vmu_n(\mX)\vmu_n(\mX)^T,
\qquad \mbox{and} \qquad 
\vmu_n(\mX)
=
\lambda_n^{-1}\sum_{i=1}^n \vx_i.
$$

\noindent 
Then
$$
\bV_r(\Sigma_{jj})
=
\frac{
2C_{jj}^2
+
(\nu_n-p-1)d_j
}
{
(\nu_n-p-1)^2(\nu_n-p-3)
}.
$$

\noindent 
Equivalently, with $\vd=(d_1,\ldots,d_p)$,
$$
\operatorname{dg}[\bE\bW\bV_r(\mSigma)]
=
\frac{
2\operatorname{dg}(\mC)^{\odot 2}
+
(\nu_n-p-1)\vd
}
{
(\nu_n-p-1)^2(\nu_n-p-3)
}.
$$

It remains to compute each of the $d_j$. Under the current approximation to the missing data,
$\vx_i$ has mean $\widehat \vx_i$ and covariance $\mB_i$, independently across $i$. For a fixed
coordinate $j$, define
$\mX_j=(x_{1j},\ldots,x_{nj})$,
$\va_j=(\widehat x_{1j},\ldots,\widehat x_{nj})$,
and
$\vb_j=(\{\mB_1\}_{jj},\ldots,\{\mB_n\}_{jj})$.
Then
$\mX_j\sim N_n[\va_j,\operatorname{diag}(\vb_j)].$ Let
$\mQ
=
\mI_n-\lambda_n^{-1}\mathbf 1_n\mathbf 1_n^T$.
Ignoring the fixed prior term 
$\{\mPsi_0\}_{jj}$, the $j$-th diagonal element of
$\mPsi_n(\mX)$ is $\mX_j^T \mQ\mX_j$. Therefore,
$d_j
=
\bV_q(\mX_j^T \mQ\mX_j)$.
Using the Gaussian quadratic-form variance identity (see Section \ref{sec:mvn_moments}),
$$
d_j
=
2\operatorname{tr}[\mQ\operatorname{diag}(\vb_j)\mQ\operatorname{diag}(\vb_j)]
+
4\va_j^T \mQ\operatorname{diag}(\vb_j)\mQ\va_j.
$$

\noindent Equivalently,
$$
d_j
=
2\left(1-\frac{2}{\lambda_n}\right)\mathbf 1_n^T \vb_j^{\odot 2}
+
\frac{2}{\lambda_n^2}(\mathbf 1_n^T \vb_j)^2
+
4\vb_j^T
\left(
\va_j-\frac{\mathbf 1_n^T \va_j}{\lambda_n}\mathbf 1_n
\right)^{\odot 2}.
$$

\noindent Finally, we match the corresponding moments of the normal-inverse-Wishart approximation
using
$$
\bE_q(\vmu)=\widetilde\vmu,
\qquad\mbox{and} \qquad 
\bV_q(\vmu)=\frac{\bE_q(\mSigma)}{\widetilde\lambda},
$$

Matching $\bE_r(\vmu)=\bE_q(\vmu)$,
$\mbox{tr}[\bV_r(\vmu)]=\mbox{tr}[\bV_q(\vmu)]$,
$\bE_r(\mSigma) = \bE_q(\mSigma)$,
and $\vone_p^T\mbox{dg}[\bE\bW\bV_r(\mSigma)] = \vone_p^T\mbox{dg}[\bE\bW\bV_q(\mSigma)]$
and solving for $\widetilde{\vmu}$, $\widetilde{\lambda}$,
$\widetilde{\mPsi}$ and $\widetilde{\nu}$ we obtain the updates:
$$
\widetilde{\vmu} \leftarrow \bE_r(\vmu);
\quad 
\widetilde{\lambda}  \leftarrow \frac{\mbox{tr}[\bE_r(\mSigma)]}{\mbox{tr}[\bV_r(\vmu)]};
\quad 
\ds \widetilde{\nu}  
    \leftarrow  \frac{
        2\,\vone_p^T[\mbox{dg}(\bE_r(\mSigma))^{\odot 2}]
    }{
        \vone_p^T\mbox{dg}[\bE\bW\bV_r(\mSigma)] 
    } + p + 3; 
\quad 
\widetilde{\mPsi} 
    \leftarrow \ds ( \widetilde{\nu} - p  - 1) \, \bE_r(\mSigma).
    $$

When there is no missing data, $\mA=\vzero$ and $\vd=\vzero$, and the update reduces to the
complete-data normal-inverse-Wishart update. When missing data are present, 
$\mA$ and $\vd$
propagate uncertainty from the missing components into the moment-matching updates for
$\vmu$ and $\mSigma$.
Algorithm \ref{alg:mp_mvn_missing_blocked} summarizes the MFMP approximation.

\newpage

\begin{algorithm}[!ht]
\caption[Algorithm 1]{Blocked MFMP algorithm for the MVN model with missing data}
\label{alg:mp_mvn_missing_blocked}
{\footnotesize
\begin{algorithmic}[1]
		
\REQUIRE{$\mX\in \bR^{n\times p}$, $\mPsi_0 \in \bS_+^p$, $\lambda_0>0$, $\nu_0 + n > p + 3$. }

\STATE Initialise: $\widetilde{\vmu}$, $\widetilde{\lambda}$, $\widetilde{\mPsi}$ and $\widetilde{\nu}$ and set $\lambda_n \leftarrow \lambda_0 + n; 
\quad \nu_n \leftarrow \nu_0 + n$;

\STATE For each $i$ determine $\vc_i$, $\mR_i$, $O_i$ and $M_i$ from $\mX$ and set $\widehat{\vx}_i\leftarrow \vx_i$, and $\mB_i \leftarrow \vzero$ for all $i$ s.t. $M_i=\varnothing$.

	\REPEAT  

        \FOR{$i$ such that $M_i \ne \varnothing$}

        \IF{$O_i=\varnothing$}

            \STATE Approximate posterior moments of $\vz_i$ when $O_i=\varnothing$:   
            $$\bE_r(\vz_i) \leftarrow\widetilde{\vmu}; \qquad \bV_r(\vz_i) \leftarrow \left( 1 + \frac{1}{\widetilde{\lambda}}\right) \frac{\widetilde{\mPsi}}{\widetilde{\nu} -p - 1};
            $$

        \ELSE

            \STATE Approximate posterior moments of $\vz_i$ when $O_i\ne\varnothing$:   
     $$
\begin{array}{l}
\mL_i \ds \leftarrow \widetilde{\mPsi}_{M_iO_i}\widetilde{\mPsi}_{O_iO_i}^{-1}; \qquad 
      \widetilde{\mPsi}_{M_i|O_i} \leftarrow \widetilde{\mPsi}_{M_iM_i} - \widetilde{\mPsi}_{M_iO_i}\widetilde{\mPsi}_{O_iO_i}^{-1}\widetilde{\mPsi}_{O_iM_i};
\\ [1ex]
 \bE_r(\vz_i) 
    \ds \leftarrow \widetilde{\vmu}_{M_i} + \mL_i(\vx_{i,O_i} - \widetilde{\vmu}_{O_i});
\\
\bV_r(\vz_i) 
    \ds \leftarrow \left[ 1 + \widetilde{\lambda}^{-1} + (\vx_{i,O_i} - \widetilde{\vmu}_{O_i})^T\widetilde{\mPsi}_{O_iO_i}^{-1}(\vx_{i,O_i} - \widetilde{\vmu}_{O_i}) \right] \frac{\widetilde{\mPsi}_{M_i|O_i}}{\widetilde{\nu} - |M_i|-1};
\end{array}
$$
            
        \ENDIF

        \STATE Perform MP update for $q(\vz_i)= N(\widetilde{\vm}_i,\widetilde{\mV}_i)$ via moment matching
        $$\widetilde{\vm}_i \leftarrow \bE_r(\vz_i);
        \qquad \widetilde{\mV}_i \leftarrow \bV_r(\vz_i);$$

        \STATE and update $\bE_q(\vx_i)$ and $\bV_q(\vx_i)$: $\widehat{\vx}_i \leftarrow \vc_i + \mR_i\widetilde{\vm}_i;
        \quad \mB_i \leftarrow \mR_i \widetilde{\mV}_i\mR_i^T;$
    \ENDFOR

    \STATE Calculate intermediate quantities:
    $$
\widehat{\vmu}_n \leftarrow \sum_{i=1}^n\widehat{\vx}_i/\lambda_n;
\quad 
    \mA \leftarrow \sum_{i=1}^n\mB_i;
    \quad 
\mC \leftarrow \mPsi_0 + \sum_{i=1}^n \widehat{\vx}_i\widehat{\vx}_i^T + \left( 1 - \lambda_n^{-1}\right)\mA - \lambda_n \widehat{\vmu}_n\widehat{\vmu}_n^T
$$ 

\noindent and for $j=1,\ldots,p$ set $\va_j = (\widehat{x}_{1j},\ldots,\widehat{x}_{nj})$
and $\vb_j = (\{ \mB_1\}_{jj},\ldots, \{ \mB_n\}_{jj})$ and
$$
\ds d_{j} \leftarrow 2 \left( 1 - \frac{2}{\lambda_n}\right) \vone^T\vb_j^{\odot 2} + \frac{2}{\lambda_n^2}(\vone^T\vb_j)^2 + 4 \vb_j^T\left( \va_j - \frac{\vone^T\va_j}{\lambda_n} \vone_n \right)^{\odot 2};
$$

    \STATE Calculate local likelihood based target moments:
    $$
\begin{array}{c}
\bE_r(\vmu) 
    \leftarrow \widehat{\vmu}_n; 
\quad 
\ds \bE_r(\mSigma) 
    \leftarrow \frac{\mC}{\nu_n - p - 1};
\quad 
\bV_r(\vmu) 
\ds \leftarrow \frac{\bE_r(\mSigma)}{\lambda_n}
+ \frac{\mA}{\lambda_n^2};
\\ [2ex]
    \ds \mbox{dg}[\bE\bW\bV_r(\mSigma)]
    \ds \leftarrow \frac{2\,\mbox{dg}(\mC)^{\odot 2} + (\nu_n - p - 1)\,\vd }{ (\nu_n - p - 1)^2(\nu_n - p   - 3)};
\end{array}
$$

\STATE Set $\bE_r(\vmu)=\bE_q(\vmu)$,
$\mbox{tr}[\bV_r(\vmu)]=\mbox{tr}[\bV_q(\vmu)]$,
$\bE_r(\mSigma) = \bE_q(\mSigma)$,
and $\vone_p^T\mbox{dg}[\bE\bW\bV_r(\mSigma)] = \vone_p^T\mbox{dg}[\bE\bW\bV_q(\mSigma)]$
to determine $\widetilde{\vmu}$, $\widetilde{\lambda}$,
$\widetilde{\mPsi}$ and $\widetilde{\nu}$ via:
$$
\widetilde{\vmu} \leftarrow \bE_r(\vmu);
\quad 
\widetilde{\lambda}  \leftarrow \frac{\mbox{tr}[\bE_r(\mSigma)]}{\mbox{tr}[\bV_r(\vmu)]};
\quad 
\ds \widetilde{\nu}  
    \leftarrow  \frac{
        2\,\vone_p^T[\mbox{dg}(\bE_r(\mSigma))^{\odot 2}]
    }{
        \vone_p^T\mbox{dg}[\bE\bW\bV_r(\mSigma)] 
    } + p + 3; 
\quad 
\widetilde{\mPsi} 
    \leftarrow \ds ( \widetilde{\nu} - p  - 1) \, \bE_r(\mSigma);
    $$

\UNTIL{convergence criteria are met}

\end{algorithmic}
}

\end{algorithm}

\section{Details for probit regression example}
\label{supp:probit}

\subsection{Derivations for MFVB}

The derivation, except for very minor changes can be found in \textcite{Ormerod2010} (setting $\vy=\vone_n$, and using $\mZ = \mbox{diag}(2\vy - \vone_n)\mX$ in place of $\mX$). We refer the interested reader there.
The MFVB approximation corresponding to $q(\vbeta,\va) = q(\vbeta)\,q(\va)$ is
$$
\begin{array}{lcl}
q(\vbeta) \ds = N( \widetilde{\vmu}, \widetilde{\mSigma})
&
\quad \mbox{and} \quad 
&
q(a_i)  \ds = \mbox{TN}_+(\widetilde{m}_{i}, 1 ), \qquad i=1,\ldots,n,
\end{array}
$$ 

\noindent where the update for $q(\vbeta)$ is given by
$\widetilde{\vmu} \leftarrow  \mS\mZ^T\bE_q(\va)$,
and 
$\widetilde{\mSigma}\leftarrow \mS$.
Algorithm \ref{alg:vb_probit} summarizes the MFVB algorithm for the probit regression model. Algorithm \ref{alg:vb_probit} is very fast (per iteration) due to the fact that the matrix $\mS$ needs to be only calculated once outside the main loop.

\begin{algorithm}[!ht]
\caption[Algorithm 6]{MFVB for probit regression}
\label{alg:vb_probit}
\begin{algorithmic}[1]
		
%\REQUIRE{$\vy\in \bR^n$, $\mX\in \bR^{n\times p}$ and initial value for $\widetilde{\vmu}$.}  

\REQUIRE{$\vy\in\{0,1\}^n$, $\mX\in\bR^{n\times p}$, prior precision
matrix $\mD$, and initial value for 
$\widetilde\vmu$.}

\STATE Set $\mZ = \mbox{diag}(2\vy - \vone_n)\mX$, and $\mS = (\mZ^T\mZ + \mD)^{-1}$

\REPEAT  

	\STATE Update $q(\va)$ via
	$\widetilde{\vm} \leftarrow  \mZ\widetilde{\vmu}$
	
	\STATE Update $q(\vbeta)$ via
	$\widetilde{\vmu} \leftarrow \mS\mZ^T(\widetilde{\vm} + \zeta_1(\widetilde{\vm})); \, \widetilde{\mSigma} \leftarrow \mS;$

\smallskip 
\UNTIL{convergence criteria are met}
		
\end{algorithmic}
\end{algorithm}

\newpage 
\subsection{Additional details for MP}

Derivations for the MP approximation for probit regression outline in Section~\ref{sec:probit_example} are relatively complete. A few details are included here for completeness.

Since $a_i| \vbeta,y_i$ is a unit-variance normal distribution truncated below at zero,
$$
\bE_q(a_i)
=
\widetilde m_{i}
+
\zeta_1(\widetilde m_{i}),
\qquad \mbox{where} \qquad 
\zeta_k(t)=\frac{d^k}{dt^k}\log\Phi(t).
$$

Algorithm~\ref{alg:mp_probit} uses
the function $\xi_k(m,v)$ which is defined 
as follows.
For $k=1,2$, 
$$
\xi_k(m,v)
=
\bE[\zeta_k(T)],
\qquad \mbox{where} \qquad 
T\sim N(m,v),
$$

\noindent with the convention that for vectors $\vm$ and $\vv$, $\xi_k(\vm,\vv)$ is evaluated component-wise.
The full algorithm is listed as Algorithm~\ref{alg:mp_probit}.

%\subsection{Algorithm: MP for probit regression}

\begin{algorithm}[!ht]
\caption[Algorithm 6]{MP for probit regression}
\label{alg:mp_probit}
\begin{algorithmic}[1]
		
% \REQUIRE{$\vy\in \bR^n$, $\mX\in \bR^{n\times p}$}  
% \STATE Set $\mZ = \mbox{diag}(2\vy - \vone_n)\mX$, and initialize $\widetilde{\vmu}$.

\REQUIRE{$\vy\in\{0,1\}^n$, $\mX\in\bR^{n\times p}$, prior precision
matrix $\mD$, and initial value for 
$\widetilde\vmu$.}

\STATE Set Set $\mZ = \mbox{diag}(2\vy - \vone_n)\mX$, $\mS = (\mZ^T\mZ + \mD)^{-1}$
and initialize 
$\widetilde{\mSigma}=\mS$.

\REPEAT  

	\STATE Calculate $\mZ\widetilde{\vmu}$ and $\mbox{dg}(\mZ\widetilde{\mSigma} \mZ^T)$
	
	\STATE Approximate $\xi_1(\mZ\widetilde{\vmu},\mbox{dg}(\mZ\widetilde{\mSigma} \mZ^T))$   and
	$\xi_2(\mZ\widetilde{\vmu},\mbox{dg}(\mZ\widetilde{\mSigma} \mZ^T))$ via quadrature.

	\STATE Update $q(\va)$ via
 $\widetilde{\vm} \leftarrow \mZ\widetilde{\vmu} + \xi_1(\mZ\widetilde{\vmu},\mbox{dg}(\mZ\widetilde{\mSigma} \mZ^T))$.	
	
	\STATE Update $q(\vbeta)$ via
	$$
	\begin{array}{rl}
	\widetilde{\vmu} 
	    & \ds \leftarrow \mS\mZ^T\widetilde{\vm}
	\\ [1ex]
	\widetilde{\mSigma} 
	    & \ds \leftarrow  \mS + \mS\mZ^T \,[\, \mI_n + \mbox{diag}\{  \xi_2(\mZ\widetilde{\vmu},\mbox{dg}(\mZ\widetilde{\mSigma} \mZ^T)) \}\,]\,\mZ\mS \\ [1ex]
	    & \ds \qquad + \mS\mZ^T\,[\,\mI_n + \mbox{diag}\{\zeta_2(\mZ\widetilde{\vmu})\}\,]\,\mZ\widetilde{\mSigma}\mZ^T\,[\,\mI_n + \mbox{diag}\{\zeta_2(\mZ\widetilde{\vmu})\}\,]\,\mZ\mS
	\end{array} 
	$$

\UNTIL{convergence criteria are met}
		
\end{algorithmic}
\end{algorithm}

\newpage
 
\subsection{Proof of Theorem \ref{Th:Theorem3}}
\label{sec:probit_proof} 
 
\paragraph{Regularity conditions.}
Assume that $p$ is fixed and that the following conditions hold.
\begin{enumerate}
\item[(A1)] The prior precision matrix 
$\mD$ is fixed and positive definite.

\item[(A2)] 
%The posterior mode 
% $\widehat\vbeta$ exists, is unique, and lies in the interior of
% $\bR^p$. The MP fixed point
% $(\widetilde\vmu,\widetilde\mSigma)$ exists and is locally stable.
%
The posterior mode $\widehat\vbeta$ exists, is unique, and lies in the interior of
$\bR^p$. The MP fixed point $(\widetilde\vmu,\widetilde\mSigma)$ exists.

\item[(A3)] The design matrix satisfies
$n^{-1}\mX^T \mX = n^{-1}\mZ^T\mZ \to \mSigma_x$, where
$\mSigma_x\in\sS_p^+$.
In particular, the eigenvalues of 
$\mX^T\mX$ are of order $n$.

\item[(A4)] At the MP fixed point,
$\widetilde\mSigma=O_p^m(n^{-1})$,
where $O_p^m(r_n)$ means that every element of the matrix is $O_p(r_n)$.

\item[(A5)] Let
$\vs^2=\operatorname{dg}(\mZ\widetilde\mSigma \mZ^T)$.
The Taylor series remainders in the Gaussian-smoothed probit derivatives satisfy
$\mZ^T \mDelta_1=O_p^m(1)$,
where
$\xi_k(\mZ\widetilde\vmu,\vs^2)
=
\zeta_k(\mZ\widetilde\vmu) + \mDelta_k$, $k=1,2$, and 
$\mZ^T\operatorname{diag}(\mDelta_2)\mZ=O_p^m(1)$. In addition, if 
$\widetilde\vmu-\widehat\vbeta=O_p^m(n^{-1})$, assume
$\mZ^T
[
\operatorname{diag}\{\zeta_2(\mZ\widetilde\vmu)\}
-
\operatorname{diag}\{\zeta_2(\mZ\widehat\vbeta)\}
]
\mZ
=
O_p^m(1)$.

\item[(A6)] For every $\overline\vbeta$ on the line segment between
$\widehat\vbeta$ and $\widetilde\vmu$, the matrix
$$
\mD-\mZ^\top\operatorname{diag}[\zeta_2(\mZ\overline\vbeta)]\mZ
$$

\noindent 
has eigenvalues of order $n$, bounded away from zero after scaling by $n^{-1}$.

% The curvature matrices
% $$
% \mD-\mZ^T\operatorname{diag}
% [\zeta_2(\mZ\widehat\vbeta)]\mZ
% \qquad \mbox{and} \qquad 
% \mD-\mZ^T\operatorname{diag}[\zeta_2(\mZ\widetilde\vmu)]\mZ
% $$

% have eigenvalues of order $n$, bounded away from zero after scaling by $n^{-1}$.
\end{enumerate}

Conditions (A3)--(A6) are the main order conditions used in the proof. Condition (A3) controls the scale of $\mZ^T \mZ$, while (A6) ensures that the posterior curvature matrix and the corresponding MP curvature matrix are invertible with inverse of order $O_p(n^{-1})$. Condition (A5) controls the Taylor remainders arising from replacing the Gaussian-smoothed quantities $\xi_j(\mZ\widetilde\vmu,\vs^2)$ by their zero-variance limits 
$\zeta_j(\mZ\widetilde\vmu)$. A simple sufficient condition for (A5) is a bounded-design assumption together with 
$\widetilde\mSigma^\ast=O_p^m(n^{-1})$. For random designs, suitable moment and tail conditions on the covariates also imply these bounds.

Before we begin the proof, the following two lemmas are needed.

\begin{Lemma}\label{lemma1}
{\bf [Inverse perturbation bound].}
Let $\mA_n$ and $\mE_n$ be square matrices of fixed dimension. Suppose that
$\mA_n$ and $\mA_n+\mE_n$ are nonsingular, and that for any fixed matrix norm,
$$
\|\mA_n^{-1}\|=O_p(a_n),
\qquad
\|(\mA_n+\mE_n)^{-1}\|=O_p(a_n),
\qquad
\|\mE_n\|=O_p(b_n).
$$

\noindent 
Then $(\mA_n+\mE_n)^{-1}-\mA_n^{-1}
=
O_p(a_n^2 b_n)$.
In particular, if $\mA_n$ has eigenvalues of order $n$, so that
$\|\mA_n^{-1}\|=O_p(n^{-1})$, and if 
$\|\mE_n\|=O_p(1)$, then
$$(\mA_n+\mE_n)^{-1}
=\mA_n^{-1}+O_p(n^{-2}).
$$
\end{Lemma}
 
\paragraph{Proof:} First we use the exact identity
$(\mA_n+\mE_n)^{-1}-\mA_n^{-1}
=
-(\mA_n+\mE_n)^{-1}\mE_n\mA_n^{-1}$.
Taking norms gives
$$\|(\mA_n+\mE_n)^{-1}-\mA_n^{-1}\|
\le
\|(\mA_n+\mE_n)^{-1}\|\,\|\mE_n\|\,\|\mA_n^{-1}\|.
$$

\noindent 
By the assumed stochastic orders, the right-hand side is
$O_p(a_n)\,O_p(b_n)\,O_p(a_n)
=
O_p(a_n^2b_n)$.
The special case follows by taking 
$a_n=n^{-1}$ and $b_n=1$.

\begin{Lemma}\label{lemma2}
{\bf [Inverse bound for the covariance fixed-point operator].}
Let $\mA_n$, $\mB_n$, and $\mC_n$ be symmetric 
$p\times p$ matrices, where $p$ is fixed. Suppose that
$\mA_n=\mB_n+\mC_n$,
that $\mB_n$ is positive semidefinite, and that $\mA_n$ and $\mC_n$ are positive definite with
eigenvalues of order $n$. Define the linear operator
$$
\mathcal L_n(\mE)
=
\mA_n\mE\mA_n-\mB_n\mE\mB_n .
$$

\noindent 
Then $\mathcal L_n$ is invertible on the space of $p\times p$ matrices and
$\|\mathcal L_n^{-1}\|=O_p(n^{-2})$.
Consequently, if
$\mathcal L_n(\mE_n)=O_p^m(1)$,
then $\mE_n=O_p^m(n^{-2})$.
\end{Lemma}

\paragraph{Proof:}
Since $\mA_n=\mB_n+\mC_n$,
$$
\mathcal L_n(\mE)
=
\mA_n\mE\mA_n-\mB_n\mE\mB_n
=
\mC_n\mE\mA_n+\mB_n\mE\mC_n.
$$

\noindent 
Vectorizing gives
$\operatorname{vec}\{\mathcal L_n(\mE)\}
=
\left(\mA_n\otimes \mC_n+\mC_n\otimes \mB_n\right)\operatorname{vec}(\mE)$,
where we have used symmetry of 
$\mA_n$, $\mB_n$, and $\mC_n$. 

Let
$\mK_n=\mA_n\otimes \mC_n+\mC_n\otimes \mB_n$. Since $\mA_n$ and $\mC_n$ are positive definite and $\mB_n$ is positive semidefinite,
$\mK_n$ is positive definite. Moreover,
$$
\lambda_{\min}(\mK_n)
\ge
\lambda_{\min}(\mA_n\otimes \mC_n)
=
\lambda_{\min}(\mA_n)\lambda_{\min}(\mC_n).
$$

\noindent 
By assumption, $\lambda_{\min}(\mA_n)$ and $\lambda_{\min}(\mC_n)$ are bounded below by a positive multiple of $n$, with probability tending to one. Hence,
$\lambda_{\min}(\mK_n)
\ge
\lambda_{\min}(\mA_n)\lambda_{\min}(\mC_n)
$
is bounded below by a positive multiple of $n^2$, with probability tending to one. Therefore,
$$\|\mK_n^{-1}\|=O_p(n^{-2}).$$

\noindent 
Thus,
$\|\mathcal L_n^{-1}\|=O_p(n^{-2})$,
and so if $\mathcal L_n(\mE_n)=O_p^m(1)$, 
then
$\operatorname{vec}(\mE_n)
=
\mK_n^{-1}\operatorname{vec}(\mathcal L_n(\mE_n))
=
O_p(n^{-2})\,O_p(1)$,
and since $p$ is fixed, this implies
$\mE_n=O_p^m(n^{-2})$.
 
\bigskip 
\noindent We are now ready to prove Theorem~\ref{Th:Theorem3}.

\paragraph{Proof of Theorem \ref{Th:Theorem3}:}
We prove the result at the fixed point of Algorithm~8. Let
$$
\mA=\mZ^T \mZ + \mD, \quad \mbox{so that}  \quad \mS=\mA^{-1}.
$$

\noindent Since $p$ is fixed and by (A3)   the design satisfies
$n^{-1}\mZ^T \mZ\to \mSigma_x\in\sS_p^+$.
Thus, $\mA = O_p^m(n)$ and 
$\mS=O_p^m(n^{-1})$. 
By (A4), at the fixed point,
$\widetilde\mSigma=O_p^m(n^{-1})$.
The weighted empirical sums appearing below are controlled by the empirical order conditions in (A5).
% By (A4) at the fixed point,  
% $\widetilde\mSigma=O_p^m(n^{-1})$, 
% and that the weighted empirical sums appearing
% below have the stated orders. 
These conditions hold, for example, under standard bounded
design conditions, or under random-design conditions with sufficiently many moments.

The posterior mode $\widehat\vbeta$ satisfies
$$
\mZ^T\zeta_1(\mZ\widehat\vbeta)-\mD\widehat\vbeta=\vzero
\qquad \mbox{or equivalently} \qquad 
\widehat\vbeta
=
\mS \mZ^T[\mZ\widehat\vbeta + \zeta_1(\mZ\widehat\vbeta)].
$$

Let $\vs^2=\operatorname{dg}(\mZ\widetilde\mSigma \mZ^T)$.
Since $\widetilde\mSigma=O_p^m(n^{-1})$, we have $s_i^2 = O_p(n^{-1}\|\vz_i\|^2)$. By a Taylor series
expansion of $\xi_1$ 
around $s_i^2=0$,
$$
\xi_1(\mZ\widetilde\vmu,\vs^2)
=
\zeta_1(\mZ\widetilde\vmu)+\mDelta_1,
$$

\noindent 
where
$\mZ^T \mDelta_1=O_p^m(1)$ by (A5).
At the fixed point of Algorithm~8 we have
$\widetilde\vmu
=
\mS \mZ^T
[\mZ\widetilde\vmu+\xi_1(\mZ\widetilde\vmu,\vs^2)]$.
Subtracting the corresponding equation for 
$\widehat\vbeta$ gives
$$
\mA(\widetilde\vmu-\widehat\vbeta)
=
\mZ^T \mZ(\widetilde\vmu-\widehat\vbeta)
+
\mZ^T[\zeta_1(\mZ\widetilde\vmu)-\zeta_1(\mZ\widehat\vbeta)]
+
\mZ^T \mDelta_1.
$$

By the mean value theorem,
$\zeta_1(\mZ\widetilde\vmu)-\zeta_1(\mZ\widehat\vbeta)
=
\mW \mZ(\widetilde\vmu-\widehat\vbeta)$,
where $\mW$ is diagonal with entries between
$\zeta_2(\vz_i^T\widetilde\vmu)$ 
and $\zeta_2(\vz_i^T\widehat\vbeta)$. 
Hence,
$$
(\mD-\mZ^T \mW\mZ)(\widetilde\vmu-\widehat\vbeta)
=
\mZ^\top \mDelta_1.
$$

\noindent 
Since $-\zeta_2(x)>0$ for all $x$ the matrix 
$\mD-\mZ^T \mW\mZ$ has eigenvalues of order $n$ under the
regularity conditions above, in particular (A6). Therefore,
$\widetilde\vmu-\widehat\vbeta
=
O_p^m(n^{-1})$.

We now turn to the covariance. At the fixed point, define
$$
\begin{array}{c}
\mW_\zeta=\operatorname{diag}
[\zeta_2(\mZ\widetilde\vmu)],
\quad 
\mB_\zeta = \mZ^T(\mI_n+\mW_\zeta)\mZ,
\qquad
\mC_\zeta = \mA - \mB_\zeta = \mD - \mZ^T \mW_\zeta \mZ 
\qquad \mbox{and}
\\ 
\mW_\xi=\operatorname{diag}
[\xi_2(\mZ\widetilde\vmu,\vs^2)],
\quad 
\mB_\xi = \mZ^T(\mI_n+\mW_\xi)\mZ,
\qquad
\mC_\xi = \mA - \mB_\xi = \mD - \mZ^T \mW_\xi \mZ.

\end{array}
$$

\noindent Note that
the matrix $\mB_\zeta$ is positive semidefinite because
$1+\zeta_2(x)>0$ for all $x$ (this is the variance of a unit-variance normal variable truncated below at zero).

Using another Taylor series expansion in $s_i^2$ we have 
$$
\xi_2(\mZ\widetilde\vmu,\vs^2)
=
\zeta_2(\mZ\widetilde\vmu)+\mDelta_2,
$$

\noindent 
where
$\mZ^T\operatorname{diag}(\mDelta_2)\mZ=O_p^m(1)$.
Thus, $\mB_\xi = \mB_\zeta+O_p^m(1)$.
The covariance fixed-point equation in Algorithm~8 can be rewritten as
$$
\widetilde\mSigma
=
\mS+\mS \mB_\xi \mS+\mS \mB_\zeta\widetilde\mSigma \mB_\zeta \mS.
$$

\noindent
Multiplying on the left and right by 
$\mA=\mS^{-1}$, using $\mB_\xi = \mB_\zeta+O_p^m(1)$, and rearranging gives
$$
\mA\widetilde\mSigma \mA
-
\mB_\zeta\widetilde\mSigma \mB_\zeta
=
\mA + \mB_\zeta + O_p^m(1).
$$

Now, since $\mC_\zeta = \mA-\mB_\zeta$ direct expansion shows that
$\mA \mC_\zeta^{-1} \mA -\mB_\zeta \mC_\zeta^{-1}\mB_\zeta
=
\mA + \mB_\zeta$.
Also, since $\mA=\mB_\zeta+\mC_\zeta$,
$(\mB_\zeta+\mC_\zeta)\mC_\zeta^{-1}(\mB_\zeta+\mC_\zeta)-\mB_\zeta\mC_\zeta^{-1}\mB_\zeta
=
\mB_\zeta+\mC_\zeta+\mB_\zeta
=
\mA+\mB_\zeta$.
Therefore, if $\mE=\widetilde\mSigma-\mC_\zeta^{-1}$, then
$\mA \mE \mA - \mB_\zeta \mE \mB_\zeta
=
O_p^m(1)$.
By Lemma~ \ref{lemma2}, the associated linear operator
$\mathcal L_n(\mE)=\mA\mE\mA-\mB_\zeta \mE \mB_\zeta$
has inverse of order $O_p(n^{-2})$, since 
$\mA=\mB_\zeta+\mC_\zeta$, $\mB_\zeta$ is positive semidefinite, and $\mA$ and 
$\mC_\zeta$ are positive definite with eigenvalues of order $n$.
% Using Lemma \ref{lemma2}
% the associated linear operator
% $\mathcal L_n(\mE)=\mA\mE\mA-\mB_\zeta\mE\mB_\zeta$
% has inverse of order $O_p(n^{-2})$ under the same local stability conditions, because
% $\mA$, $\mB_\zeta$, and $\mC_\zeta=\mA-\mB_\zeta$ are all of order $n$, with $\mC_\zeta$ being positive definite. 
Hence,
$\mE=O_p^m(n^{-2})$,
and therefore
$\widetilde\mSigma
= \mC_\zeta^{-1}+O_p^m(n^{-2})$.

It remains to replace $\mC_\zeta$ by the curvature matrix evaluated at the posterior mode. Since,
$\widetilde\vmu-\widehat\vbeta=O_p^m(n^{-1})$, another Taylor series expansion gives
$$
\mZ^T
\left[
\operatorname{diag}\{\zeta_2(\mZ\widetilde\vmu)\}
-
\operatorname{diag}\{\zeta_2(\mZ\widehat\vbeta)\}
\right]
\mZ
=
O_p^m(1).
$$

\noindent 
This follows from the preceding mean bound under bounded-design conditions; for random
designs it is included in the empirical order assumptions. Thus,
$\mC_\zeta
=
\mD-\mZ^T\operatorname{diag}[\zeta_2(\mZ\widehat\vbeta)]\mZ
+
O_p^m(1)$.
Since the leading matrix has eigenvalues of order $n$, the inverse perturbation bound lemma above gives
$$
\mC_\zeta^{-1}
=
[\mD-\mZ^T\operatorname{diag}\{\zeta_2(\mZ\widehat\vbeta)\}\mZ]^{-1}
+
O_p^m(n^{-2}).
$$

\noindent 
Combining this with the previous display leads to
$$
\widetilde\mSigma
=
[\mZ^T\operatorname{diag}\{-\zeta_2(\mZ\widehat\vbeta)\}\mZ+\mD]^{-1}
+
O_p^m(n^{-2}).
$$

\noindent 
Since $p$ is fixed, element-wise 
$O_p(n^{-2})$ convergence implies Frobenius convergence of
the same order, i.e.,
$$
\|\widetilde\mSigma-\mV\|_F=O_p(n^{-2}),
$$

\noindent 
where
$\mV=
[\mZ^T\operatorname{diag}\{-\zeta_2(\mZ\widehat\vbeta)\}\mZ+\mD]^{-1}$.
Together with the mean result above, this proves the theorem.

\newpage 
\section{Additional numerical results for logistic regression}
\label{sec:additional}

This section reports the logistic regression analogue of the probit regression results in Section~5.2. The same datasets, prior specification, performance metrics, and computational setup are used. The purpose of these results is to verify that the behaviour observed for probit regression is not specific to the probit likelihood.

\begin{table}[!h]
\begin{center}
{\footnotesize
\centering
\begin{tabular}{l|l|ccccccr}
Data & Method  & RMSE($\mu$) & RMSE($\sigma$) & MIN & Q1. & Q2. & Frob. & Time (s) \\
\hline
Breastfeed 
& stan  & 0.007 & {\bf 0.0108} & 97.6 & 97.9 & 98.1 & {\bf 0.016} &       0.493 \\
& Laplace   & 0.258 & 0.0705 & 81.5 & 88.1 & 90.8 & 0.058 &       0.003 \\
& MFVB  & 0.154 & 0.2600 & 77.6 & 78.9 & 84.5 & 0.185 &       0.002 \\
%& IL    & 0.011 & 0.0101 & 98.7 & 99.1 & 99.1 &       &       0.010 \\
%& EP    & 0.015 & 0.0313 & 95.0 & 96.7 & 98.5 & 0.031 &       0.124 \\
& EP    & 0.008 & 0.0297 & 94.8 & 96.5 & 98.4 & 0.029 &       0.004 \\
& GVB   & 0.006 & 0.0346 & 94.6 & 96.2 & 98.3 & 0.033 &       0.008 \\
%& DMVB  & 0.014 & 0.0144 & 93.9 & 96.1 & 97.9 & 0.018 &       0.005 \\
%& PHA-L & 0.258 & 0.0533 & 96.3 & 96.6 & 97.0 & 0.058 &       0.003 \\
%& PHA-G & 0.006 & 0.0182 & 98.8 & 99.1 & 99.3 & 0.033 &       0.003 \\
%& PHA-E & 0.015 & 0.0196 & 98.7 & 99.0 & 99.2 & 0.031 &       0.012 \\
& GMP-L & 0.038 & 0.0313 & 97.9 & 98.3 & 98.5 & 0.025 &       0.022 \\
& GMP-G & {\bf 0.004} & 0.0184 & {\bf 98.8} & {\bf 99.1} & {\bf 99.3} & 0.022 &       0.004 \\
& GMP-E & 0.006 & 0.0188 & {\bf 98.8} & {\bf 99.1} & 99.2 & 0.020 &       0.029 \\
\hline 
Ion.
& stan  & 0.012 & 0.007 & {\bf 96.9} & 97.8 & 98.0 & {\bf 0.005} &       1.90 \\
& Laplace   & 0.319 & 0.068 & 45.1 & 70.5 & 82.1 & 0.020 &       0.01 \\
& MFVB  & 0.152 & 0.241 & 56.7 & 68.0 & 70.9 & 0.057 &       0.14 \\
%& IL    & 0.057 & 0.015 & 91.8 & 94.7 & 96.2 &       &       0.89 \\
%& EP    & 0.036 & 0.021 & 93.9 & 97.1 & 97.7 & 0.009 &       0.82 \\
& EP   & {\bf 0.008} & {\bf 0.015} & 95.9 & {\bf 98.0} & {\bf 98.6} & 0.007 &       0.10 \\
& GVB   & 0.012 & 0.035 & 94.3 & 96.6 & 97.0 & 0.012 &       0.11 \\
%& DMVB  & 0.102 & 0.113 & 81.2 & 87.8 & 89.4 & 0.047 &       0.04 \\
%& PHA-L & 0.319 & 0.126 & 82.1 & 85.1 & 86.9 & 0.020 &       0.03 \\
%& PHA-G & 0.012 & 0.033 & 95.1 & 97.0 & 97.2 & 0.012 &       0.06 \\
%& PHA-E & 0.036 & 0.042 & 94.5 & 95.9 & 96.3 & 0.009 &       0.15 \\
& GMP-L & 0.047 & 0.058 & 90.8 & 93.6 & 94.4 & 0.017 &       0.31 \\
& GMP-G & 0.012 & 0.033 & 95.2 & 96.9 & 97.3 & 0.011 &       0.04 \\
& GMP-E & 0.010 & 0.035 & 95.0 & 96.7 & 97.0 & 0.014 &       0.25 \\
\hline 
Spam
& stan  & 0.007 & {\bf 0.004} & 97.2 & 97.8 & 98.0 & {\bf 0.0018} &      44.4 \\
& Laplace   & 0.089 & 0.023 & 75.0 & 89.3 & 94.1 & 0.0039 &       0.3 \\
& MFVB  & 0.200 & 0.225 & 11.8 & 57.1 & 71.2 & 0.0445 &      12.9 \\
%& IL    & 0.017 & 0.008 & 95.3 & 98.9 & 99.3 &        &      57.1 \\
%& EP    & 0.005 & 0.017 & 81.3 & 95.7 & 97.5 & 0.0036 &      34.5 \\
& EP   & {\bf 0.004} & 0.017 & 81.0 & 95.8 & 97.4 & 0.0035 &       4.4 \\
& GVB   & 0.005 & 0.022 & 78.8 & 95.6 & 97.3 & 0.0045 &       4.4 \\
%& DMVB  & 0.039 & 0.047 & 74.8 & 94.6 & 97.0 & 0.0113 &       5.5 \\
%& PHA-L & 0.089 & 0.024 & 91.8 & 97.1 & 97.9 & 0.0039 &       0.5 \\
%& PHA-G & 0.005 & 0.007 & 97.9 & 99.4 & 99.6 & 0.0045 &       0.5 \\
%& PHA-E & 0.005 & 0.009 & 97.4 & 99.4 & 99.6 & 0.0036 &       2.6 \\
& GMP-L & 0.013 & 0.014 & 96.5 & 98.6 & 99.0 & 0.0046 &       4.1 \\
& GMP-G & 0.005 & 0.007 & {\bf 97.9} & {\bf 99.4} & {\bf 99.6} & 0.0026 &       0.8 \\
& GMP-E & 0.006 & 0.009 & 97.5 & 99.2 & 99.5 & 0.0025 &       5.1 \\
\hline 
Caravan
& stan   & {\bf 0.015} & {\bf 0.016} & {\bf 96.9} & 97.8 & 98.1 & {\bf 0.025} &    1020.1 \\
& Laplace    & 0.341 & 0.074 & 63.3 & 94.5 & 96.9 & 0.033 &       0.6 \\
& MFVB   & 0.082 & 0.390 & 52.4 & 71.7 & 73.4 & 0.138 &      36.3 \\
%& IL     & 0.091 & 0.025 & 91.9 & 98.7 & 99.0 &       &     299.0 \\
%& EP     & 0.043 & 0.083 & 82.7 & 97.5 & 98.8 & 0.035 &     158.5 \\
& EP    & {\bf 0.015} & 0.072 & 83.1 & 97.4 & 98.9 & 0.032 &      21.6 \\
& GVB    & 0.235 & 0.159 & 69.9 & 94.5 & 97.8 & 0.072 &      21.5 \\
%& DMVB   & 0.180 & 0.196 & 72.0 & 96.0 & 98.3 & 0.109 &      21.0 \\
%& PHA-L  & 0.341 & 0.170 & 44.7 & 95.2 & 96.9 & 0.033 &       0.9 \\
%& PHA-G  & 0.235 & 0.134 & 79.8 & 94.0 & 97.0 & 0.072 &       0.9 \\
%& PHA-E  & 0.043 & 0.063 & 89.6 & 98.6 & 99.1 & 0.035 &       4.7 \\
& GMP-L  & 0.100 & 0.073 & 89.4 & 96.6 & 97.7 & 0.034 &      15.6 \\
& GMP-G  & 0.138 & 0.110 & 81.1 & 96.5 & 98.2 & 0.054 &      39.4 \\
& GMP-E  & 0.041 & 0.059 & 91.2 & {\bf 98.6} & {\bf 99.1} & 0.036 &      62.2 \\

\end{tabular}
}
\end{center}
\caption{Performance metrics for Logistic regression. }
\label{tab:logistic_reg}
\end{table}

Table~\ref{tab:logistic_reg} shows broadly similar behaviour to the probit results in Table~2 of the main text. MFVB tends to underestimate posterior uncertainty and gives substantially lower marginal density accuracies on several datasets. EP and the GMP variants are generally the most accurate methods. In particular, GMP often improves the marginal density accuracy of the approximation used for initialization, while remaining much faster than the short-run MCMC benchmark.

As in the probit case, the quality of GMP depends on the initialization, and no single initialization dominates across all datasets and metrics.

Unlike the probit results in the main text, there is no auxiliary-variable MP entry for logistic regression; the MP-based entries here are the GMP variants. We leave such a variant for future work.

\newpage 

\section{Bayesian Lasso details}\label{Suppl:Sec_Bayes_Lasso}

This appendix gives the MFVB and GVB approximations used to initialize the GMP updates for the Bayesian Lasso example in Sections~\ref{sec:gmp_lasso} and~\ref{sec:BayesLasso_examples}.

\subsection{Model and latent-variable representation}

We consider the model
%\begin{equation}\label{eq:response}
%\begin{array}{c}
$\displaystyle y_i \mid \vbeta,\sigma^2 \sim N(\vx_i^T\vbeta,\sigma^2)$
%\end{array}
%\end{equation}
with
\begin{equation}\label{eq:lasso}
\beta_j \mid \sigma^2 \sim \mbox{Laplace}\left( 0, \frac{\sigma}{\lambda} \right) 
\qquad \mbox{that is}
\qquad 
p(\beta_j \mid \sigma^2) =  \frac{\lambda}{2\sigma} \exp\left(-\frac{\lambda}{\sigma}  |\beta_j|\right).
\end{equation}

\noindent
We treat $\lambda^2 > 0$ as fixed to simplify the analysis. The parameters of interest are $\boldsymbol{\theta} = (\vbeta, \sigma^2)$. We assume that $\vy$ and $\mX$ are centered so that, without loss of generality, the intercept can be ignored. Since (\ref{eq:lasso}) is difficult to handle directly, we instead adopt an auxiliary representation. This entails replacing (\ref{eq:lasso}) with
\begin{equation}\label{eq:lasso_rep_a}
\begin{array}{c}
\ds \beta_j \mid \sigma^2, a_j \sim N\left( 0, \frac{\sigma^2}{a_j\lambda^2}  \right),
\qquad 
\ds a_j \stackrel{\mbox{\scriptsize iid}}{\sim} \mbox{IG}(1,1/2), \quad  1\le j\le p,
\end{array}
\end{equation}

\noindent \textcite{ParkCasella2008} show that we need to condition $\beta_j$ on $\sigma^2$ for the regression coefficient posteriors to be unimodal. The above is equivalent to \textcite{ParkCasella2008}, where we have shifted the dependence of $a_j$ on $\lambda^2$ to having $\beta_j$ depending on $\lambda^2$.
\noindent Finally, we use $\sigma^2 \sim \mbox{IG}(A,B)$ as the (conjugate) prior for $\sigma^2$.

\subsection{Mean Field Variational Bayes}

The VB approximation corresponding to the factorization
$q(\vbeta,\sigma^2,\va) = q(\vbeta)q(\sigma^2)q(\va)$
leads to 
$$
\begin{array}{rl}
q(\vbeta) 
& \ds = N\left( \vmu\equiv 
\mQ\mX^T\vy,
\mSigma \equiv (\widetilde{B}/\widetilde{A}) \mQ
\right) \qquad \mbox{and}

\\ [2ex]

q(\sigma^2) 
& \ds= \mbox{IG}\left( 
\widetilde{A} \equiv A + \tfrac{n+p}{2},
\widetilde{B} \equiv B + \tfrac{1}{2}\|\vy - \mX\vmu\|^2 
+ \tfrac{\lambda^2}{2}\vmu^T\mD\vmu 
+ \tfrac{1}{2}\mbox{tr}\left( \mSigma(\mX^T\mX + \lambda^2 \mD)\right) 
\right). 

%\\ [2ex]
% \ds q(\lambda^2) 
% & \ds = \mbox{Gamma}\left( 
% \widetilde{u} \equiv u + \frac{p}{2},
% \widetilde{v} \equiv v + \frac{\widetilde{a}}{2\widetilde{b}} (\vmu^T\mD\vmu+\mbox{tr}(\mD\mSigma)) \right) 
\end{array} 
$$

\noindent Since $p(\va|\sD,\vbeta,\sigma^2) = \prod_{j=1}^p p(a_j|\sD,\beta_j,\sigma^2)$
we have $q(\va) = \prod_{j=1}^p$
and so we have independently
$$
\begin{array}{rl}
% q(a_j) 
% & \ds = \mbox{GIG}\left( 
% A_j \equiv 1, 
% B_j \equiv \bE_q(\lambda^2) \frac{\widetilde{a}(\mu_j^2 + \sigma_j^2)}{\widetilde{b}}, 
% P_j \equiv \frac{1}{2} \right)

% \\ [2ex]

% \mbox{or} \qquad 
q(a_j) & \ds = \mbox{Inverse-Gaussian}\left( 
d_j \equiv \sqrt{ \frac{\widetilde{B}/\widetilde{A}}{\lambda^2(\mu_j^2 + \Sigma_{jj}^2)}}, 
\lambda_j \equiv 1 \right)
 
 \\ [2ex]

\end{array} 
$$

\noindent where $\mQ = (\mX^T\mX + \lambda^2\mD)^{-1}$, 
%$$
%d_j  = \bE_q(a_j^{-1}) =  \sqrt{\frac{\widetilde{b}}{\bE_q(\lambda^2)\widetilde{a} (\mu_j^2 + \sigma_j^2)}},
%$$
%
%\noindent and 
$\mD = \mbox{diag}(\bE_q(\va)) = \mbox{diag}(\vd)$.
Here we use the inverse-Gaussian parameterization with density proportional to
$$
x^{-3/2}\exp\left\{-\frac{\lambda(x-m)^2}{2m^2x}\right\},
\qquad x>0,
$$

\noindent 
where $m$ is the mean and $\lambda$ is the shape.

\subsection{Gaussian Variational Bayes}

Consider the model representation (\ref{eq:lasso}) and suppose that $q(\vbeta,\sigma^2) = q(\vbeta)q(\sigma^2)$
with $q(\vbeta) = N(\vmu,\mSigma)$. Then (ignoring additive constants)
$$
\begin{array}{rl}
\ds \log p(\vy,\vbeta,\sigma^2)
& \ds = - \frac{1}{2\sigma^2} \|\vy - \mX\vbeta\|_2^2 
        - \frac{\lambda}{\sigma}\|\vbeta\|_1  
    - \left( A + \frac{n+p}{2} + 1 \right) \log(\sigma^2)
    - B\sigma^{-2}.

\end{array} 
$$

\noindent For fixed $q(\sigma^2)$, given $q(\vbeta) = N_p(\vmu,\mSigma)$ the optimal values of $\vmu$ and $\mSigma$ maximize (ignoring additive constants) 
$$
\mbox{ELBO} = - \tfrac{1}{2} c_1 \left[ \|\vy - \mX\vmu\|_2^2 + \mbox{tr}(\mX^T\mX\mSigma) \right]
        - c_2 \bE_q\left( \|\vbeta\|_1 \right)  + \tfrac{1}{2}\log|\mSigma|
$$

\noindent were $c_1 = \bE_q(\sigma^{-2})$ and $c_2 = \lambda \bE_q(\sigma^{-1})$.
For $\beta_j\sim N(\mu_j,\Sigma_{jj})$, let
$s_j=\sqrt{\Sigma_{jj}}$, and 
$r_j=\mu_j/s_j$ then
$$
\bE_q|\beta_j|
=
\mu_j\{\Phi(r_j)-\Phi(-r_j)\}
+
2s_j\phi(r_j).
$$

% noting that
% $$
% \bE_q(|\beta_j|) = \mu_i \mbox{erf}\left( \frac{\widetilde{\mu}_i}{\sqrt{2}} \right)
% + 2\sigma_i \phi(\widetilde{\mu}_i)
% = \mu_i \left[ 
% \Phi(\mu_i\Sigma_{ii}^{-1/2}) - \Phi(-\mu_i \Sigma_{ii}^{-1/2})
% \right] + 2\Sigma_{ii}^{1/2} \phi(\mu_i \Sigma_{ii}^{-1/2})
% $$

\noindent and
$$
\frac{\partial \bE_q|\beta_j|}{\partial \mu_j} = \Phi(r_j) - \Phi(-r_j).
$$

Then
$$
\begin{array}{rl}
\ds \frac{\partial 
\mbox{ELBO}
}{
    \partial \vmu
}
    & \ds = c_1\mX^T(\vy - \mX\vmu) - c_2\left[ \Phi(\vr) - \Phi(-\vr)  \right] \qquad \mbox{and}
\\ [2ex]

\ds \frac{\partial^2 
    \mbox{ELBO}
}{
    \partial \vmu \partial\vmu^T
}
& \ds = - c_1\mX^T\mX - c_2 \mbox{diag}\left[ 
2 \mbox{dg}(\mSigma)^{-1/2} \odot \phi(\vr) 

\right].
\end{array}
$$

\noindent We also have
$$
\ds \frac{\partial \mbox{ELBO}}{\partial \Sigma_{ij}} = \tfrac{1}{2}\mbox{tr}\left[ 
\left\{  \mSigma^{-1} 
- c_1\mX^T\mX 
- c_2\,\mbox{diag}\left( 2 \mbox{dg}(\mSigma)^{-1/2}\odot \phi(\vr)
\right) \right\} \mE_{ij} \right]. 
$$

\noindent This leads to the updates
\begin{itemize}
    \item $\mSigma \leftarrow \left[ c_1\mX^T\mX 
+ c_2\,\mbox{diag}\left( 2 \mbox{dg}(\mSigma)^{-1/2}\odot \phi(\vr)
\right)  \right]^{-1}$

\item $\vmu \leftarrow \vmu + \mSigma \left[ c_1\mX^T(\vy - \mX\vmu) - c_2\left\{ \Phi(\vr) - \Phi(-\vr)  \right\} \right]$.
\end{itemize}

\noindent 
These equations are solved by fixed-point iteration, using the current values of
$\vmu$ and $\mSigma$ on the right-hand side.
We found that this fixed point iteration can be unstable and usually requires the algorithm to be initialized at good starting values. However,
this fixed point approach is much faster than
optimizing the ELBO directly with respect to $\mSigma$.

The optimal $q(\sigma^2)$ is of the form
$$
\begin{array}{rl}
q(\sigma^2)
%& \ds \propto \exp\left[ - \left( A + \tfrac{n+p}{2} + 1 \right) \log(\sigma^2)
%- \left\{ \lambda \cdot \bE_q\left( \|\vbeta\|_1 \right) \right\} \sigma^{-1} 
%- \left\{ B + \tfrac{1}{2} \bE_{q(\vbeta)}\left( \|\vy - \mX\vbeta\|_2^2 \right) 
%\right\} \sigma^{-2} 
        
%\right] 

%\\  

& \ds = Z^{-1} \exp\left[ - (\widetilde{A} + 1)\log(\sigma^2)  - \widetilde{B}\sigma^{-1} - \widetilde{C}\sigma^{-2} \right]
\end{array}
$$

\noindent where 
$\widetilde{A} = A + \frac{n + p}{2}$, 
$\widetilde{B} = \lambda \cdot \bE_q\left( \|\vbeta\|_1 \right)$,
%\qquad % \mbox{and} \qquad 
$\widetilde{C} = B + \tfrac{1}{2} \bE_{q(\vbeta)}\left( \|\vy - \mX\vbeta\|_2^2 \right)$,
and 
$$
\begin{array}{rl}
Z
& \ds = \int_0^\infty \exp\left[ - (\widetilde{A} + 1)\log(t) - \widetilde{B}t^{-1/2} - \widetilde{C}t^{-1}  \right] dt
\end{array}
$$

\noindent with $t=\sigma^2$. The $q$-density is not the usual inverse-Gamma distribution due to the $\sigma^{-1}$ term appearing and needs to be handled numerically.
The normalizing constant $Z$ and the moments
$\bE_q(\sigma^{-2})$, $\bE_q(\sigma^{-1})$,  and $\bE_q(\sigma^2)$
are evaluated numerically using a composite trapezoid rule.

% intentionally left blank

\printbibliography[title={Supplementary References}]

\end{refsection}

\end{document}